\pdfoutput=1
\documentclass[%
 aip,
 jmp,
 amsmath,amssymb,
 reprint,%
]{revtex4-1}

\usepackage{graphicx}
\usepackage{dcolumn}
\usepackage{bm}
\usepackage{tikz}
\usepackage[utf8]{inputenc}
\usepackage[T1]{fontenc}
\usepackage{mathptmx}

\usepackage{amsthm}
\theoremstyle{definition}

\newtheorem{theorem}{Theorem}[section]
\theoremstyle{remark}

\newtheorem{proposition}{Proposition}[section]
\newtheorem{lemma}{Lemma}[section]

\begin{document}

\preprint{AIP/123-QED}

	\title{Graph rules for the linked cluster expansion  
		of the Legendre effective action}

\author{R. Banerjee}
\email{rub18@pitt.edu.}
\author{M.Niedermaier}%
 \email{mnie@pitt.edu.}
\affiliation{ 
Department of Physics and Astronomy\\
		University of Pittsburgh, 100 Allen Hall\\
		Pittsburgh, PA 15260, USA
}%


\date{\today}

\begin{abstract}
Graph rules for the linked cluster expansion of the 
		Legendre effective action $\Gamma[\phi]$ are derived and proven
		in $D \geq 2$ Euclidean dimensions. 
		A key aspect is the weight assigned to articulation 
		vertices which is itself shown to be computable from 
		labeled tree graphs. The hopping interaction is allowed to 
		be long ranged and scale dependent, thereby producing an in 
		principle exact solution of $\Gamma[\phi]$'s functional 
		renormalization group equation.    
\end{abstract}

\maketitle

	\section{Introduction} 
	
	The Legendre effective action is a central quantity in 
	all areas of many-body quantum physics. In particular, it
	features prominently in the functional renormalization 
	group approach based on the flow equation 
	\begin{equation} 
	\label{i1} 
	\partial_k \Gamma_k = \frac{1}{2} {\rm Tr}\big\{\partial_k R_k 
	[ \Gamma^{(2)}_k + R_k]^{-1}\big\}\,, 
	\end{equation}
	describing its response to a modulation of the system's mode content 
	set by the kernel $R_k$. The flow equation (\ref{i1})  
	is now being used in fields as diverse as: solid state physics, 
	statistical physics, and quantum gravity, see \cite{FRGbook1,FRGbook2,FRGbook3} for 
	book-sized accounts. The response (\ref{i1}) is itself kinematical 
	in nature, dynamical information in injected exclusively through 
	initial conditions. As a consequence, the results 
	obtained are only as non-perturbative as the initial conditions are. 
	An especially interesting choice of initial conditions are ultralocal ones 
	as they can in a lattice formulation be be computed exactly
	from single site integrals \cite{DupuisMachado}. A solution of (\ref{i1}) 
	with such initial data, if feasible, will emulate a linked cluster or hopping 
	expansion but with a scale dependent long-ranged interaction 
	\begin{equation} 
	\label{i2}
	S[\chi] = \sum_x s[\chi_x] + \frac{\kappa}{2} 
	\sum_{x,y} \chi_y \ell_{xy}(k) \chi_x\,. 
	\end{equation}
	For definiteness we consider a self-interacting 
	scalar field theory on a $D$-dimensional hypercubic lattice
	(identified with $\mathbb{Z}^D$) in a dimensionless formulation. 
	Here, $s:\mathbb{R} \rightarrow \mathbb{R}$ is a function bounded from below that collects 
	all terms from the original lattice action referring to a 
	single site. The hopping parameter $\kappa>0$ arises 
	as a dimensionless combination of the original mass and coupling 
	parameters and the lattice spacing. A fundamental 
	lattice action would only connect nearest neighbors
	on the lattice through $\ell_{xy}$. In order to obtain a 
	solution of (\ref{i1}) we allow $\ell_{xy}$ to be long-ranged      
	and be modulated by the control parameter $k$. The 
	details of the modulation are inessential in the following as we 
	take $\kappa$ as the control parameter and replace 
	(\ref{i1}) by 
	\begin{eqnarray}
	\label{i3}
	\partial_{\kappa} \Gamma_{\kappa} = 
	\frac{1}{2} \sum_{x,y} \ell_{xy} \big[ \Gamma_{\kappa}^{(2)} + \kappa \ell 
	\big]_{xy}^{-1} \,,\quad 
	\Gamma_{\kappa} = \Gamma_0 + \sum_{l \geq 2} \kappa^{l} 
	\Gamma_{l}[\phi]\,.
	\end{eqnarray}
	Here $\Gamma_0[\phi] = \sum_x \gamma(\phi_x)$, where $\gamma$ and its derivatives $\gamma_n$
	are computable at a single site from $s$ only. The $O(\kappa)$ 
	term vanishes, $\Gamma_2[\phi] 
	= -\frac{1}{4} \sum_{x,y} (\ell_{xy})^2 \gamma_2(\phi_x)^{-1} \gamma_2(\phi_y)^{-1}$, and all 
	$\Gamma_{l},\, l\! \geq \!3$, are then determined recursively;
	see (\ref{rec2}) in Appendix A.  Importantly the series can be expected to have finite 
	radius of convergence $\kappa < \kappa_c$; see the discussion below. Once the series (\ref{i3}) has been constructed, 
	an in principle exact solution of (\ref{i1}) arises 
	simply by substitution, $\Gamma_k = \Gamma_{\kappa}|_{\ell \mapsto \ell(k)}$. 
	
	The direct iteration (\ref{rec2}) becomes, however, impractical 
	beyond $O(\kappa^6)$ or so (both in manual computations and in 
	automated symbolic implementations). The repeated functional 
	differentiations of $\Gamma_0[\phi]$ leads to site identifications 
	whose combinatorics is best recast in graph theoretical terms. 
	The graph theoretical analysis of hopping expansions 
	of course has a long history, see \cite{Wortis,Clusterbook1,Clusterbook2} 
	and the references therein. The convergence proofs of generalized 
	Mayer expansions typically rely on tree graph bounds \cite{BK}.
	In the computational uses of linked cluster expansions, the
	focus is normally on nearest neighbor interactions and 
	quantities of direct interest for critical behavior
	like generalized susceptibilities \cite{Clusterbook1,Clusterbook2}; 
	a convergence proof for them in scalar quantum field theory
	can be reduced to tree graph bounds \cite{ReiszLRH}. 
	The effective action $\Gamma_{\kappa}$ can alternatively 
	be defined as a (slightly modified) Legendre transform of 
	the free energy functional $W_{\kappa}$. Graph theoretical rules 
	for the linked cluster expansion of $W_{\kappa}$ have originally 
	been presented by Wortis \cite{Wortis} and will be briefly 
	reviewed in Section 2. Graph theoretical expansions for $\Gamma_{\kappa}[\phi]$ have
	been discussed previously but do not cover the material presented here: 
	the rules and results of \cite{GammIsing1,GammIsing2} hinge on specific 
	features of the Ising model which do not generalize. Hybrid 
	perturbative expansions have been considered in \cite{GamFeyn}. 
	A relevant combinatorial Legendre transform has 
	been studied in \cite{Jackson,Faris} in a setting that emulates
	perturbation theory. Some of the results of \cite{Jackson} will 
	reoccur in our analysis of $\Gamma_0$ in Appendix B. There are also 
	abstract variants of a Legendre transform  formulated in terms of 
	combinatorial species \cite{BrydgesLeroux,Faris}. None of these 
	seem to bear an obvious relation to our result.
	
	We present the  solution of the recursion implied by (\ref{i3}) in graph theoretical 
	terms. Let $\mathcal{L}_{l}$ be the set of one-line irreducible connected 
	graphs with $l$ edges. For any $L = (V,E) \in \mathcal{L}_{l}$ and 
	any vertex $v \in V$ consider the decomposition of $L$ into 
	 one-vertex irreducible subgraphs, $|I(v)|$ of which each contain a copy of 
	$v$. The set $B(v)$ of copies is used to label a class of {\it tree} 
	graphs $\mathcal{T}(B(v),n), n=1,\ldots, |I(v)|$. To each $T \in \mathcal{T}(B(v),n)$ 
	two integers $s(T)$ and ${\rm Perm}B(v)/{\rm Sym}(T)$ are assigned,
	as detailed in Section III. Then:

	{\bf Theorem.} For any $l \geq 2$ the exact solution of the recursion implied by
	(\ref{i3}), i.e. (\ref{rec2}), is given by 
	\begin{eqnarray}
	\label{i4}
	\nonumber
	\Gamma_{l}[\phi] & = & \sum_{L= (V,E) \in \mathcal{L}_{l}} \frac{(-)^{l+1}}{{\rm Sym}(L)} 
	\prod_{e \in E} \ell_{\theta(e)} \prod_{v \in V}  \mu^{\Gamma}(v|B)
	\\
	\mu^{\Gamma}(v|B) & = &  \sum_{n=1}^{|I(v)|} \sum_{T \in \mathcal{T}(B(v),n)} (-)^{s(T)} 
	\frac{|{\rm Perm}(B(v))|}{{\rm Sym}(T)} \mu(T)\,.
	\end{eqnarray}
	In the first line an unconstrained sum over the lattice points associated 
	with the vertices is tacit. Further $E$ is the edge list with 
	$\theta(e)$ the pair of vertices connected by $e$,  and 
	${\rm Sym}(L)$ is the symmetry factor of $L$.    
	In the second line, $\mu(T)$ is a weight 
	depending only on the value of $\phi$ at $v$. 
	
	The paper is organized as follows. In Section II.A we summarize 
	known graph rules for the free energy functional and set the terminology. A mixed recursion relation (\ref{GamvsW4}), equivalent to the  one implied by (\ref{i3}) is derived in Section II.B, and used to derive the first line of (\ref{i4}). The relevant class of labeled
	tree graphs is introduced in Section III.A, the graph rules for 
	$\mu^{\Gamma}(v|L)$ are formulated and illustrated in Section III.B, 
	and an all-order proof for their validity is given in Section III.C.
	A simplified version of $\mu^{\Gamma}(v|B)$ obtained by 
	performing subsums with fixed $\mu(T)$ is derived in Section IV.
	Appendix A presents explicit and independently computed results 
	for $\Gamma_2,\ldots, \Gamma_5$. Appendix B discusses the 
	single site data and their combinatorics. 
	
	\newpage
	
	\section{From connected graphs to articulation vertices} 
	
	For convenience we refer to expansions in powers of $\kappa \ell_{xy}$ (with 
	$\ell_{xx} =0$ but $\ell_{xy} \neq 0$ for ${\rm dist}(x,y) \geq 1$) as 
	a long range hopping (LRH) expansion. The graph expansions considered
	have two main ingredients: first, a class of graphs with some partial 
	order consistent with the order in $\kappa$. Second, a weight function 
	that assigns to each graph of the class a numerical value depending 
	on certain input data. In addition to $\ell_{xy}$ itself, the input data 
	are always the derivatives $\omega_m(h) = \partial^m \omega/\partial h^m$ and/or 
	$\gamma_m(\varphi) = \partial^m \gamma/\partial \varphi^m$ of the single site 
	functions described in Appendix B. The class of graphs and the 
	weight functions will depend on the quantity considered.   
	The goal of this section is to reduce the problem of identifying 
	the graph rules for $\Gamma_{\kappa}$'s LRH expansion to the 
	determination of the weight associated to articulation vertices.
	
	\subsection{Basics} 
	
	The $\Gamma_k$ flow equation (\ref{i1}) can be obtained as the 
	Legendre transform of a Polchinski-type flow equation for $W_k$,
	the mode modulated free energy functional. For an action of the 
	form (\ref{i2}) one may again take $\kappa$ as the control parameter
	to obtain along the familiar lines
	\begin{equation}
	\label{wrec1}
	\partial_{\kappa} W_{\kappa}[H] = -\frac{1}{2} \sum_{x,y} \ell_{xy} 
	\bigg\{ \frac{\delta^2 W_{\kappa}[H]}{\delta H_x \delta H_y} + 
	\frac{\delta W_{\kappa}[H]}{\delta H_x} 
	\frac{\delta W_{\kappa}[H]}{\delta H_y} 
	\bigg\}\,.
	\end{equation}
	Here we impose ultralocal initial data $W_0[H] = \sum_x \omega(H_x)$, 
	where $\omega(h)$  is determined by the single site action $s$ in (\ref{i2}). 
	The ansatz $W_{\kappa}[H] = W_0[H] + \sum_{l \geq 1} \kappa^{l} W_{l}[H]$
	converts (\ref{wrec1}) into the recursive system 
	\begin{equation}
	\label{wrec2}
	W_{l+ 1}[H] = -\frac{1}{2(l\!+\!1)} \sum_{x,y} \ell_{xy} 
	\bigg\{ \frac{\delta^2 W_{l}[H]}{\delta H_x \delta H_y} + 
	\sum_{k=0}^{l} \frac{\delta W_{k}[H]}{\delta H_x} 
	\frac{\delta W_{l-k}[H]}{\delta H_y} 
	\bigg\}\,,\quad l \geq 0\,.
	\end{equation}
	Explicitly, the first two orders read 
	\begin{eqnarray}
	\label{wrec3}
	W_{1}[H]&=& -\frac{1}{2} \sum_{x,y} \ell_{xy}  \frac{\delta W_{0}[H]}{\delta H_x} 
	\frac{\delta W_{0}[H]}{\delta H_y} \,,
	\nonumber\\
	W_{2}[H]&=& \frac{1}{2} \sum_{x,y,z,w} \ell_{xy} \ell_{zw}\,
	\frac{\delta^2 W_{0}[H]}{\delta H_y \delta H_w}\,\bigg\{ 
	\frac{1}{2}\, \frac{\delta^2 W_{0}[H]}{\delta H_x \delta H_z}+
	\frac{\delta W_{0}[H]}{\delta H_x} 
	\frac{\delta W_{0}[H]}{\delta H_z} 
	\bigg\}\,.
	\end{eqnarray}
	The repeated $H_x,H_y, \ldots$, functional derivatives of $W_0[H]$ produce point 
	identifications and coefficients that are source-dependent derivatives of the 
	single-site generating function $\omega(h)$. The combinatorics of these 
	point identifications is best formulated in graph theoretical terms.  
	Such rules have been formulated and proven by Wortis \cite{Wortis};
	the relation to a Polchinski-type flow equation was noted in 
	\cite{BK} where subject to additional conditions also a convergence proof 
	is given. 
	
	{\bf Graph rules for W[H]:} 
	\vspace{-2mm} 
	\begin{itemize} 
		\item[(a)] At order $l\geq 1$ in $\kappa$ draw all topologically distinct connected 
		graphs $C = (V,E) \in \mathcal{C}_{l}$ with $l = |E|$ edges connecting 
		$2, \ldots, l+1$ vertices. Assign a dummy label to each vertex. 
		\item[(b)] Divide by the symmetry factor 
		${\rm Sym}(C)$ of the graph. 
		\item[(c)] To each graph a weight $\mu^W(C)$ is assigned as follows: a vertex $i$ 
		of degree $n$ is attributed a weight $\omega_n(H_i)$, an edge 
		connecting $i,j$ is attributed a factor $-\ell_{ij}$. 
		\item[(d)] Embed the graph into the lattice $\mathbb{Z}^D$ by associating each vertex 
		with a unique lattice point, $i \mapsto x_i$, $i =1,\ldots, |V|$, the same 
		lattice point may occur several times. Perform an unconstrained sum over all 
		$x_1, x_2, \ldots, x_{|V|}$. 
	\end{itemize} 
	
	For illustration consider the graphs in (a) divided by their 
	symmetry factors in (b) to $O(\kappa^3)$:
	\begin{eqnarray}
	\label{wgraph1} 
	W_{\kappa}[H]& \stackrel{(a),(b)}{=}&\,
	\begin{tikzpicture}
	\filldraw[black] (0,0) circle (2pt) ;
	\end{tikzpicture}\,
	-\frac{1}{2}\,
	\begin{tikzpicture}
	\draw[black, thick] (0,0) -- (1,0);
	\filldraw[black] (0,0) circle (2pt) ;
	\filldraw[black] (1,0) circle (2pt) ;
	\end{tikzpicture}
	+\frac{1}{4}\,
	\begin{tikzpicture}
	\draw[black, thick] (0,0) edge[bend left=30] (1,0);
	\draw[black, thick] (0,0) edge[bend right=30] (1,0);
	\filldraw[black] (0,0) circle (2pt) ;
	\filldraw[black] (1,0) circle (2pt) ;
	\end{tikzpicture}\,
	+\frac{1}{2}\,
	\begin{tikzpicture}
	\draw[black, thick] (0,0) -- (0.5,0.5);
	\draw[black, thick] (0.5,0.5) -- (1,0);
	\filldraw[black] (0,0) circle (2pt) ;
	\filldraw[black] (0.5,0.5) circle (2pt) ;
	\filldraw[black] (1,0) circle (2pt) ;
	\end{tikzpicture}\,
	\nonumber	\\[2mm]
	&-&\frac{1}{12}\,
	\begin{tikzpicture}
	\draw[black, thick] (0,0) -- (1,0);
	\draw[black, thick] (0,0) edge[bend left=30] (1,0);
	\draw[black, thick] (0,0) edge[bend right=30] (1,0);
	\filldraw[black] (0,0) circle (2pt) ;
	\filldraw[black] (1,0) circle (2pt) ;
	\end{tikzpicture}\,
	-\frac{1}{2}\,
	\begin{tikzpicture}
	\draw[black, thick] (0,0) edge[bend left=30] (0.5,0.5);
	\draw[black, thick] (0,0) edge[bend right=30] (0.5,0.5);
	\draw[black, thick] (0.5,0.5) -- (1,0);
	\filldraw[black] (0,0) circle (2pt) ;
	\filldraw[black] (0.5,0.5) circle (2pt) ;
	\filldraw[black] (1,0) circle (2pt) ;
	\end{tikzpicture}\,
	-\frac{1}{6}\,
	\begin{tikzpicture}
	\draw[black, thick] (0,0) -- (0.5,0.5);
	\draw[black, thick] (0.5,0.5) -- (1,0);
	\draw[black, thick] (0,0) -- (1,0);
	\filldraw[black] (0,0) circle (2pt) ;
	\filldraw[black] (0.5,0.5) circle (2pt) ;
	\filldraw[black] (1,0) circle (2pt) ;
	\end{tikzpicture}\,
	-\frac{1}{2}\,
	\begin{tikzpicture}
	\draw[black, thick] (0,0) -- (0.5,0.5);
	\draw[black, thick] (0.5,0.5) -- (1,0);
	\draw[black, thick] (1,0)-- (1.5,0.5);
	\filldraw[black] (0,0) circle (2pt) ;
	\filldraw[black] (0.5,0.5) circle (2pt) ;
	\filldraw[black] (1,0) circle (2pt) ;
	\filldraw[black] (1.5,0.5) circle (2pt) ;
	\end{tikzpicture}\,
	-\frac{1}{6}\,\begin{tikzpicture}[scale=1.1]
	\draw[black, thick] (0,0) -- (0.25,0.25);
	\draw[black, thick] (0.25,0.25) -- (0.5,0);
	\draw[black, thick] (0.25,0.25)-- (0.25,0.6);
	\filldraw[black] (0,0) circle (2pt) ;
	\filldraw[black] (0.25,0.25) circle (2pt) ;
	\filldraw[black] (0.5,0) circle (2pt) ;
	\filldraw[black] (0.25,0.6) circle (2pt) ;
	\end{tikzpicture}\,
	+O(\kappa^4).
	\\\nonumber
	\end{eqnarray}
	Upon application of parts (c),(d) this matches the recursively computed result.
	Generally, the graph rule can be recast symbolically as: 
	\begin{equation} 
	\label{wgraph2} 
	W_{l}[H] = 
	\sum_{C = (V,E) \in \mathcal{C}_{l}}
	\frac{(-)^{l}}{{\rm Sym}(C)}  
	\prod_{e \in E} \ell_{\theta(e)} \prod_{v \in V}
	\omega_{ d(v)}(H_v) \,,
	\end{equation}
	where the lattice summations from step (d) are tacit 
	and the double product comprises $\mu^W(C)$.    
	The graph sum is over all connected graphs $C = (V,E)$ with 
	$|E| = l$ edges, $d(v)$ is the degree of the vertex $v$,
	$\theta(e)$ is the pair of vertices $e\in E$ connects. A recent algorithm that generates these graphs can be found in \cite{MO}.
	The symmetry factor ${\rm Sym}(C)$ of $C$ is defined below. Since 
	also the graph terminology is not entirely standardized we 
	compiled a brief glossary at the end of this subsection. 
	
	Once $W[H]$ is known to some order, the connected 
	correlation functions (or cumulants) can be obtained by 
	differentiation. It is plain from (\ref{wgraph2}) that the   
	cumulants $(W^{(k)}[H])_{y_1,\ldots, y_k}$, $y_1\neq \ldots \neq y_k$,
	also have a graph expansion and that the contributing  graphs 
	are $k$-rooted, i.e.~have $k$ external vertices eventually 
	labeled by $y_1,\ldots, y_k$. The relevant symmetry factor
	thus is that of the $k$-rooted graph, where the isomorphisms 
	have to leave the external vertices individually invariant.    
	The edges are assigned a $-\ell_{ij}$ factor as before, also 
	for edges where one of the vertices is an external vertex. 
	The vertex weight can always be obtained by differentiation
	from the $\omega_{d(v)}(H_v)$ product in (\ref{wgraph2}).
	
	{\bf A brief graph glossary:} 
	
	A {\it graph} is a pair $G = (V,E)$ of nonempty disjoint sets 
	equipped with a map $\theta$ that associates to each  $e\in E$ an unordered pair $\theta(e)=\{v,w\}$,  $v,w\in V$. The elements of $V$ are called vertices (or nodes), those of $E$ are called edges (or links, or lines). This definition allows for several edges to be mapped into the same unordered pair of vertices, in which case the edges are called multiple edges. Otherwise the graph is called simple, in which case we shall identify $E$ with a subset of  $V_2:=\big\{ \{v,w\}:\, v,w\in V\big\}$. The degree (or valency or number of 
	incident lines) $d(v)$ 
	of a vertex $v \in V$ is the cardinality of the set 
	$\{ e \in E:\,   v \in \theta(e)\}$. 
	If $|V|_k$ is the cardinality of $\{ v \in V: \, d(v) = k\}$ 
	one has $2 |E| = \sum_k k |V|_k$. 
	
	Let $(V,E)$ be a graph. A {\it trail} from $v$ to $w$,
	$v,w \in V$ is a sequence $v_0, e_1, \ldots, e_n, v_n$ 
	with $v_0 = v,\, v_n = w$, such that the edges $e_i$ are distinct and $\theta(e_i)=\{v_{i-1}, v_{i}\}$.
	A graph is connected if for every pair of its vertices $v,w$   
	there is a trail from $v$ to $w$. A connected component 
	of $G$ is a maximal connected subgraph of $G$.
	A trail from $v$ to $w$ such that $v$ and $w$ coincide is called
	a cycle. The cyclomatic number $c(G)$ is the number of 
	cycles of a graph $G$. The Euler relation states
	\begin{equation}
	\label{Euler} 
	c(G) = |E| - |V| + 1\,.      
	\end{equation}
	A {\it tree} $T$ is a connected simple graph without cycles;
	in particular $|V| - |E| =1$ holds.  
	
	Two graphs $(V,E)$ and $(V',E')$, with respective maps $\theta,\,\theta'$, are called isomorphic
	(or {\it topologically equivalent}) if there exist bijections 
	$\pi_1 : V \rightarrow V'$, $\pi_2:E\rightarrow E'$ such 
	that $\theta(e)=\{v,w\}$ {\it iff}  $\theta'(\pi_2(e))=\{\pi_1(v),\pi_1(w)\}$. These isometries form a group, ${\rm Aut}(G)$, which with the above definition included permutations of multiple edges. The {\it symmetry factor} of $G$ is defined by
	\begin{equation}
	\label{Symdef}
	{\rm Sym}(G) = |{\rm Aut}(G)|\,. 
	\end{equation}
	\noindent Often the automorphism group refers to the corresponding simple graph only, in which case the permutation of multiple edges occurs as an extra factor in the definition of the symmetry factor \cite{Wortis,ReiszLRH}.
	
	The same notion of isometry  applies if the elements of a 
	subset $R \subset V$, called 
	the rooted vertices,  are left individually invariant 
	by the bijection. The elements of $R$ can be viewed as 
	distinguishable and labeled, $R = \{r_1,\ldots , r_k\}$, 
	in which case $G$ is called $k$-rooted. 
	
	A graph $G' = (V',E')$ is called a subgraph of $G = (V,E)$ 
	if $V'\subset V$ and $E' \subset E$. For a graph $G$ let 
	$G\backslash \{v\}$ be the subgraph obtained by deleting $v$ 
	and all edges containing $v$. For a connected graph $G = (V,E)$ 
	a vertex $v \in V$ is called an {\it articulation point} if the 
	$G\backslash \{v\}$ is disconnected. A connected graph without 
	articulation points is called one-vertex irreducible ({\bf 1VI})
	(or two-connected). For a connected graph $G$ a {\it block} $G'$ 
	is a maximal 1VI subgraph, i.e.~a graph $G'\subset G$ that 
	is 1VI and such that for any 1VI subgraph $G''$ the inclusion 
	$G' \subset G'' \subset G$ entails $G'' = G'$. The set of 
	blocks $\{G_1,\ldots, G_k\}$, $G_i = (V_i,E_i)$, $i=1,\ldots,k$, 
	of a connected graph $G = (V,E)$ is referred to as $G$'s block 
	decomposition \cite{Viral}. The blocks induce a partition of the edge set 
	$E = E_1 \cup \ldots \cup E_k$, with $E_i \cap E_j = \emptyset$, 
	$i \neq j$. Each articulation point belongs to more than one $V_i$ 
	while non-articulation vertices belong to exactly one.

	A bridge in a connected graph is an edge whose omission produces 
	a disconnected graph. A one-line irreducible ({\bf 1LI}) graph is a 
	bridgeless connected graph. A one-line irreducible graph may still 
	get disconnected upon removal of a vertex. The block decomposition 
	of 1LI graphs will be central later on.   
	
	\subsection{The role of one-line and one-vertex irreducible graphs}

	Our task will be to convert the above $W$-graph rules into ones directly 
	applicable to the $\Gamma_{\kappa}$ expansion defined by (\ref{i3}). 
	Both functionals are related by the following modified Legendre transform 
	\begin{equation}
	\label{GamvsW1}  
	\Gamma_{\kappa}[\phi] := \phi\cdot H_{\kappa}[\phi] - W_{\kappa}[H_{\kappa}[\phi]] - 
	\kappa \mathcal{V}[\phi]\,,  
	\quad 
	\frac{\delta W_{\kappa}}{\delta H}\big( H_{\kappa}[\phi] \big) = \phi\,,
	\end{equation}
	for a $\kappa$-independent mean field $\phi$. The modification by 
	the $\mathcal{V}[\phi] := \frac{1}{2}\sum_{x,y} \phi_x \ell_{xy} \phi_y$ term  
	is introduced so as to obtain the closed flow equation (\ref{i3}). 
	Differentiating (\ref{GamvsW1}) with respect to $\kappa$ gives 
	$\partial_\kappa\Gamma_\kappa= -(\partial_\kappa W_{\kappa})[H_{\kappa}[\phi]] 
	- \mathcal{V}[\phi]$. Inserting the series expansions
	\begin{eqnarray}
	\label{GamvsW2} 
	W_\kappa[H]&=&\sum_{l\geq 0}\kappa^l W_{l}[H]\,,\quad
	\Gamma_{\kappa}[\phi]=\sum_{l\geq 0}\kappa^l \Gamma_{l}[\phi]\,,\;\; 
	\Gamma_1[\phi] \equiv 0\,,
	\nonumber\\
	H_\kappa[\phi]&=&\sum_{l\geq 0}\kappa^l H_{l}[\phi]\,,
	\quad \quad H_{l} = \Gamma_{l}^{(1)} + \delta_{l,1} \mathcal{V}^{(1)}\,,\;\;l \geq 0\,,
	\end{eqnarray}
	one obtains $\Gamma_0[\phi] = \phi\cdot H_0[\phi] - W_0[H_0[\phi]]$, 
	$\Gamma_1[\phi] = - W_1[H_0[\phi]] - \mathcal{V}[\phi] \equiv 0$, and for $l \geq 2$ 
	\begin{equation}
	\label{GamvsW3}
	\Gamma_{l}[\phi]= - W_l [H_0[\phi]] -
	\sum_{m=1}^{l-1}\sum_{k=1}^{m} \frac{l\!-\!m}{l k!}
	\!\!\!
	\sum_{ \begin{array}{l}
			\mbox{\tiny 
				$m_1 + ...+ m_k =m$}\\[-2mm] 
			\mbox{\tiny $m_j \geq 1$} 
	\end{array}} \!\!\!\!
	W^{(k)}_{l-m}[H_0[\phi]]\cdot H_{m_1}[\phi]\ldots H_{m_k}[\phi]\,.
	\vspace{-5mm} 
	\end{equation}
	Note that $W^{(1)}_{\kappa}[H_{\kappa}[\phi]] = \phi$ still 
	enters (\ref{GamvsW3}) implicitly in defining the $H_m[\phi]$. 
	Upon expansion one finds $W_0^{(1)}[H_0[\phi]] = \phi$, 
	$W_0^{(2)}[H_0[\phi]] \cdot H_1[\phi]  + W_1^{(1)}[H_0[\phi]] =0$, 
	and for $l \geq 2$  
	\begin{equation}
	\label{GamvsW5} 
	H_{l}[\phi] \cdot 
	W_0^{(2)}[H_0[\phi]] + W_{l}^{(1)}[H_0[\phi]] 
	+ \frac{\delta}{\delta H_0} 
	F_{l-1}[H_0,H_1,\ldots, H_{l-1}] \Big|_{H_m = H_m[\phi]} =0\,,
	\end{equation}
	where 
	\begin{eqnarray} 
	\label{GamvsW6}
	&& F_{l-1}[H_0, \ldots, H_{l-1}] := \sum_{m=1}^{l-1} 
	\sum_{k=1}^m \check{B}_{mk}(H_1, \ldots, H_{m-k+1})\cdot W_{l-m}^{(k)}[H_0]
	\nonumber\\
	&& \quad + \sum_{k=2}^{l} \check{B}_{l k}(H_1, \ldots, H_{l +1 -k}) 
	\cdot W_0^{(k)}[H_0]\,, 
	\end{eqnarray}
	and the $\check{B}_{m,k}$ are modified Bell polynomials,
	$k! \check{B}_{mk}(H_1,\ldots, H_{l -k +1}) :=
	\sum_{m_1 +... + m_k = m, m_j \geq 1}$ $H_{m_1} H_{m_2} \ldots H_{m_k}$. 
	These relations can be solved iteratively 
	for the $H_{l}[\phi]$ and also show inductively that $H_{l}[\phi] = 
	\Gamma^{(1)}_{l}[\phi]$.
	
	In (\ref{GamvsW3}), (\ref{GamvsW5}) and similar relations later on  there are 
	tacit summations over lattice sites, summarily indicated by a
	``$\,\cdot\,$''.  A contraction of 
	$(W_{l-m}^{(k)})_{y_1 \ldots y_k}$ may contain subsums where 
	where one or more lattice points coincide. The graph rules for the 
	cumulants outlined after (\ref{wgraph2}) then change slightly. 
	Since multiple $h$ derivatives can act on the same $\omega_{d(v)}(h)$, the 
	number of rooted vertices $r$ can be $r = 1,\ldots ,k$.  
	The tacit lattice sums ensure that all possible 
	combinations will occur, so that $W^{(k)}_{l-m}$
	expands into a sum of $r$-rooted connected graphs with 
	$l\!-\!m$ edges; we write $\mathcal{C}_{l-m}^{\bullet r}$ for the 
	set of such graphs. The topology of each graph in $\mathcal{C}_{l-m}^{\bullet r}$ 
	is the same as its counterpart in $\mathcal{C}_{l -m}$, only the rooted 
	vertices have their $\omega_n$ weight shifted from $n=d(v)$ to 
	$n= d(v) + \# \mbox{of $h$-derivatives}$, and the symmetry factor changes. The contracted lattice sums 
	in (\ref{GamvsW3}), (\ref{GamvsW5}) ensure that each graph 
	in $\mathcal{C}_{l-m}^{\bullet r}$ is paired with an $r$-rooted product 
	of $H_{m_1}, \ldots, H_{m_k}$'s graph expansions, 
	such that a term 
	corresponding to an unrooted $\mathcal{C}_{l}$ graph arises. 
	This graph expansion of (\ref{GamvsW3})'s right hand side allows for 
	many cancellations. In order to identify the underlying pattern we 
	derive a property of the LRH expansion of the effective action well-known 
	for its perturbative expansion but not limited to it: 
	\begin{lemma} 
		\label{lemma1LI} 
		The graphs contributing to $\Gamma_{\kappa}[\phi]$'s LRH expansion 
		are 1LI, i.e.~remain $\ell$-connected even when any one $\ell$-line is cut. 
	\end{lemma} 
	\begin{proof} 
		The proof is an adaptation of the argument familiar for the 
		Feynman diagrams occurring in a perturbative expansion. 
		In a first step one computes the linear response of 
		$\Gamma_{\kappa}[\phi]$ under a replacement of the hopping 
		matrix 
		\begin{equation} 
		\label{lemma1LIa} 
		\ell_{xy} \mapsto \ell_{xy} + \epsilon e_x e_y\,,
		\end{equation}
		where $e_x$ is a vector and $\epsilon\geq 0$. We momentarily change 
		notation and write $W_{\epsilon}[H]$, $\Gamma_{\epsilon}[\phi]$ for the 
		functionals obtained by the replacement (\ref{lemma1LIa})
		and $W[H], \Gamma[\phi]$ for the original ones, 
		without indicating the $\kappa$-dependence. Starting from the 
		functional integral realization 
		\begin{eqnarray}
		&& \exp W[H] :=\int\! \prod_x 
		d \chi_x \exp\{- S[\chi] + \sum_x H_x \chi_x\}\,. 
		\nonumber \\
		&& S[\chi] = S_0[\chi] + \kappa \mathcal{V}[\chi]\,, 
		\quad 
		S_0[\chi] = \sum_x s(\chi_x) \,,\quad 
		\mathcal{V}[\chi] = \frac{1}{2} \sum_{x,y} \chi_x \ell_{xy} \chi_y\,. 
		\end{eqnarray}
		and expanding in powers of $\epsilon$ one finds to linear order 
		\begin{equation}
		\label{lemma1LIb}  
		W_{\epsilon}[H] = W[H] - \epsilon \frac{\kappa}{2} 
		\sum_{x,y} e_x e_y \bigg( 
		\frac{\delta^2 W}{\delta H_x \delta H_y} + 
		\frac{\delta W}{\delta H_x} 
		\frac{\delta W}{\delta H_y} \bigg) + O(\epsilon^2)\,.  
		\end{equation}
		For the altered functionals the definition of the modified 
		Legendre transform (\ref{GamvsW1}) reads 
		\begin{equation} 
		\label{lemma1LIc}
		\Gamma_{\epsilon}[\phi] = \phi\cdot H_{\epsilon}[\phi] - W_{\epsilon}[H_{\epsilon}[\phi]] 
		- \frac{\kappa}{2} \phi \cdot(\ell \!+ \! \epsilon e \otimes e) \cdot \phi\,, 
		\quad \frac{\delta W_{\epsilon}}{\delta H}\big( H_{\epsilon}[\phi] \big) = \phi\,.    
		\end{equation}
		Differentiating with respect to $\epsilon$ gives 
		$\partial_{\epsilon} \Gamma_{\epsilon}[\phi] = \frac{\kappa}{2} (e \otimes e)\! \cdot \!
		W^{(2)}(H[\phi]) + O(\epsilon)$. Since $W^{(2)}(H[\phi]) = (\Gamma^{(2)} + \kappa v)^{-1}$ 
		one obtains 
		\begin{equation} 
		\label{lemma1LId} 
		\Gamma_{\epsilon}[\phi] = \Gamma[\phi] + \epsilon \frac{\kappa}{2} 
		\sum_{x,y} e_x e_y \big[ \Gamma^{(2)} + \kappa \ell\big]^{-1}_{xy} + 
		O(\epsilon^2)\,. 
		\end{equation} 
		The replacement (\ref{lemma1LIa}) emulates the effect of cutting $\ell$-lines 
		and to linear order in $\epsilon$ the effect of cutting precisely one
		$\ell$-line is traced. Viewed as a function of $H$ the response,
		being proportional to $W^{(2)}[H]$,  expands into $\ell$-connected 
		LRH graphs by Section II.A. The recursion (\ref{GamvsW5}) shows that the 
		$\kappa$ expansion of $H[\phi]$ produces contracted functional derivatives 
		of the $W_0,W_1,\ldots , W_{l}$ evaluated at $H_0[\phi]$ for all $H_{l}[\phi]$, 
		$l \geq 1$. The $W_m^{(k)}$ derivatives correspond to $r\leq k$-rooted 
		$\ell$-connected diagrams and the contractions are pointwise with analogous terms. 
		Hence, also as a functional of $\phi$ the linear response (\ref{lemma1LId}) expands 
		into $\ell$-connected LRH graphs only.          
	\end{proof}

	The graph expansion of the right hand of (\ref{GamvsW3}) contains a large 
	number of terms associated with one-line reducible graphs. By Lemma \ref{lemma1LI} 
	these must cancel which allows one to simplify the right hand side considerably. 
	In the sum each $W_{l -m}^{(k)}$ expands into $r$-rooted, $r = 1,\ldots ,k$, 
	connected diagrams many of which are one-line reducible. The rooted vertices are 
	directly (without extra $\ell_{xy}$ link) attached to (and summed over 
	the lattice point associated with) possibly multiple copies of a $1$-rooted 
	graph representing a $\Gamma^{(1)}_m +\delta_{m,1}\mathcal{V}^{(1)}$. A term occurring in the graph
	expansion of $W^{(k)}_{l-m}[H_0[\phi]]\cdot 
	H_{m_1}[\phi]\ldots H_{m_k}[\phi]$ will be one-line reducible if 
	(i) $m_i =1$ for one or more $i \in \{1,\ldots,k\}$, 
	since $H_1[\phi]_x = \sum_y \ell_{xy} \phi_y$. (ii) the $r$-rooted
	$W$-graph stemming from  $W^{(k)}_{l-m}$ is one-line reducible. 
	(iii) if a $W_1^{(k)}$ term enters, as $W_1[H] = -\frac{1}{2} \sum_{x,y} 
	\ell_{xy} \omega_1(H_x) \omega_1(H_y)$. All these terms must cancel against the one-line 
	reducible terms in $-W_{l}$. We write $W_{l}^{(k)}[\Gamma_0^{(1)}]|_{\rm 1LI}$
	for the quantity obtained from $W_{l}^{(k)}[\Gamma_0^{(1)}]$'s graph
	expansion by omitting all terms corresponding to one-line 
	reducible graphs and $[m]$ for the integer part of $m \in \mathbb{R}_+$. 
	Then $\Gamma_2[\phi] = - W_2[\Gamma_0^{(1)}]|_{\rm 1LI}$, 
	$\Gamma_3[\phi] = - W_3[\Gamma_0^{(1)}]|_{\rm 1LI}$,
	and for $l \geq 4$ the following simplified version of  
	(\ref{GamvsW3}) holds 
	\begin{equation}
	\label{GamvsW4}
	\Gamma_{l}[\phi]= - W_l [\Gamma^{(1)}_0]\Big|_{\rm 1LI}  -
	\sum_{m=2}^{l-2}\sum_{k=1}^{[m/2]} \!\!\!
	\sum_{ \begin{array}{l}
			\mbox{\tiny 
				$m_1 + \ldots + m_k =m$}\\[-2mm] 
			\mbox{\tiny $m_i \geq 2$} 
	\end{array}} \!\!\!\!
	\frac{(l\!-\!m)}{l k!}W^{(k)}_{l-m}[\Gamma^{(1)}_0] \Big|_{\rm 1LI} \cdot 
	\Gamma_{m_1}^{(1)} \ldots \Gamma_{m_k}^{(1)}\,.
	\end{equation}
	\vspace{-5mm} 
	
	An immediate consequence of (\ref{GamvsW4}) is:
	\smallskip 
	
	\begin{lemma} 
		\label{lemmanoart} 
		Let $L$ be a 1LI graph without articulation points 
		and let $\mu^W(L),\mu^{\Gamma}(L)$ be the weight (including sign and symmetry 
		factors) with which it occurs in the expansion of $W,\Gamma$, respectively. 
		Then 
		\begin{equation} 
		\label{noartpoint} 
		\mu^{\Gamma}(L) = - \mu^W(L)|_{H_0 = \Gamma_0^{(1)}}\,.
		\end{equation}
	\end{lemma} 
	\begin{proof} It suffices to show that all terms in the sum on 
		the right hand side of (\ref{GamvsW4}) expand into graphs with 
		articulation points. As seen above, each $W_{l -m}^{(k)}|_{\rm 1LI}$ 
		expands into $r$-rooted, $r =1,\ldots, k$, 1LI graphs that are directly 
		(without extra $\ell_{xy}$ link) attached to an $r$-rooted product 
		$\Gamma_{m_1}^{(1)} \ldots \Gamma_{m_k}^{(1)}$ (with the same $\Gamma^{(1)}_m, m \geq 2$, 
		possibly occurring several times) where each factor expands into  
		$1$-rooted 1LI graphs. Each of the rooted vertices therefore is an 
		articulation point and the graphs contributing to a $W_{l -m}^{(k)}|_{\rm 1LI}$ 
		term in the sum have at least one articulation point. 
	\end{proof}

	On account of the previous results the problem of finding 
	a graph rule for the LRH expansion of $\Gamma_{\kappa}$ has 
	been reduced to understanding the weight $\mu^{\Gamma}(v)$ 
	that ought to be assigned to articulation points: by 
	Lemma \ref{lemma1LI} we know that the graphs contributing to 
	$\Gamma_{l}[\phi]$ are one-line irreducible (1LI). 
	As long as the 1LI graph considered has no articulation 
	points Lemma \ref{lemmanoart} straightforwardly provides the 
	weight. The same reasoning shows that the maximal number of 
	articulation vertices in some $L \in \mathcal{L}_{l}$ is 
	$[(l\! -\!2)/2]$. One may anticipate a trade-off to occur: 
	the vastly reduced 
	number of graphs to be considered (compared to $W$) will be 
	compensated in part by a more complicated weight assignment 
	for articulation vertices. Overall a very significant simplification 
	is found to occur already at low orders; see Table \ref{graphcount}.

	\begin{table}[h]
		\begin{center}
			\begin{tabular}{|c | c | c | c|}
				\hline	
				$l$ & $|\mathcal{C}_{l}|$ & $|\mathcal{L}_{l}|$ & $\#$ art.~vert. 
				\\
				\hline	
				& & & \\[-2mm] 
				\hspace{1mm}3 \hspace{2mm}& 5 & 2 & 0\\
				4 & 12 & 4 & 1\\
				5 & 33 & 8 & 2 \\
				6 & 100 & 22 & 8,1\\
				\hline 		
			\end{tabular}
		\end{center}
		\vspace{-3mm}
		\caption{Number of connected, one-line irreducible, one-line irreducible 
			graphs with $1,2,\ldots$ articulation points, respectively, and $l$
			edges.}
		\label{graphcount} 
	\end{table}
	
	Up to $l =3$ all 1LI graphs are also 1VI, so that the  graph rules for 
	$W_{l}$ (with vertex 
	weights $\omega_m(\varphi) := \omega_m(h)|_{h = h(\varphi)}$, $m \geq 1$) gives the correct 
	answer for $l \leq 3$. In the figure below the weights from the $W$ 
	rule match the terms in the directly computed result (\ref{gammresults234}):   
	\begin{eqnarray}
	\label{grule3.1}
	\Gamma[\phi]&=&\,
	\begin{tikzpicture}
	\filldraw[black] (0,0.5) circle (2pt) ;
	\end{tikzpicture}\,
	-\frac{1}{4}\,
	\begin{tikzpicture}
	\draw[black, thick] (0,0) edge[bend left=30] (1,0);
	\draw[black, thick] (0,0) edge[bend right=30] (1,0);
	\filldraw[black] (0,0) circle (2pt) ;
	\filldraw[black] (1,0) circle (2pt) ;
	\end{tikzpicture}\,
	+\frac{1}{12}\,
	\begin{tikzpicture}
	\draw[black, thick] (0,0) edge (1,0);
	\draw[black, thick] (0,0) edge[bend left=30] (1,0);
	\draw[black, thick] (0,0) edge[bend right=30] (1,0);
	\filldraw[black] (0,0) circle (2pt) ;
	\filldraw[black] (1,0) circle (2pt) ;
	\end{tikzpicture}\,
	+\frac{1}{6}\,
	\begin{tikzpicture}
	\draw[black, thick] (0,0) -- (0.5,0.5);
	\draw[black, thick] (0.5,0.5) -- (1,0);
	\draw[black, thick] (0,0) -- (1,0);
	\filldraw[black] (0,0) circle (2pt) ;
	\filldraw[black] (0.5,0.5) circle (2pt) ;
	\filldraw[black] (1,0) circle (2pt) ;
	\end{tikzpicture}\,
	+O(\kappa^4).
	\end{eqnarray}
	For $l =4$ the same works for all but the second to last term,
	which corresponds to a ``pair of glasses'' graph. 
	The vertex in the middle is an articulation point and by inspection of 
	(\ref{gammresults234}) one reads off the weight that should be attributed 
	to it:
	
	\begin{eqnarray}
	\label{grule4.1} 
	\begin{tikzpicture}
	\hspace*{-0.2cm}
	\draw[black, thick] (0,0) edge[bend left=30] (1,0);
	\draw[black, thick](0,0) edge[bend right=30] (1,0);
	\draw[black, thick] (1,0) edge[bend left=30] (2,0);
	\draw[black, thick] (1,0) edge[bend right=30] (2,0);
	\filldraw[black] (0,0) circle (2pt) ;
	\filldraw[black] (1,0) circle (2pt) node[right,above]{\small$v$\normalsize};
	\filldraw[black] (2,0) circle (2pt) ;
	\end{tikzpicture}
	&& 
	\nonumber
	\\[-1cm]
	&& \makebox[1.9cm]{}\textrm{Symmetry factor}=2^3
	\nonumber\\
	&&\makebox[1.9cm]{}
	\mu^{\Gamma}(v) = \omega_4(\varphi_v) -\gamma_2(\varphi_v)  \omega_3(\varphi_v)^2\,.
	\end{eqnarray}  
	
	In each case we also note the symmetry factor of the full graph 
	next to it. For $l =5$ there are two graphs with articulation 
	points for which the explicitly computed weights are: 
	
	\begin{eqnarray} 
	\label{grule5.1}
	\begin{tikzpicture}
	\hspace*{0.7cm}
	\draw[black, thick] (0,0) edge[bend left=30] (1,0);
	\draw[black, thick](0,0) edge[bend right=30] (1,0);
	\draw[black, thick] (1,0) edge[bend left=30] (2,0);
	\draw[black, thick] (1,0) edge[bend right=30] (2,0);
	\draw[black, thick] (1,0) -- (2,0);
	\filldraw[black] (0,0) circle (2pt) ;
	\filldraw[black] (1,0) circle (2pt) node[right,above]{\small$v$\normalsize};
	\filldraw[black] (2,0) circle (2pt) ;
	\end{tikzpicture} 
	&& 
	\nonumber
	\\[-1cm]
	&&\makebox[2.7cm]{}\textrm{Symmetry factor}=2\times 3!
	\nonumber\\
	&& \makebox[2.7cm]{}
	\mu^{\Gamma}(v) =  \omega_5(\varphi_v) - \gamma_2(\varphi_v)\omega_3(\varphi_v) \omega_4(\varphi_v)\,. 
	\end{eqnarray}
	
	\begin{eqnarray}
	\label{grule5.2}
	\begin{tikzpicture}
	\hspace*{-0.2cm}
	\draw[black, thick] (1,0) edge[bend left=30] (2,0);
	\draw[black, thick] (1,0) edge[bend right=30] (2,0);
	\draw[black, thick] (2,0) edge (2.87,0.5);
	\draw[black, thick] (2,0) edge (2.87,-0.5);
	\draw[black, thick] (2.87,-0.5) edge (2.87,0.5);
	\filldraw[black] (1,0) circle(2pt);
	\filldraw[black] (2,0) circle (2pt)node[right,above]{\small$v$\normalsize};
	\filldraw[black] (2.87,0.5) circle (2pt);
	\filldraw[black] (2.87,-0.5) circle (2pt);
	\end{tikzpicture} 
	&& 
	\nonumber
	\\[-1.5cm]
	&&\makebox[2cm]{}\textrm{Symmetry factor}=2^2
	\nonumber\\
	&&\makebox[2cm]{}
	\mu^{\Gamma}(v) = \omega_4(\varphi_v) - \gamma_2(\varphi_v)\omega_3(\varphi_v)^2\,. 
	\end{eqnarray}
	
	Clearly, the first term in the weight associated to an 
	articulation point is the one expected from the $W$ graph rules;
	it is the systematics of the additional terms that need to be 
	understood. 
\vspace{-0.5cm}

	\subsection{Recursive computation of the weights of articulation points} 
	
	Our guiding principle in pinning down these systematics will be 
	the relation (\ref{GamvsW4}). It expresses $\Gamma_{l}$'s graph 
	expansion in terms of those of $\Gamma_2^{(1)}, \ldots, \Gamma_{l -2}^{(1)}$,
	modulo pieces known from the $W$-graph rules. By construction (\ref{GamvsW4}) is equivalent to the closed 
	recursion (\ref{rec2}). In contrast to (\ref{rec2})  the mixed recursion (\ref{GamvsW4}) allows one to isolate directly contributions 
	from individual graphs, in particular those with articulation points. 
	For example, for $l\! =\!4$ one has $\Gamma_4 = - W_4|_{\rm 1LI} - 
	\frac{1}{2} W_2^{(1)}|_{\rm 1LI} \cdot \Gamma_2^{(1)}$.
	Applying the graphical differentiation rules to $W_2$ and 
	$\Gamma_2$ one quickly recovers (\ref{grule4.1}). Similarly, for $l =5$ 
	one obtains from (\ref{GamvsW4}) $\Gamma_5 = - W_5|_{\rm 1LI} 
	- \frac{3}{5} W_3^{(1)}|_{\rm 1LI} \cdot \Gamma_2^{(1)} - \frac{2}{5} 
	W_2^{(1)}|_{\rm 1LI} \cdot \Gamma_3^{(1)}$, and (\ref{grule5.1}), 
	(\ref{grule5.2}) 
	can be confirmed graphically. With $\Gamma_{l},\,l =2,\ldots, 5$, known 
	explicitly from Appendix A the same procedure allows one to obtain the 
	weights of all $l=6,7$ graphs with articulation points. At $l =6$ 
	there are $8$ graphs with one articulation vertex and $1$ with 
	two articulation vertices, see Table 1. The $l=6$ graph with 
	two articulation vertices is the ``triple bubble'' graph
	and both have the same weight associated to them as $v$ in 
	(\ref{grule4.1}).  
	
	More interesting are the $l=7$ graphs for which we present  
	three examples: 
	\begin{eqnarray}
	\label{grule7.1a}
	\begin{tikzpicture}
	\hspace*{-1.5cm}
	\draw[black, thick] (0,0) edge[bend left=30] (1,0);
	\draw[black, thick] (0,0) edge[bend right=30] (1,0);
	\draw[black, thick] (1,0) edge[bend left=30] (2,0);
	\draw[black, thick] (1,0) edge[bend right=30] (2,0);
	\draw[black, thick] (1,0) edge (1,1);
	\draw[black, thick] (1,0) edge[bend left=30] (1,1);
	\draw[black, thick] (1,0) edge[bend right=30] (1,1);
	\filldraw[black] (0,0) circle (2pt);
	\filldraw[black] (1,0) circle(2pt)node[right,below]{\small$v$\normalsize};
	\filldraw[black] (1,1) circle (2pt);
	\filldraw[black] (2,0) circle (2pt);
	\end{tikzpicture}
	\nonumber
	\\[-1.8cm]
	&& \makebox[1.5cm]{}
	 \textrm{Symmetry factor}=2^3\times 3!
	\nonumber\\
	&&  \makebox[1.5cm]{}
	 \mu^{\Gamma}(v) =  \omega_7 - 2 \gamma_2\omega_3\omega_6- 
	\gamma_2\omega_4\omega_5 
	\nonumber\\
	&& \makebox[1.5cm]{}
	+ \gamma_2^{2}\omega_3^2\omega_5
	+ 2 \gamma_2^{2}\omega_3\omega_4^{2}+ \gamma_3\omega_3^2\omega_4
	\\[0.5cm]
	\label{grule7.1c}
	\begin{tikzpicture}
	\hspace*{-0.5cm}
	\draw[black, thick] (-0.5,0.87) edge[bend left=30] (0.5,0.87);
	\draw[black, thick] (-0.5,0.87) edge[bend right=30] (0.5,0.87);
	\draw[black, thick] (1.5,0.87) edge[bend left=30] (2.5,0.87);
	\draw[black, thick] (1.5,0.87) edge[bend right=30] (2.5,0.87);
	\draw[black, thick] (0.5,0.87) edge (1.5,0.87);
	\draw[black, thick] (1,0) edge (0.5,0.87);
	\draw[black, thick] (1,0) edge (1.5,0.87);
	\filldraw[black] (-0.5,0.87) circle(2pt);
	\filldraw[black] (1,0) circle(2pt);
	\filldraw[black] (0.5,0.87) circle (2pt)node[right,above]{\small$v$\normalsize};
	\filldraw[black] (1.5,0.87) circle (2pt)node[right,above]{\small$v'$\normalsize};
	\filldraw[black] (2.5,0.87) circle (2pt);
	\end{tikzpicture}
	\nonumber
	\\[-1.6cm]
	&& \makebox[1.5cm]{} \textrm{ Symmetry factor}=2^3
	\nonumber\\
	&& \makebox[1.5cm]{}
	\mu^{\Gamma}(v)=\mu^{\Gamma}(v')= \omega_4 - \gamma_2\omega_3^2\\
	\nonumber
	\\[0.5cm]
	\label{grule7.2} 
	\begin{tikzpicture}
	\hspace*{-0.5cm}
	\draw[black, thick] (-0.5,0.87) edge[bend left=30] (0.5,0.87);
	\draw[black, thick] (-0.5,0.87) edge[bend right=30] (0.5,0.87);
	\draw[black, thick] (0.5,0.87) edge(1.5,0.87);
	\draw[black, thick] (0.5,0.87) edge[bend left=30] (1.5,0.87);
	\draw[black, thick] (0.5,0.87) edge[bend right=30] (1.5,0.87);
	\draw[black, thick] (1.5,0.87) edge[bend left=30] (2.5,0.87);
	\draw[black, thick] (1.5,0.87) edge[bend right=30] (2.5,0.87);
	\filldraw[black] (-0.5,0.87) circle(2pt);
	\filldraw[black] (0.5,0.87) circle (2pt)node[right,above]{\small$v$\normalsize};
	\filldraw[black] (1.5,0.87) circle (2pt)node[right,above]{\small$v'$\normalsize};
	\filldraw[black] (2.5,0.87) circle (2pt);
	\end{tikzpicture}
	\nonumber
	\\[-1cm]
	&& \makebox[1.5cm]{} \textrm{ Symmetry factor}=2^3\times 3!
	\nonumber\\
	&& \makebox[1.5cm]{}
	\mu^{\Gamma}(v) = 
	\mu^{\Gamma}(v') = 
	\omega_5 - \gamma_2\omega_3\omega_4\,
	\\
	\nonumber
	\end{eqnarray}
	Here and below we omit the $\varphi$ arguments of the
	$\omega_m$'s. 
	Note that the weight in (\ref{grule7.1a}) is new while 
	those in (\ref{grule7.1c}) and (\ref{grule7.2}) are recycled 
	from (\ref{grule4.1}), (\ref{grule5.2}) and (\ref{grule5.1}), 
	respectively. 

	So far the graph expansion of the explicitly computed 
	$\Gamma_2,\ldots, \Gamma_5$ from Appendix A could be used as
	an input to obtain the results for all $l =6,7$ graphs.
	The recursion (\ref{GamvsW4}) also allows one compute the weights 
	of individual higher order graphs {\it without} knowing the   
	full results for the $\Gamma_m$'s at lower orders. 
	We illustrate this with two $l=9$ graphs chosen 
	so that the $l=7$ input graphs are among the ones 
	preciously displayed. 
	
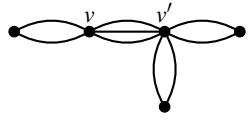
\begin{eqnarray}
	\label{grule9.1}
	\begin{tikzpicture}
	\hspace*{2.25cm}
	\draw[black, thick] (-0.5,0.87) edge[bend left=30] (0.5,0.87);
	\draw[black, thick] (-0.5,0.87) edge[bend right=30] (0.5,0.87);
	\draw[black, thick] (0.5,0.87) edge (1.5,0.87);
	\draw[black, thick] (0.5,0.87) edge[bend left=30] (1.5,0.87);
	\draw[black, thick] (0.5,0.87) edge[bend right=30] (1.5,0.87);
	\draw[black, thick] (1.5,0.87) edge[bend left=30] (2.5,0.87);
	\draw[black, thick] (1.5,0.87) edge[bend right=30] (2.5,0.87);
	\draw[black, thick] (1.5,0.87) edge[bend left=30] (1.5,-0.13);
	\draw[black, thick] (1.5,0.87) edge[bend right=30] (1.5,-0.13);
	\filldraw[black] (-0.5,0.87) circle(2pt);
	\filldraw[black] (1.5,-0.13) circle(2pt);
	\filldraw[black] (0.5,0.87) circle (2pt)node[right,above]{\small$v$\normalsize};
	\filldraw[black] (1.5,0.87) circle (2pt)node[right,above]{\small$v'$\normalsize};
	\filldraw[black] (2.5,0.87) circle (2pt);
	\end{tikzpicture}
	\nonumber
	\\[-2cm]
	&&\makebox[4cm]{}  \textrm{ Symmetry factor}=2^4\times 3!\,;\;\;
	\textrm{input}\;(\ref{grule7.1a}), (\ref{grule7.2}) 
	\nonumber\\
	&& \makebox[4cm]{} 
	\mu^{\Gamma}(v) =  \omega_5 -\gamma_2 \omega_3 \omega_4
	\nonumber\\
	&& \makebox[4cm]{} 
	\mu^{\Gamma}(v') = \omega_7 - 2 \gamma_2 \omega_3 \omega_6 - \gamma_2 \omega_4\omega_5
	\\[2mm]
	&& \makebox[4cm]{} 
	+ \gamma_2^{2}\omega_3^2\omega_5 +2 \gamma_2^{2}\omega_3 \omega_4^{2} +
	\gamma_3 \omega_3^2\omega_4\,
	\nonumber
	\\[5mm]
	
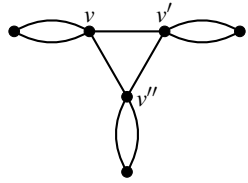
\label{grule9.4}
	\begin{tikzpicture}
	\hspace*{2.25cm}
	\draw[black, thick] (-0.5,0.87) edge[bend left=30] (0.5,0.87);
	\draw[black, thick] (-0.5,0.87) edge[bend right=30] (0.5,0.87);
	\draw[black, thick] (1.5,0.87) edge[bend left=30] (2.5,0.87);
	\draw[black, thick] (1.5,0.87) edge[bend right=30] (2.5,0.87);
	\draw[black, thick] (0.5,0.87) edge (1.5,0.87);
	\draw[black, thick] (1,0) edge[bend left=30] (1,-1);
	\draw[black, thick] (1,0) edge[bend right=30] (1,-1);
	\draw[black, thick] (1,0) edge (0.5,0.87);
	\draw[black, thick] (1,0) edge (1.5,0.87);
	\filldraw[black] (-0.5,0.87) circle(2pt);
	\filldraw[black] (1,0) circle (2pt)node[right]{\small$v''$\normalsize};
	\filldraw[black] (1,-1) circle(2pt);
	\filldraw[black] (0.5,0.87) circle (2pt)node[right,above]{\small$v$\normalsize};
	\filldraw[black] (1.5,0.87) circle (2pt)node[right,above]{\small$v'$\normalsize};
	\filldraw[black] (2.5,0.87) circle (2pt);
	\end{tikzpicture}
	\nonumber
	\\[-25mm]
	&&\makebox[4cm]{}  \textrm{ Symmetry factor}=2^3\times 3!\,;\;\;
	\textrm{input}\;\;(\ref{grule7.1c})
	\nonumber\\
	&& \makebox[4cm]{}  
	\mu^{\Gamma}(v) = \mu^{\Gamma}(v') = \mu^{\Gamma}(v'') = \omega_4 
	- \gamma_2 \omega_3^2\,.
	\\[10mm]
	\nonumber
	\end{eqnarray}
    
	These examples illustrate a pattern that holds generally. 
	To formulate it we introduce a natural grading for the quantities 
	considered. It is induced by the derivatives of the single site 
	functions $\omega(h), \gamma(\varphi)$ and their interrelations 
	discussed in Appendix B. 
	\medskip
	
	\begin{lemma} For a monomial in $\omega_m, m\geq 3$, $\omega_2^{-1}$, 
		define its degree by: ${\rm deg} \,\omega_m = m, {\rm deg} \,\omega_2^{-1} = -2$, and 
		extended additively to products. Then:
		\begin{itemize} 
			\item[(a)] ${\rm deg} \gamma_m = - m$, $m \geq 2$. 
			\item[(b)] ${\rm deg} W_{l}^{(k)} = 2 l\! +\!k$, ${\rm deg} \Gamma_{l}^{(k)} = 
			2l\! -\! k$, $k\geq 0$, to all orders $l \geq 1$ of the LRH expansion. 
			\item[(c)] The weight $\mu^{\Gamma}(v)$ assigned to an articulation vertex $v$  
			has homogeneous degree ${\rm deg}[\mu^{\Gamma}(v)]$ which coincides with its degree 
			in the $W$-graph rule, i.e. ${\rm deg}[\mu^W(v)] = l$, for an $l$-valent vertex. 
			\item[(d)] The weight $\mu^{\Gamma}(v)$ assigned to an articulation vertex $v$ 
			of degree $d(v) = m \geq 4$ can be normalized such that 
			\begin{equation}
			\label{GamEuler} 
			\mu^{\Gamma}(v) = \omega_m -\!\!\!  
			\sum_{ 3i_3 + \ldots + (m\!-\!1) i_{m-1} = m + 2 i_2} 
			\!\!\!d_{i_3\ldots i_{m-1}} \,(\omega_2^{-1})^{i_2} \omega_3^{i_3} 
			\ldots \omega_{m-1}^{i_{m-1}} \,,
			\end{equation}
			and analogously in any mixed $\omega_m,\gamma_m$ form. 
		\end{itemize}
	\end{lemma} 
	
	\begin{proof} 
		(a) manifest from (\ref{0lege2}). (b) $W_1[H] = -\frac{1}{2} 
		\sum_{x,y} \ell_{xy} \omega_1(H_x) \omega_1(H_y)$
		gives ${\rm deg} W_1 =2$, each $H$ derivative raises the degree by $1$, 
		so ${\rm deg} W_{l} = 2 l$ follows from the recursion (\ref{wrec2}).  
		Similarly, ${\rm deg}\Gamma_2 = 4$ from (\ref{gammresults234}), each $\phi$ derivative 
		lowers the degree by $1$ (as $\partial_{\varphi} = \gamma_2 \partial_h$), and 
		${\rm deg} \Gamma_{l} = 2 l$ follows from the recursion 
		(\ref{rec2}). Since ${\rm deg} W_{l -m}^{(k)} = 2(l -m) + k$, 
		${\rm deg}[ \Gamma_{m_1}^{(1)} \ldots \Gamma_{m_k}^{(1)}] = 2(m_1 + \ldots + m_k) - k$,
		compatibility with (\ref{GamvsW3}) is ensured. (c) The weight 
		$\mu^{\Gamma}(v)$ is in principle determined by the recursion (\ref{GamvsW3}),
		(\ref{GamvsW4}). By (b) these relations preserve homogeneity which implies 
		${\rm deg}[\mu^{\Gamma}(v)] = {\rm deg}[\mu^W(v)]$. (d) is a consequence of (c)
		and the gross structure of (\ref{GamvsW4}). 
	\end{proof} 
	
	In summary, let $\mathcal{L}_{l}$ be the set of one-line irreducible 
	graphs with $l=|E|$ links. Then
	\begin{equation}
	\label{Gamschematic}
	\Gamma_{l}[\phi] = \sum_{L= (V,E) \in \mathcal{L}_{l}} \frac{(-)^{l+1}}{{\rm Sym}(L)} 
	\prod_{e \in E} \ell_{\theta(e)} \prod_{v \in V}  \mu^{\Gamma}(v|L)\,,
	\end{equation}
	with a tacit unconstrained sum over the lattice points associated with 
	the vertices upon embedding. Here $\mu^{\Gamma}(v|L)$ is as in (\ref{GamEuler})
	where only the coefficients $d_{i_3,\ldots, i_{m-1}}$ remain to be 
	determined. These coefficients depend on the 1VI subgraphs that are 
	joined at the articulation vertex, not just on the degree 
	of the vertex; so we write $\mu^{\Gamma}(v|L)$ from now on. 
	
	For completeness' sake we justify in detail why the  
	weights  $\mu^{\Gamma}(v|L)$, $L \in \mathcal{L}_{l}$, are determined 
	recursively by (\ref{GamvsW4}). For the graphical evaluation of 
	(\ref{GamvsW4}), graph rules for $\Gamma_m^{(1)}, m =1, \ldots, l\!-\!2$,
	are needed. Differentiating (\ref{Gamschematic}) produces an analogous 
	expansion in terms of $1$-rooted one-line irreducible graphs
	for which we write $\mathcal{L}_m^{\bullet 1}$ at order $|E| =m$.  
	The product over $\mu^{\Gamma}(v|L)$ extends over all but the 
	rooted vertex, where $\partial \mu^{\Gamma}/\partial \varphi$ occurs. 
	In the context of (\ref{GamvsW4}) the coefficients 
	$d_{i_1 \ldots i_{m-1}}$ entering the $\Gamma_m^{(1)}$, 
	$m =1, \ldots, l\!-\!2$, are assumed to be known and 
	those for the graphs in $\mathcal{L}_{l}$ are to be determined. 
	The additional piece of information entering are 
	the graph rules for $W^{(k)}_{l-m}|_{\rm 1LI}$, $1 \leq k \leq [m/2]$. 
	These can be inferred from (\ref{wgraph2}). Since multiple 
	$h$ derivatives can act on the same $\omega_{d(v)}(h)$, the 
	number of rooted vertices $r$ can be $r = 1,\ldots , k$.  
	The tacit lattice sums ensure that all possible 
	combinations will occur, so that $W^{(k)}_{l-m}|_{\rm 1LI}$ 
	expands into a sum of $r$-rooted 1LI graphs with 
	$l\!-\!m$ edges, the set of which we denote by $\mathcal{L}_{l-m}^{\bullet r}$. 
	The topology of each graph in $\mathcal{L}_{l-m}^{\bullet r}$ is the same as 
	its counterpart in $\mathcal{L}_{l -m}$, only the rooted 
	vertices have their $\omega_m$ weight shifted 
	from $m=d(v)$ to $m= d(v) + \# \mbox{of $h$-derivatives}$, and the symmetry factor changes.  
	Each term in the graph expansion of $W^{(k)}_{l-m}[\Gamma_0^{(1)}]|_{\rm 1LI}
	\cdot \Gamma_{m_1}^{(1)} \ldots \Gamma_{m_k}^{(1)}$, then  
	has the rooted subgraphs joined at the roots so that 
	an unrooted graph in $\mathcal{L}_{l}$ arises. In any concrete 
	instance the procedure is evident and has been used 
	to work out the previous examples. The formulation 
	of the general evaluation principle for (\ref{GamvsW4})'s
	right hand side justifies that the recursion works
	generally and just needs to be `solved'.   
	

	\newpage 
	\section{Graph implementation of the $\Gamma_{\kappa}[\phi]$ LRH expansion} 
	
	So far each $\Gamma_{l}$ is known to expand into 1LI irreducible graphs $L$ 
	whose weights in (\ref{Gamschematic}) are known modulo the coefficients 
	$d_{i_3,\ldots i_{m-1}}$ in (\ref{GamEuler}). These coefficients depend on 
	the decomposition of $L$ into 1VI subgraphs, turn out to be integers, 
	and can be understood in terms a separate set of {\it tree} graphs. 
	To preclude a possible confusion let us stress that these tree graphs 
	are conceptually and technically different from the ones 
	governing the interplay between vertex functions and 
	connected correlation functions, see Appendix B for the latter. 
	
	\subsection{Labeled tree graphs} 
	
	We begin by introducing a class of unlabeled tree graphs called `dashed', 
	which get labeled in a second step.  
	
	{\bf Definition:} The `dashed' graphs are tree graphs 
	where two types of vertices are connected by dashed lines. 
	The set of ``open circle'' vertices is denoted by $\nu_0$,
	the set of ``dashed'' vertices is denoted by $\nu_1$,
	and the edge list $\epsilon \subset (\nu_0 \cup \nu_1)_2$ is constrained as follows. The valency
	of an open circle vertex is $1,2,\ldots$, dashed vertices 
	have valency $3,4,\ldots$, and no two dashed-vertices are 
	connected by a single dashed line. The Euler relation for tree 
	graphs then holds in the form $|\nu_0| + |\nu_1| = |\epsilon| +1$. 
	We write $\mathcal{T}_n$ for the set of topologically distinct such 
	graphs with $n = |\nu_0|$ open circle vertices.
	\medskip
	
	For example the graphs in $\mathcal{T}_1,\ldots,\mathcal{T}_4$ are
	\begin{eqnarray}
	\label{ddeg0} 
	&&\makebox[3mm]{}   \mathcal{T}_1\makebox[2cm]{} \;\; \begin{tikzpicture}
	\draw[fill=white] (0,0) circle (3pt);
	\end{tikzpicture}\,\,
	\\[2mm] 
	&&\makebox[3mm]{}    \mathcal{T}_2 \makebox[2cm]{} \;\; \begin{tikzpicture}
	\draw[black, dashed] (0.5,0) -- (1.5,0);
	\draw[fill=white] (0.5,0) circle (3pt);
	\draw[fill=white] (1.5,0) circle (3pt);
	\end{tikzpicture}\,\,
	\nonumber\\
	&&\makebox[3mm]{}    \mathcal{T}_3\makebox[2cm]{} \;\; \begin{tikzpicture}
	\draw[black, dashed] (0,0) -- (0.5,0.5);
	\draw[black, dashed] (0.5,0.5) -- (1,0);
	\draw[fill=white]  (0,0) circle (3pt) ;
	\draw[fill=white] (0.5,0.5) circle (3pt) ;
	\draw[fill=white]  (1,0) circle (3pt) ;
	\end{tikzpicture}\,,\,
	\begin{tikzpicture}
	\draw[black, dashed]  (0,0) -- (0.5,0.5);
	\draw[black, dashed]  (0.5,0.5) -- (1,0);
	\draw[black, dashed]  (0.5,0.5)-- (0.5,1);
	\draw[fill=white]  (0,0) circle (3pt) ;
	\draw[fill=white]  (1,0) circle (3pt) ;
	\draw[fill=white]  (0.5,1) circle (3pt) ;
	\end{tikzpicture}\,\,
	\nonumber\\
	&&\makebox[3mm]{}    \mathcal{T}_4\makebox[2cm]{} \;\; \begin{tikzpicture}
	\draw[black, dashed] (0,0) -- (0.5,0.5);
	\draw[black, dashed] (0.5,0.5) -- (1,0);
	\draw[black, dashed] (1,0) -- (1.5,0.5);
	\draw[fill=white]  (0,0) circle (3pt) ;
	\draw[fill=white] (0.5,0.5) circle (3pt) ;
	\draw[fill=white]  (1,0) circle (3pt) ;
	\draw[fill=white]  (1.5,0.5) circle (3pt) ;
	\end{tikzpicture}\,,\,
	\begin{tikzpicture}
	\draw[black, dashed]  (0,0) -- (0.5,0.5);
	\draw[black, dashed]  (0.5,0.5) -- (1,0);
	\draw[black, dashed]  (0.5,0.5)-- (0.5,1);
	\draw[fill=white]  (0,0) circle (3pt) ;
	\draw[fill=white]  (0.5,0.5) circle (3pt) ;
	\draw[fill=white]  (1,0) circle (3pt) ;
	\draw[fill=white]  (0.5,1) circle (3pt) ;
	\end{tikzpicture}\,,\,
	\begin{tikzpicture}
	\draw[black, dashed]  (0,0) -- (0.5,0.5);
	\draw[black, dashed]  (0.5,0.5) -- (1,0);
	\draw[black, dashed]  (0.5,0.5)-- (0.5,1);
	\draw[black, dashed]  (1,0)--(2,0);
	\draw[fill=white]  (0,0) circle (3pt) ;
	\draw[fill=white]  (1,0) circle (3pt) ;
	\draw[fill=white]  (0.5,1) circle (3pt) ;
	\draw[fill=white](2,0)circle (3pt) ;
	\end{tikzpicture}\,,\,
	\begin{tikzpicture}
	\draw[black, dashed] (0,0)--(0.5,0);
	\draw[black, dashed](0.5,0)-- (1,0);
	\draw[black, dashed] (0.5,0)--(0.5,0.5);
	\draw[black, dashed] (0.5,0)--(0.5,-0.5);
	\draw[fill=white]   (0,0) circle (3pt) ;
	\draw[fill=white] (1,0) circle (3pt) ;
	\draw[fill=white] (0.5,0.5) circle (3pt) ;
	\draw[fill=white] (0.5,-0.5) circle (3pt) ;
	\end{tikzpicture}.
	\nonumber
	\end{eqnarray}
	The restriction that  no two dashed-vertices can be connected by a 
	single dashed line eliminates from consideration graphs of the form
	\begin{eqnarray}
	\nonumber
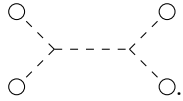
	\begin{tikzpicture}
	\draw[black, dashed] (0.5,0) -- (1.5,0);
	\draw[black, dashed] (0.5,0) -- (0,0.5);
	\draw[black, dashed] (0.5,0) -- (0,-0.5);
	\draw[black, dashed] (1.5,0) -- (2,0.5);
	\draw[black, dashed] (1.5,0) -- (2,-0.5);
	\draw[fill=white] (0,0.5)circle (3pt) ;
	\draw[fill=white] (0,-0.5) circle (3pt) ;
	\draw[fill=white] (2,0.5) circle (3pt) ;
	\draw[fill=white] (2,-0.5) circle (3pt) ;
	\end{tikzpicture}.
	\end{eqnarray}
	The graphs in $\mathcal{T}_{n+1}$ can be obtained from those in $\mathcal{T}_n$ by 
	adding one dashed leg with an open circle in all topologically inequivalent 
	ways to an open circle, a dashed line, or a dashed vertex. 
	Further, the constituents of a dashed graph can be attributed a 
	``dashed degree {\rm ddeg}" as follows: 
	\begin{equation}
	\label{ddeg1} 
	\begin{array}{cl} 
		\mbox{$|o|$-valent open circle vertex $o$, $|o| \geq 1$}& \quad {\rm ddeg} = 
		|o| + d_j\,,
		\\[2mm]
		\mbox{dashed line connecting two open circles}& 
		\quad {\rm ddeg} = -2\,,
		\\[2mm]
		\mbox{$m$-valent dashed vertex, $m \geq 3$} &
		\quad {\rm ddeg} = -m\,.
	\end{array}
	\end{equation}
	Here $d_j, j=0,\ldots,n\!-\!1$, are integers whose significance will 
	become clear shortly. Then: 
	\begin{equation} 
	\label{ddeg2}
	{\rm ddeg}(t) = \sum_{j=0}^{n-1} d_j \,, 
	\quad \mbox{for any dashed graph} \; t \in \mathcal{T}_n,\,n\geq 2\,.
	\end{equation}
	This can be seen by induction on $n$ using the before mentioned 
	recursive generation. Any of the three
	operations generating a graph in $\mathcal{T}_{n+1}$ from one in $\mathcal{T}_n$ is 
	readily seen to preserve ddeg. By inspection of (\ref{ddeg0}) the 
	assertion holds for $n=1,2,3$ and (\ref{ddeg2}) follows.   
	Note that this gives a more fine grained invariant than merely the 
	Euler relation (\ref{Euler}) for tree graphs. Instead of 
	viewing the $d_j$ as parameterizing the ddeg function one 
	may also regard them as labels for the dashed graphs themselves.
	We then write $\mathcal{T}_n^D$ for the set of dashed graphs $\mathcal{T}_n$ 
	with an integer from the $n$-tuple $D = (d_0,d_1,\ldots, d_{n-1})$ 
	assigned to each open circle vertex. The use of an $n$-tuple
	is natural in the iteration of the map (\ref{ddeg4}) below.
	Later on we use the same notation $\mathcal{T}_n^D$ when $D$ is a multiset of 
	integers of cardinality $n$. 
	
	{\bf Compatibility with differentiation:} Each $\tau \in \mathcal{T}_n^D$ can 
	be assigned a weight $\mu(\tau)$ as follows
	\begin{equation}
	\label{ddeg3} 
	\begin{array}{cl} 
		\mbox{$\omega_{|o| + d_j}$} & \mbox{to an $|o|$-valent open circle vertex $o$, 
			$|o| \geq 1$}\,,
		\\[2mm]
		\mbox{$\gamma_2$} & \mbox{to a dashed line connecting two open circles}\,,
		\\[2mm]
		\mbox{$\gamma_m$} & \mbox{to an $m$-valent dashed vertex, $m \geq 3$}\,. 
	\end{array}
	\end{equation}
	The degree of each factor in $\mu(\tau)$ equals the ddeg of the 
	underlying graph, $d(\omega_{|o| + d_j}) = |o| + d_j =$ ddeg($|o|$-valent
	open circle vertex), etc.  Hence $d (\mu(\tau)) = \sum_{j=0}^{n-1} d_j$,
	for all $\tau \in \mathcal{T}_n^D$. We write $\mu(\mathcal{T}_n^D)$ for the span 
	of all $\mu(\tau), \tau \in \mathcal{T}_n^D$. Augmenting a $(n\!+\!1)$-st 
	integer $d_n$ we claim that 
	\begin{equation}
	\label{ddeg4} 
	\omega_{d_n +1}(\varphi) \partial_{\varphi} : \mu(\mathcal{T}_n^D) \rightarrow 
	\mu(\mathcal{T}_{n+1}^D)\,,
	\end{equation}
	with the understanding that $D= (d_0,\ldots d_{n-1}, d_n)$ 
	in the range. This follows from the basic differentiation rules
	$\partial_{\varphi} \omega_m = \gamma_2 \omega_{m+1}$, $\partial_{\varphi} \gamma_m = 
	\gamma_{m+1}$ and the way $\omega_{d_n +1} \partial_{\varphi}$ acts on 
	the three types of factors in each $\mu(\tau)$:  acting on 
	$\omega_{d_j +1}$ the operator produces $\omega_{d_n +1} \gamma_2 \omega_{d_j +1}$,
	equivalent to adding a dashed edge with an open circle vertex 
	to an existing open circle vertex. Acting on $\gamma_2$ 
	it produces $\omega_{d_j +1} \gamma_3$, which adds a 
	a dashed edge with an open circle to a dashed 
	line. Finally, acting on $\gamma_m,\,m \geq 3$,
	gives $\omega_{d_j +1} \gamma_{m+1}$, which adds a 
	line with an open circle to an existing dashed vertex. 
	These basic operations are in one-to-one correspondence 
	to those generating the unlabeled graphs $\mathcal{T}_{n+1}$ from 
	$\mathcal{T}_n$, verifying that (\ref{ddeg4}) has the correct range. 
	Starting at $n=1$ with an $\omega_{d_0}$
	assigned to the open circle vertex, one may verify directly that repeated 
	action of (\ref{ddeg4}) produces a sum of terms 
	whose underlying graphs match those in (\ref{ddeg0})
	but with integers $|o|+d_j$, $j=0,1,2 \ldots$, assigned
	to their open circle vertices.  
	The map (\ref{ddeg4}) provides the raison d'\^{e}tre  for 
	the dashed graphs. 
	
	\begin{lemma} 
		\label{difflemma} 
		The recursion (\ref{GamvsW4}) generates only 
		vertex weights $\mu^{\Gamma}(v|L)$ in (\ref{Gamschematic}) 
		that lie in the ($v$-dependent) direct sum of 
		$\mu(\mathcal{T}_n^D)$, for $n=1,\ldots,n_{\rm max}$, $n_{\rm max} \leq d(v)-3$,
		for some integer multiset $D_n$.   
	\end{lemma} 
	
	\begin{proof}  We proceed by induction on $l$ with $L \in \mathcal{L}_{l}$. 
	The assertion holds by inspection of (\ref{grule4.1}), 
	(\ref{grule5.1}), (\ref{grule5.2}) for $l = 4,5$. For 
	the $l -1 \mapsto l$ step in the recursion (\ref{GamvsW4}) 
	we denote by $L_j \in \mathcal{L}_{m_j}$ one of the 1LI graphs in 
	$\Gamma_{m_j}$'s graph expansion and by $L^W \in \mathcal{L}_{l-m}$ 
	one of the 1LI graphs in $W_{l-m}$'s expansion. We focus 
	on one of the vertices $v$ where the graphs are joined and 
	write $v_0$ for $v$'s copy in $L^W$ and $v_j$ for $v$'s copy in $L_j^{\Gamma}$, 
	$j =1,\ldots, k$. By the $W$-graph rule the structure 
	of $L^W$ is irrelevant only the weight $\omega_{d(v_0)}(h)|_{h = H_v}$ 
	(and the inverse symmetric factor ${\rm Sym}(L^W)$
	irrelevant here) enters. If $r$ of the $k$ functional 
	differentiations with respect to some $H_i$ act on the 
	chosen $H_v$ site the weight will be shifted to 
	$\omega_{d(v_0) +r}(h)|_{h = h(\varphi_{v})}$. The associated graph 
	will still be denoted by $L^W$; it now has one rooted vertex 
	$v_0$ to which we attribute multiplicity $r$. 
	The single differentiation 
	of $\Gamma_{m_j}$ with respect to some $\phi_i$ will always produce 
	$1$-rooted graphs, and for the ones rooted at $v_j$ we
	write $L_j^{\Gamma} \in \mathcal{L}_{m_j}^{1 \bullet}$. In any 
	one term contributing to (\ref{GamvsW4}) at $v$, a 
	$v_0$ of multiplicity $r$ will have $r$ 1LI graphs attached, which are
	selected from the $L_j^{\Gamma} \in \mathcal{L}_{m_j}^{1 \bullet}$, 
	$j=1,\ldots,k$. Without loss of generality we take 
	$L_j^{\Gamma} \in \mathcal{L}_{m_j}^{1 \bullet}$, 
	$j=1,\ldots,r$, as the graphs attached to $v_0$. 
	For fixed $r$ the weight associated with $v$ is 
	by (\ref{Gamschematic}) 
	\begin{equation}
	\label{ddeg5} 
	\omega_{d(v_0) +r } \prod_{j=1}^r \partial_{\varphi} \mu^{\Gamma}(v_j|L_j)\,. 
	\end{equation}
	By the induction hypothesis all $\mu^{\Gamma}(v_j|L_j)$
	have an expansion in the ($v_j$-dependent) direct sum of 
	$\mu(\mathcal{T}_n^D), n =1,\ldots, n_{\rm max}$. Focus on a 
	term with $n_j$ open circle vertices in
	$\mu^{\Gamma}(v_j|L_j)$.  For $r=1$ the product (\ref{ddeg5}) 
	is directly of the form (\ref{ddeg4}) and the assertion 
	follows. For $r \geq 2$ one notices that each 
	$\partial_{\varphi} \mu^{\Gamma}(v_j|L_j)$ has one degree
	lower and can be attributed to tree graphs with 
	$n_j$ open circle vertices and an extra edge
	without open circle (so that these trees are 
	not dashed graphs as defined above). The product 
	(\ref{ddeg5}) contains $1\!+\! n_1\! + \ldots \!+ \!n_r$ 
	$\omega_m$ factors and expands into terms that can again be attributed to 
	dashed graphs. This is because the extra $r$ dashed edges are
	joined at the extra open circle vertex 
	associated with $\omega_{d(v_0) + r}$. In other words, the 
	set of dashed graphs is closed under a gluing operation that 
	first creates an external dashed edge and then joins any number 
	of such trees at an extra open circle vertex. 
	As $1 \leq r \leq k \leq [m/2]$ runs through all possible 
	values allowed by (\ref{GamvsW4}) only terms that can be associated 
	with dashed graphs $\mathcal{T}_n^D$, for some $n$ are generated. 
	Comparison with (\ref{GamEuler}) shows that the maximal $n$ that can 
	occur in a normalized weight $\mu^{\Gamma}(v|L)$ is 
	$n_{\rm max} \leq d(v) -3$ (while the actual $n_{\rm max}$ turns 
	out to be much smaller). 
	\end{proof}
	
	It remains to understand the coefficients with which the various 
	dashed graphs occur in $\mu^{\Gamma}(v|L)$.  To this end a different 
	type of labeling turns out to be useful.

	{\bf Assignment of labels:} The labels are set partitions of 
	vertices as frequently used in other contexts. In the situation at 
	hand, the vertex set $\{b_1,\ldots, b_I\}$ will later be identified 
	with the one associated with an articulation vertex $v$ in the block 
	decomposition (as defined at the end of Section II.A) 
	of the underlying one-line irreducible graph $L$. For now we ignore the 
	origin of the set $B = \{b_1,\ldots, b_I\}$ and consider its set
	partitions. If all elements of $B$ are distinct, a set partition 
	of $B$ is a set of non-empty disjoint subsets of $B$ whose union 
	is $B$. An element of a partition is called a cell; we write 
	$\mathcal{S}(B,k)$ for the set partitions of $B$ with $k$ cells.  
	The number of partitions of a set $B$ with 
	$I$ distinct elements into $n$ cells is given by $S(I,n)$, the 
	Stirling number of the second kind. The total number of 
	set partitions is given by the Bell number $B(I) = \sum_{n=1}^I 
	S(I,n)$. A convenient generating function is 
	$\exp\{ y(e^x\! -\!1)\} = \sum_{I,n \geq 0} S(I,n)\, y^n x^I/I!$. Generalizations have been considered in \cite{multisets1}.
	
	The same concept applies to multisets, i.e.~sets of pairs
	$(b_i, m_i)$ where $m_i \in \mathbb{N}$, specifies the multiplicity 
	with which $b_i$ occurs. We write 
	
	\begin{equation} 
	\label{multiset1} 
	B = \{ b_1^{m_1}, \ldots, b_J^{m_J}\} = \{b_i:\, i \in I\} \,,\quad 
	I = \{ \underbrace{1,..., 1}_{m_1}, \ldots, 
	\underbrace{J,...,J}_{m_J}\} \,,
	\end{equation}

	\noindent for the multiset with multiplicities $(m_1,\ldots, m_J)\in \mathbb{N}^J$. 
	In the alternative notation with explicitly repeated elements the 
	indexing $I$ is viewed as a multiset. 
	Two multisets are identical iff they contain the same elements 
	with the same multiplicities.  
	The partitions of a multiset are defined as the set partitions of 
	a $|I|=\sum_j m_j$ element set where $m_j$ copies of $b_j$ are 
	identified afterwards and `duplicates' are omitted from the list. 
	There are several notions of `duplicates' one can use; we allow 
	both repeated cells and repeated elements within a cell 
	but eliminate duplicates of the same partition.  
	For example, $\mathcal{S}(\{a,b,c^2\}, 3)$ has $4$ elements, 
	$\{\{a\}, \{b\}, \{c^2\}\}$, $\{\{a\}, \{b,c\}, \{c\}\}$,  
	$\{\{a,b\}, \{c\}, \{c\}\}$, $\{\{a,c\}, \{b\}, \{c\}\}$,  
	as opposed to $|\mathcal{S}(\{a,b,c,d\}, 3)| =6$.
	
	Unless specified otherwise we allow $B$ in the following to 
	be a multiset of the form (\ref{multiset1}). A partition of $B$ with 
	$n$ cells is then used to label the open circle vertices of 
	a graph in $\mathcal{T}_n$. One may think of each open circle vertex as a `bag' 
	that contains a cell. Technically, the labeling map is 
	for each partition $\pi \in\mathcal{S}(B,n)$, a bijection 
	$\nu_0 \mapsto \nu_0^{\pi}$, of sets of cardinality $n$. 
	While the vertices $\nu_0$ of the unlabeled graph 
	may be assigned `dummy' labels that can be freely permuted 
	in probing for isomorphisms the elements of 
	$\nu_0^{\pi}$ can only be permuted if their labels coincide. 
	Isomorphic labeled graphs are defined as 
	in Section II.A, with $V = \nu_0^{\pi}$ as vertex set. 
	We write $\mathcal{T}(B,n)$, $1 \leq n \leq I$, for the set of topologically 
	inequivalent dashed graphs with $n$ open circle vertices labeled by 
	$\mathcal{S}(B,n)$. Further, for some unlabeled $t \in \mathcal{T}_n$ we write 
	$T \in \mathcal{T}(B,n)$ for one of its labeled counterparts.  
	
	As an illustration consider $n=3$. The set partitions of 
	$B= \{b_1,b_2,b_3\}$ are  
	\begin{eqnarray}
	\label{dashed1}
	&& \Big\{\{b_1,b_2,b_3\}\Big\}, 
	\nonumber\\
	&& \Big\{\{b_1\},\{b_2,b_3\}\Big\},\,\Big\{\{b_2\},\{b_1,b_3\}\Big\},\,\Big\{\{b_3\},\{b_1,b_2\}\Big\},
	\nonumber\\
	&& \Big\{ \{b_1\}, \{b_2\}, \{b_3\}\Big\}\,.
	\end{eqnarray}
	These are then assigned as labels to the open circle vertices of the graphs 
	in $\mathcal{T}_1,\mathcal{T}_2,\mathcal{T}_3$: 
	\begin{eqnarray}
	\label{dashed2} 
	&\!\!&\!\! \{b_1,b_2,b_3\} \in \begin{tikzpicture}
	\draw[fill=white] (-0.3,0.)circle (3pt) ;
	\end{tikzpicture} 
	\nonumber\\
	&\!\!&\!\! \{b_1\} \in \begin{tikzpicture}
	\draw[fill=white] (-0.2,0.)circle (3pt) ;
	\draw[black, dashed] (0,0) -- (0.6,0);
	\draw[fill=white] (0.8,0)circle (3pt) ;
	\end{tikzpicture} \ni \{b_2,b_3\}\,,\;\;\;
	\{b_2\} \in \begin{tikzpicture}
	\draw[fill=white] (-0.2,0.)circle (3pt) ;
	\draw[black, dashed] (0,0) -- (0.6,0);
	\draw[fill=white] (0.8,0)circle (3pt) ;
	\end{tikzpicture}\ni \{b_1,b_3\}\,,\;\;\;
	\{b_3\} \in \begin{tikzpicture}
	\draw[fill=white] (-0.2,0.)circle (3pt) ;
	\draw[black, dashed] (0,0) -- (0.6,0);
	\draw[fill=white] (0.8,0)circle (3pt) ;
	\end{tikzpicture} \ni \{b_1,b_2\}\,,
	\nonumber\\
	&\!\!& \!\!
	\{b_1\} \in \begin{tikzpicture}
	\draw[black, dashed] (0,0) -- (0.5,0.5);
	\draw[black, dashed] (0.5,0.5) -- (1,0);
	\draw[fill=white]  (0,0) circle (3pt) ;
	\draw[fill=white] (0.5,0.5) circle (3pt) ;
	\draw[fill=white]  (1,0) circle (3pt) ;
	\node at (0.5,0.9) { $\{b_2\} $};
	\end{tikzpicture}
	\ni \{b_3\} \,,\quad\hspace*{0.5cm}
	\{b_2\} \in \begin{tikzpicture}
	\draw[black, dashed] (0,0) -- (0.5,0.5);
	\draw[black, dashed] (0.5,0.5) -- (1,0);
	\draw[fill=white]  (0,0) circle (3pt) ;
	\draw[fill=white] (0.5,0.5) circle (3pt) ;
	\draw[fill=white]  (1,0) circle (3pt) ;
	\node at (0.5,0.9) { $\{b_1\} $};
	\end{tikzpicture}
	\ni \{b_3\} \,,\quad\hspace*{0.5cm}
	\{b_1\} \in \begin{tikzpicture}
	\draw[black, dashed] (0,0) -- (0.5,0.5);
	\draw[black, dashed] (0.5,0.5) -- (1,0);
	\draw[fill=white]  (0,0) circle (3pt) ;
	\draw[fill=white] (0.5,0.5) circle (3pt) ;
	\draw[fill=white]  (1,0) circle (3pt) ;
	\node at (0.5,0.9) { $\{b_3\} $};
	\end{tikzpicture}
	\ni \{b_2\} \,,
	\nonumber\\
	&\!\!& \!\! \{b_1\} \in 
	\begin{tikzpicture}
	\draw[black, dashed]  (0,0) -- (0.5,0.5);
	\draw[black, dashed]  (0.5,0.5) -- (1,0);
	\draw[black, dashed]  (0.5,0.5)-- (0.5,1);
	\draw[fill=white]  (0,0) circle (3pt) ;
	\draw[fill=white]  (1,0) circle (3pt) ;
	\draw[fill=white]  (0.5,1) circle (3pt) ;
	\node at (0.5,1.4) { $\{b_2\} $};
	\end{tikzpicture}
	\ni \{b_3\} 
	\end{eqnarray}
	Clearly, none of the labeled graphs (\ref{dashed2}) allows 
	for nontrivial automorphisms. This may change when multisets are 
	used to generate the labels.
	
	{\bf Symmetry factors:} Each unlabeled $t \in \mathcal{T}_n$ has an 
	automorphism group which we define in the obvious way:
	let $\nu_0$ be the set of open circle vertices,
	$\nu_1$ the set of dashed vertices, and $\epsilon \subset 
	(\nu_0 \cup \nu_1)_2$ the 
	edge list, subject to the constraints in the definition. An  
	automorphism of $t$ is a permutation of $\nu_0 \cup \nu_1$ 
	that leaves $\nu_0$, $\nu_1$ and the edge list separately 
	invariant. These form a group for which we write 
	${\rm Aut}(t)$. 
	
	The labeling process described above precludes nontrivial 
	permutations except when $B$ has repeated elements. 
	For a multiset (\ref{multiset1}) the $m_i$ copies
	of $b_i$ can be permuted giving rise to a direct 
	product ${\rm Perm}(B) := S_{m_1} \times \ldots \times S_{m_J}$    
	of symmetric groups acting on $B$. For  a partition  
	$\pi \in \mathcal{S}(B,n)$ let $\nu_0^{\pi} = \{ (o_i, c_i):\,
	i =1,\ldots, n\}$ be the set of $n$ labeled open circle vertices.
	Note that each dummy labeled $o_i$ is paired with non-dummy $c_i$, 
	the index $i$ merely enumerates the list of pairs. 
	Each cell $c_i, i =1,\ldots,n$ may again be a multiset 
	$\{b_1^{c_{i,1}}, \ldots , b_J^{c_{i,J}}\}$ with multiplicities 
	$c_{i,1}, \ldots, c_{i,J} \in \mathbb{N}_0$, where we set the multiplicity 
	to zero if the element does not occur.  A subgroup ${\rm Perm}(c_i) = 
	S_{c_{i,1}} \times \ldots \times S_{c_{i,J}} \subset {\rm Perm}(B)$ 
	(with factors absent whose multiplicity is zero) will still 
	permute copies of the same elements that $c_i$ may contain. 
	By the very process of forming set partitions
	the direct product ${\rm fix}(\nu_0^B):= {\rm Perm}(c_1) \times 
	\ldots \times {\rm Perm}(c_n) \subset {\rm Perm}(B)$ 
	is still a subgroup, with $S_{c_{1,1}} \times \ldots \times 
	S_{c_{n,1}} \subset S_{m_1}$, etc.  In other words, ${\rm fix}(\nu_0^B)$ 
	is the subgroup of ${\rm Perm}(B)$ that maps individual labels of 
	$T \in \mathcal{T}(B,n)$ into themselves. 
	Writing $|{\rm fix}(\nu_0^B)|$
	for its order, one has by Lagrange's theorem $|{\rm Perm}(B)|/
	|{\rm fix}(\nu_0^B)| \in \mathbb{N}$. 
	
	An automorphism of $T \in \mathcal{T}(B,n)$ is defined as 
	in the unlabeled 
	case, except that the unlabeled set $\nu_0$ of $t \in \mathcal{T}_n$ 
	is replaced with the labeled one $\nu_0^{\pi} = 
	\{ (o_i,c_i):\, i=1,\ldots, n\}$, $\pi \in \mathcal{S}(B,n)$. Since 
	the $o_i$ labels are dummy two elements of $\nu_0^{\pi}$
	are regarded as equal iff their cells $c_i$ are equal 
	as multisets. The $n$ labeled vertices can thus be grouped 
	into subsets with the same label. An automorphism of the underlying 
	unlabeled graph $t$ that affects only sets of 
	equally labeled vertices 
	is also an automorphism of $T$, and all automorphisms of $T$ 
	arise in that way. They form again a group, denoted by 
	${\rm Aut}(T)$, which is a subgroup of ${\rm Aut}(t)$. 
	Finally, the {\it symmetry factor} of a labeled tree graph 
	$T \in \mathcal{T}(B,n)$ is defined by 
	\begin{equation} 
	\label{SymTdef}
	{\rm Sym}(T) = |{\rm Aut}(T)| |{\rm fix}(\nu_0^B)|\,.
	\end{equation}

	As an illustration of these concepts, reconsider the graphs in 
	(\ref{dashed2}) but now labeled by the set partitions of $\{b,b,b'\}$. 
	The symmetry factors (\ref{SymTdef}) may differ from $1$ and 
	are noted to the right of each graph:
	\begin{eqnarray}
	\label{dashed3} 
	&& \{ b,b,b'\} \in \begin{tikzpicture}
	\draw[fill=white] (-0.3,0.)circle (3pt) ;
	\end{tikzpicture} 
	\quad {\rm Sym} =2
	\nonumber\\
	&& \{b\} \in \begin{tikzpicture}
	\draw[fill=white] (-0.3,0.)circle (3pt) ;
	\draw[black, dashed] (0,0) -- (0.7,0);
	\draw[fill=white] (1,0)circle (3pt) ;
	\end{tikzpicture} \ni \{b,b'\}\quad
	{\rm Sym} = 1\,, \quad 
	\{b'\} \in \begin{tikzpicture}
	\draw[fill=white] (-0.3,0.)circle (3pt) ;
	\draw[black, dashed] (0,0) -- (0.7,0);
	\draw[fill=white] (1,0)circle (3pt) ;
	\end{tikzpicture}\ni \{b,b\}\quad
	{\rm Sym} =2\,, 
	\nonumber\\
	&& 
	\{b\} \in \begin{tikzpicture}
	\draw[black, dashed] (0,0) -- (0.5,0.5);
	\draw[black, dashed] (0.5,0.5) -- (1,0);
	\draw[fill=white]  (0,0) circle (3pt) ;
	\draw[fill=white] (0.5,0.5) circle (3pt) ;
	\draw[fill=white]  (1,0) circle (3pt) ;
	\node at (0.5,0.9) { $\{b\} $};
	\end{tikzpicture}
	\ni \{b'\} \quad {\rm Sym} =1\,,\quad \hspace*{0.8cm}
	\{b\} \in \begin{tikzpicture}
	\draw[black, dashed] (0,0) -- (0.5,0.5);
	\draw[black, dashed] (0.5,0.5) -- (1,0);
	\draw[fill=white]  (0,0) circle (3pt) ;
	\draw[fill=white] (0.5,0.5) circle (3pt) ;
	\draw[fill=white]  (1,0) circle (3pt) ;
	\node at (0.5,0.9) { $\{b'\} $};
	\end{tikzpicture}
	\ni \{b\} \quad {\rm Sym} =2\,, 
	\nonumber\\ 
	&&\{b\} \in 
	\begin{tikzpicture}
	\draw[black, dashed]  (0,0) -- (0.5,0.5);
	\draw[black, dashed]  (0.5,0.5) -- (1,0);
	\draw[black, dashed]  (0.5,0.5)-- (0.5,1);
	\draw[fill=white]  (0,0) circle (3pt) ;
	\draw[fill=white]  (1,0) circle (3pt) ;
	\draw[fill=white]  (0.5,1) circle (3pt) ;
	\node at (0.5,1.4) { $\{b\} $};
	\end{tikzpicture}
	\ni \{b'\} \;\quad {\rm Sym} =2
	\end{eqnarray}
	We now claim that 
	\begin{equation}
	\label{BTratio}  
	|{\rm Perm}(B)|/{\rm Sym}(T) \in \mathbb{N}\,,
	\end{equation}
	We present a direct proof on the level of multisets here.
	In Section IV the result is recovered along different lines. 
	Suppose that $\tilde{p}$ of the elements of $\nu_0^{\pi}$ of $T$ 
	are equally labeled, and that there is a subgroup $A$ of 
	${\rm Aut}(t)$ that acts transitively on $\{o_1,\ldots, o_p\}$ 
	of $t$, with $p\leq \tilde{p}$. 
	A may act on vertices other than
	$\{o_1,\ldots,o_p\}$, once these are labeled only a subgroup 
	$A_0$ of $A$ may act on the equally 
	labeled $\{(o_j,c_j):\,j=1,\ldots,n\}$,
	$c_1 = \ldots = c_p$. Note that $p!/|A_0|$ is an integer. 
	We wish to lift this $A_0 \subset {\rm Aut}(t)$ to an 
	automorphism group of the labeled version $T$ of $t$ 
	with the $p$ equally labeled $\{(o_j,c_j):\,j=1,\ldots,p\}$.
	Focus on one of the elements of the $p$ identical cells
	$c_j$  with nonzero multiplicity, say 
	$b_1^{c_{j,1}}$, $c_{j,1} = n_1 \geq 1$, 
	$j =1,\ldots ,p$, without loss of generality. Overall
	these are $p n_1$ copies of $b_1$ which arose by 
	distributing the original $m_1 \geq p n_1$ copies 
	from (\ref{multiset1}) to the cells under consideration 
	and possibly others. Hence there is a subgroup 
	$S_{p n_1} \subset S_{m_1}$ that permutes the copies of $b_1$ 
	within each cell and mixes the $b_1$ sectors of different 
	cells. The permutations within each cell are part of 
	${\rm fix}(\nu_0^B)$ and have total order $(n_1!)^p$.  
	The other permutations implement the desired 
	automorphisms of $T$ within $S_{pn_1}$. The relevant ratio 
	thus is $p!/|A_0| \in \mathbb{N}$ times $ (p n_1)!/[p! (n_1!)^p]$. The 
	latter is indeed an integer for all $n_1 \in \mathbb{N}$. Repeating 
	the argument for 
	all elements of the $p$ identical cells with nonzero 
	multiplicities one arrives at $|{\rm Perm}(B)|/{\rm Sym}(T) 
	\in \mathbb{N}$. This argument does not rely on a group structure 
	of the permutations of elements across cells.  
	
	\subsection{Formulation and illustration of the graph rule}
	
	We now return to the previous result (\ref{Gamschematic}) and 
	provide a graph rule for the missing ingredient 
	$\mu^{\Gamma}(v|L)$, where $L \in \mathcal{L}_{l}$ is a 1LI 
	graph with $l$ edges and $v$ is one of its vertices. 
	Recall the notion of  a block decomposition from the end of 
	Section II.A. Each $L \in \mathcal{L}_{l}$ is either itself 1VI or 
	has a block decomposition $\{L_1, \ldots, L_N\}$, 
	in terms of maximal 1VI subgraphs $L_j = (B_j,E_j)$, 
	$j=1,\ldots, N$, which must also be 1LI. 
	Each articulation vertex occurs in more than one $B_j$, 
	while non-articulation vertices occur in precisely one $B_j$.  
	For a fixed articulation vertex $v$ let $B(v) = \{ L_i(v) = (B_i,E_i),
	\,i \in I(v)\}$,  with $2 \leq |I(v)| \leq N$, be the subset of 
	blocks with $v \in B_i$. Isometric blocks can be permuted, we denote 
	this permutation group by ${\rm Perm}(B(v))$ and its order by 
	$|{\rm Perm}(B(v))|$. We write $b_i(v)$ for the copy of 
	$v$ in $B_i$, $i \in I(v)$, and treat the copies $b_i(v)$ as 
	identical, $b_i(v) = b_{i'}(v)$, iff $L_i(V)$ and $L_{i'}(v)$ are 
	isomorphic. Viewed as a vertex in $L_i(v)$ each $b_i(v)$ has a 
	degree (with respect to the full lines) $d(b_i(v)) \geq 2$
	such that $\sum_{i \in I(v)} d(b_i(v)) = d(v)$.   
	Non-articulation points can formally be included in this 
	setting by allowing $|I(v)|=1$. 
	
	With this convention the edge sets are redundant and we also 
	write $B(v)= \{b_i(v):\,i \in I(v)\}$, on which the same 
	permutation group ${\rm Perm}(B(v))$ acts. In general 
	$B(v)$ will be a multiset for which use the notations 
	(\ref{multiset1}). Then ${\rm Perm}(B(v))$ is a direct product 
	$S_{m_1} \times \ldots\times S_{m_J}$ of symmetric groups. Next 
	we generate the set partitions $\mathcal{S}(B(v),n)$ with $n =1,\ldots ,|I(v)|$ 
	cells. Each cell $c_i = \{b_1^{c_{i,1}}, \ldots, b_J^{c_{i,J}} \}$ may again 
	be a multiset whose multiplicities obey $\sum_{i=1}^n c_{i,j} = m_j$
	and $\sum_{i=1}^n \sum_{j=1}^J c_{i,j} = |I(v)|$. The cells are 
	used to label the open circle vertices of the tree graphs in 
	$\mathcal{T}_n$, $1 \leq n\leq |I(v)|$, as in Section III.A; coinciding labels 
	are allowed and correspond to cells coinciding as multisets. 
	We write $\mathcal{T}(B(v),n)$ for the set of topologically 
	distinct labeled dashed graphs with $n$ open circle vertices 
	labeled by $\mathcal{S}(B(v),n)$. Individual labeled graphs are denoted by 
	$T \in \mathcal{T}(B(v),n)$, with ${\rm Sym}(T)$ the symmetry factor. 
	\medskip
	
	\begin{theorem}[Graph rules for vertex weights]
		\label{graphrule}
		The weights $\mu^{\Gamma}(v|L)$ in (\ref{Gamschematic}) depend 
		only on the block decomposition $B(v)$ of $L$ at $v$ and 
		can be obtained by the following graph rule:  
		\begin{itemize} 		
			\vspace{-2mm}
			\item[(a)] A weight $\mu(T)$ is assigned to each labeled graph 
			$T \in \mathcal{T}(B(v),n)$, $1 \leq n \leq |I(v)|$, 
			as follows: for an $|o|$-valent 
			(with respect to the dashed lines) open-circle vertex $o$ labeled by the cell 
			$c_i = \{b_1^{c_{i,1}}, \ldots, b_J^{c_{i,J}} \}$ write a factor 
			$\omega_{|o|+ d(c_i)}(\varphi_v)$, $d(c_i) := \sum_{j=1}^J c_{i,j} d(b_j)$, for 
			each dashed line connecting two open circle vertices a factor 
			$\gamma_2(\varphi_v)$, and for each vertex with $m \geq 3$ intersecting 
			dashed lines a factor $\gamma_m(\varphi_v)$ (the dashed lines 
			that intersect at the dashed-vertex do not contribute a factor). The 
			resulting monomial $\mu(T)$ in $\gamma_2,\omega_m,\gamma_m, m \geq 3$, has 
			by (\ref{ddeg2}) degree $\sum_{i=1}^n d(c_i) = \sum_{j=1}^J m_j \,d(b_j) 
			= d(v)$. 
			\item[(b)] Multiply $\mu(T)$ by 
			\begin{equation} 
			\label{grule1} 
			(-)^{s(T)} \frac{ |{\rm Perm}(B(v))|}{ {\rm Sym}(T)}\,,
			\end{equation}
			where $s(T)$ is the sum of the number of dashed lines and the number 
			of dashed vertices. Further, $|{\rm Perm}(B(v))|$ is the order of 
			the permutation group  of the blocks at $v$, and 
			${\rm Sym}(T)$ is the symmetry factor of the labeled 
			dashed graph as defined in Section III.A. 
			\item[(c)] Sum the contributions from (a),(b) over all 
			$n$ and $T \in \mathcal{T}(B(v),n)$ to obtain 
			$\mu^{\Gamma}(v|L) = \mu^{\Gamma}(v|B)$ as 
			\begin{equation} 
			\label{grule2}
			\mu^{\Gamma}(v|B) =  \sum_{n=1}^{|I(v)|} \sum_{T \in \mathcal{T}(B(v),n)} (-)^{s(T)} 
			\frac{|{\rm Perm}(B(v))|}{{\rm Sym}(T)} \mu(T)\,.
			\end{equation}
			This is normalized such that $\mu^{\Gamma}(v|L) = \omega_{d(v)}(\varphi_v)$ 
			for a non-articulation vertex ($|I(v)|=1$) of degree $d(v)$. 
		\end{itemize}
	\end{theorem} 
	
	\noindent {\bf Illustration of the graph rule: }
	
	\noindent(i) The simplest case is the ``pair of glasses'' graph in (\ref{grule4.1}). 
	It has two isomorphic blocks 
	\begin{tikzpicture}
	\draw[black, thick] (0,0) edge[bend left=30] (1,0);
	\draw[black, thick] (0,0) edge[bend right=30] (1,0);
	\filldraw[black] (0,0) circle (2pt) ;
	\filldraw[black] (1,0) circle (2pt) ;
	\end{tikzpicture}, 
	joined at the articulation point. Hence $|{\rm Perm}(B(v))|=2$. 
	The vertex set $B(v)$ contains two copies of the same element, 
	$\{b,b\}$, say, with $d(b)=2$. The set partitions of $B(v)$ are $\{\{b,b\}\}$ 
	and $\{\{b\},\{b\}\}$. Thus the labeled dashed graphs $T \in \mathcal{T}(B(v),1), 
	\mathcal{T}(B(v),2)$, are  
	\begin{eqnarray}
	\label{gill4.1}
	\{b,b\} \in \begin{tikzpicture}
	\draw[fill=white] (-0.3,0.)circle (3pt) ;
	\end{tikzpicture} \,,\quad 
	\{b\} \in \begin{tikzpicture}
	\draw[fill=white] (-0.2,0.)circle (3pt) ;
	\draw[black, dashed] (0,0) -- (0.6,0);
	\draw[fill=white] (0.8,0)circle (3pt) ;
	\end{tikzpicture} \ni \{b\}.
	\end{eqnarray}
	They have each ${\rm Sym}(T) =2$, and contribute 
	$\omega_4(\varphi)$, $-\gamma_2(\varphi) \omega_3(\varphi)^2$, respectively,
	in the sum (\ref{grule2}). This reproduces the weight in (\ref{grule4.1}). 
	
	\noindent (ii) As a more complicated exemplification consider (\ref{grule9.1}). 
	At $v'$ three block are joined: two copies of 
	\begin{tikzpicture}
	\draw[black, thick] (0,0) edge[bend left=30] (1,0);
	\draw[black, thick] (0,0) edge[bend right=30] (1,0);
	\filldraw[black] (0,0) circle (2pt) ;
	\filldraw[black] (1,0) circle (2pt) ;
	\end{tikzpicture}, 
	and 
	\begin{tikzpicture}
	\draw[black, thick] (0,0) edge[bend left=30] (1,0);
	\draw[black, thick] (0,0) edge[bend right=30] (1,0);
	\draw[black] (0,0) -- (1,0);
	\filldraw[black] (0,0) circle (2pt) ;
	\filldraw[black] (1,0) circle (2pt) ;
	\end{tikzpicture}. Hence $|{\rm Perm}(B(v'))|=2$. 
	The vertex set $B(v') = \{b,b,b'\}$ with 
	$d(b) = 2$, $d(b')=3$, gives rise to the 
	labeled tree graphs presented in (\ref{dashed3}). 
	The sum (\ref{grule2}) evaluates to 
	\vspace{-1mm}
	\begin{eqnarray}
	&& \mu(v'|L) = \frac{2}{2} \omega_7(\varphi_{v'})- \frac{2}{1} \gamma_2(\varphi_{v'}) \omega_3(\varphi_{v'}) \omega_6(\varphi_{v'})  - \frac{2}{2} \gamma_2 (\varphi_{v'})\omega_4(\varphi_{v'}) \omega_5(\varphi_{v'}) 
	\\[2mm]
	&& \quad +  
	\frac{2}{2} \gamma_2^2 (\varphi_{v'})\omega_3^2(\varphi_{v'})\omega_5(\varphi_{v'})+\frac{2}{1} \gamma_2^2 (\varphi_{v'})\omega_3 (\varphi_{v'})
	\omega_4^2(\varphi_{v'}) + \frac{2}{2} \gamma_3 (\varphi_{v'})\omega_3^2(\varphi_{v'}) \omega_4(\varphi_{v'}),
	\nonumber
	\end{eqnarray}
	in agreement with 
	(\ref{grule9.1}). At $v$ two distinct blocks are joined, 
	\begin{tikzpicture}
	\draw[black, thick] (0,0) edge[bend left=30] (1,0);
	\draw[black, thick] (0,0) edge[bend right=30] (1,0);
	\filldraw[black] (0,0) circle (2pt) ;
	\filldraw[black] (1,0) circle (2pt) ;
	\end{tikzpicture} 
	and 
	\begin{tikzpicture}
	\draw[black, thick] (0,0) edge[bend left=30] (1,0);
	\draw[black, thick] (0,0) edge[bend right=30] (1,0);
	\draw[black] (0,0) -- (1,0);
	\filldraw[black] (0,0) circle (2pt) ;
	\filldraw[black] (1,0) circle (2pt) ;
	\end{tikzpicture}. Hence $|{\rm Perm}(B(v))|=1$. 
	The vertex set is $B(v) = \{ b,b'\}$, with $d(b) =2$, 
	$d(b') =3$. The labeled tree graphs are as in (\ref{gill4.1}) 
	but with distinct elements $b,b'$. Both have symmetry factor 
	$1$ and the sum (\ref{grule2}) gives $\mu(v|L) = \omega_5(\varphi_{v}) 
	- \gamma_2(\varphi_{v})  \omega_3(\varphi_{v})  \omega_4(\varphi_{v}) $, again in 
	agreement with (\ref{grule9.1}). 
	
	
	\subsection{Proof of the graph rule} 
	
	We first bring into focus what needs to be shown. 
	By Lemma \ref{difflemma} each weight $\mu^{\Gamma}(v|L)$ lies in 
	the direct sum of $\mu(\mathcal{T}_n^D), n =1 \ldots,n_{\rm max}$, 
	$n_{\rm max} \leq d(v) -3$, for some integer multiset $D_n$. It 
	is convenient to introduce a projection operation 
	\begin{equation} 
	\label{gproof1} 
	{\rm pr}: \mathcal{T}(B(v),n) \longrightarrow \mathcal{T}(D(v),n)\,,\quad 
	T \mapsto {\rm pr}(T)\,,
	\end{equation}
	where $D(v) = d(B(v))$, as a multiset. Further, each 
	cell $c_i$ labeling $T \in \mathcal{T}(B(v),n)$ is 
	replaced by its integer degree sum $d(c_i)$. The result is 
	an element of $\mathcal{T}^D_n$, where the integers $D_n$ are 
	$d(\pi) := \{ d(c_1), \ldots, d(c_n) \}$ (viewed as a multiset), 
	if $\pi = \{ c_1,\ldots, c_n\}$ is the partition labeling $T$.
	Note that each $D_n$ is drawn from the $n$-element partitions of 
	$D(v) = \{ d(b_i):\, i \in I(v)\}$, with each integer 
	in $D_n$ equal to the sum of the integers in the cell. 
	Since distinct $b_i$'s can have the same degree and 
	many degree sums $d(c_i)$ can be equal as well, the 
	projected label set $\mathcal{S}(D(v),n)$ will in general be of 
	much smaller cardinality than $\mathcal{S}(B(v),n)$. We write 
	$\mathcal{T}(D(v),n)$ for the set of topologically inequivalent 
	labeled dashed graphs with $n$ open circle vertices 
	labeled by some $D_n \in \mathcal{S}(D(v),n)$.

	Clearly, $\mu(T) = \mu(\tau)$, for $\tau = {\rm pr} T$, if 
	$\mu(\tau)$ is formed according to (\ref{ddeg3}). 
	The graph rule is therefore compatible with Lemma 
	\ref{difflemma} and the projection (\ref{gproof1}). 
	What remains to be shown is: $n_{\rm max} = |I(v)|$, and 
	\begin{equation}
	\label{gproof2} 
	\mbox{coefficient of} \;\mu(T) = (-)^{s(T)} 
	\frac{|{\rm Perm}(B(v))|}{{\rm Sym}(T)}\,,\quad 
	T \in \mathcal{T}(B(v),n)\,,\; n = 2, \ldots n_{\rm max} \,.
	\end{equation}
	The case $n\!=\!1$ is accounted for by (\ref{GamEuler}) and 
	can be omitted. Indeed, for $n\!=\!1$ the only $T \in 
	\mathcal{T}(B(v),1)$ graph 
	is an open circle labeled by $c_1 = B(v)$, the groups 
	${\rm Perm}(B(v))$ and ${\rm fix}(\nu_0^B)$ coincide and the 
	coefficient of $\mu(T) = \omega_{d(v)}$ is $1$, in agreement 
	with (\ref{GamEuler}).  
	By Lemma \ref{lemmanoart} we can also match the situation 
	where only the $\omega_{d(v)}$ term in (\ref{gproof1}) is present 
	to graphs $L$ without articulation points, in agreement 
	with the graph rule. The key step is: 
	
	\begin{lemma} 
		\label{oneartlemma}
		For 1LI graphs with one articulation point (\ref{gproof2}) and 
		hence the graph rule (a),(b),(c) is compatible with the recursion 
		(\ref{GamvsW4}).
	\end{lemma} 
	
	\begin{proof}  We proceed by induction in $l$ by assuming that $(a),\,(b),\,(c)$ 
	of the graph rule produce the correct vertex expressions for orders 
	$1,\ldots,l-1$. Let $L\in \mathcal{L}_l$ be a 1LI graph with a single 
	articulation vertex $v$, and $L_1(v),\ldots,L_I(v)$ is its block decomposition. 
	The $L_i(v)$ are viewed as $1$-rooted graphs, with roots $b_i$ regarded 
	as identical iff the $L_i(v)$ are isomorphic as 1VI graphs. With this 
	convention $B(v)$ is a multiset of the form (\ref{multiset1}) which 
	codes the structure of $L$ at $v$. 
	In the recursion (\ref{GamvsW4}) the contribution coming from $L$ 
	is reassembled from its block decomposition by gluing together 
	various blocks arising from the graph expansion of the 
	$W^{(k)}_{l-m},\,\Gamma_{m_1}^{(1)},\ldots,\Gamma_{m_k}^{(1)}$ pieces. 
	The weights associated with non-articulation vertices are known
	from Lemma \ref{lemmanoart} so we can focus on $v$.

	At $v$ the structure of the 1LI graph $L^W$ induced by 
	$W^{(k)}_{l-m}$ is irrelevant
	(as in the proofs of Lemmas \ref{difflemma} and \ref{localitylemma}) 
	and only the shifted weight $\omega_{d_0 + k}$ enters, where $d_0=d(v_0)$ is the 
	valency (wrt the full lines) of $v$'s copy $v_0$ in the 1LI graph 
	associated with 
	$W_{l-m}$. The shift counts the number of $h$-derivatives acting on 
	$\omega_{d_0}(h)|_{h = H_v}$; since $L$ has by assumption only one 
	articulation point all $k$ derivatives in $W^{(k)}_{l-m}$ must 
	act on the same vertex weight, viz  $\omega_{d_0}(h)$. In particular $v_0$ 
	should be viewed a rooted vertex with multiplicity $k$. Attached to 
	$v_0$ will be $k$ 1LI graphs $L^{\Gamma}_j \in \mathcal{L}_{m_j}^{1 \bullet}$ 
	that arise from the graph expansion of $\Gamma_{m_j}^{(1)}$, respectively, 
	and that are rooted at some $v_j$.  Each of the $L^{\Gamma}_j$ 
	may decompose into several blocks at $v_j$, in the above convention 
	we may write $B(v_j)$ for the set of blocks stemming from 
	$L_j^{\Gamma} \in \mathcal{L}_{m_j}^{1 \bullet}$. Similarly, $L^W$ may  
	decompose into several blocks at $v_0$ and we write 
	$B(v_0)$ for their vertex set. Then $B(v)$ is the union of 
	the $B(v_j)$, $j=0,1,\ldots, k$, as $k$ runs through all possible 
	values in (\ref{GamvsW4}). For fixed $k$ the weight associated
	with $v$ can be written in terms of its copies in $B(v_j)$ 
	as 
	\begin{equation} 
	\label{gproof3}
	\omega_{d(v_0) + k} \prod_{j=1}^k \partial_{\varphi} \mu^{\Gamma}(v_j| B(v_j)) \,,
	\quad k =1,\ldots, [m/2]\,,
	\end{equation}
	where by induction hypothesis each $\mu^{\Gamma}(v_j|B(v_j))$ is given 
	by (\ref{grule2}). In particular, each $\mu^{\Gamma}(v_j|B(v_j))$
	expands into contributions associated with 
	dashed graphs in $\mathcal{T}(B(v_j),n_j)$, $n_j =1,\ldots, |I(v_j)|$,
	labeled by the set partitions of $B(v_j)$ with $n_j$ cells,
	$j=1,\ldots,k$.  The minimal number of cells is $k$, the maximal 
	number of cells is $\sum_{j=1}^k |I(v_j)| = |I(v)|- 1$. 
	The blocks in $B(v_0)$ are not subject to the $\Gamma$
	graph rule  but to the $W$ graph rule and thus only 
	give rise to the $\omega_{d(v_0) + k}$ factor in (\ref{gproof3}). 
	In terms of the integer labeled dashed graphs $d(v_0)$ labels a
	$k$-valent open circle vertex $o \in \nu_0$ and 
	conversely every $k$-valent open circle vertex in 
	an integer labeled dashed graph can be associated with 
	a $W_{l-m}^{(k)}$ induced piece. Each of the $k$-subtrees 
	joined to it has a dashed edge generated by $\partial_{\varphi}$ without 
	open circle vertex,  while all of its $n_j$ open circle vertices 
	are labeled by the set partitions in $\mathcal{S}(n_j,B(v_j))$, 
	$j=1,\ldots,k$.  The result is a dashed graph $T'$ with 
	$n =1 + \sum_{j=1}^k n_j \in \{k+1,\ldots,|I(v)|\}$ open circle 
	vertices, all but one of which are labeled by set partitions 
	drawn from those of the individual $B(v_j)$'s. In the setting 
	of (\ref{gproof2}) this fixes $n_{\rm max} = |I(v)|$. Subject to 
	the degree constraint $d(c_0) = d(v_0)$ the  so far only integer 
	labeled vertex $v_0$ can also be labeled by some vertex set $c_0$. 
	
	We are free to postulate that $T'$  ought to be relabeled -- 
	while preserving the weight --  
	by the much larger set of set partitions of $B(v)$, viewed as the 
	union of $B(v_j)$, $j=0,1,\ldots,k$, to produce a graph in 
	$T \in \mathcal{T}(B(v),n)$, $n=k+1,\ldots, |I(v)|$. For later reference 
	we mark the transition from $T'$ to $T$ with $\mu(T') = \mu(T)$ 
	by ($\ast$). Summing the contributions (\ref{gproof3}) over 
	all $k=1,\ldots, [m/2]$, $m=2, \ldots, l-2$, we know that 
	the result must be of the form (\ref{grule2}) but with a
	yet undetermined coefficient of $\mu(T)$. As noted after 
	(\ref{gproof1}) the weight only depends on the 
	projection ${\rm pr}(T)$ of the graph, but we are free 
	to stipulate that the integers occurring are the 
	degree sums of the cell partitions of $B(v)$, i.e.~$d_i = d(c_i)$. 
	Then part (a) of the graph rule holds by construction and only the 
	assertion about coefficient (\ref{gproof2}) needs to be shown.

	By the induction hypothesis each the $\Gamma_{m_j}^{(1)}$ induced 
	pieces comes with a $1/{\rm Sym}(L_j^{\Gamma})$ factor where 
	$L_j^{\Gamma}\in \mathcal{L}_{m_j}^{1 \bullet}$ is a $1$-rooted 1LI graph. 
	Similarly, by the $W$ graph rule the $W^{(k)}_{l -m}$ 
	induced piece carries a $1/{\rm Sym}(L^W)$ factor, where 
	$L^W$ is also a $1$-rooted 1LI graph but with the 
	root attributed multiplicity $k$, as seen above. Since we 
	focus on 1LI graphs that combine to $L$ (and $L$ has only 
	one articulation point) all $L^W,L_1^{\Gamma},\ldots, L_k^{\Gamma}$,  
	have a clover-like block decomposition with the blocks 
	joined at a central vertex $v_0, v_1,\ldots, v_k$, respectively.
	Each block is treated as $1$-rooted and arranged in some 
	lexicographic order that they can be identified with the 
	blocks $L_1,\ldots , L_I$ of $L$ at $v$. 
	The prefactor of a term on the right hand side of (\ref{GamvsW4}) 
	contributing to $L$ via the $k\!+\!1$ 1LI subgraphs is thus 
	\begin{equation}
	\label{gproof4}
	\frac{1}{{\rm Sym}(L^W)}\prod_{j=1}^{k}\frac{1}{{\rm Sym}(L_j^{\Gamma})}
	= \prod_{j=0}^k \frac{1}{|{\rm Perm}(B(v_j))|}
	\prod_{i=1}^I \frac{1}{{\rm Sym}(L_i)}\,.
	\end{equation}
	This collects all the pre-factors arising from 
	(\ref{wgraph2}), (\ref{Gamschematic}) and we proceed to 
	the normalized weight for the articulation vertex $v$ 
	obtained from (\ref{GamvsW4}).

	Each $\partial_{\varphi}\mu(v_j|B(v_j))$ in (\ref{gproof3}) expands into tree 
	graphs $\{T_j\}$ which we regard as 1-rooted, 
	$\{T_j\} \subset \mathcal{T}_{n_j}^{1\bullet}$, and the root as the endpoint of a 
	dashed edge without open circle. The normalized weight at $v_j$ carries 
	the coefficient
	\begin{equation}
	\label{gproof5} 
	(-)^{|\nu_1 (T_j)|+ |\epsilon(T_j)|}\frac{|{\rm Perm}(B(v_j))|}{{\rm Sym}(T_j)},
	\quad \{T_j\} \subset \mathcal{T}_{n_j}^{1\bullet}\,,
	\end{equation}
	with the symmetry factor defined in (\ref{SymTdef}). For
	$j=1,\ldots, k$ the $|{\rm Perm}(B(v_j))|$ cancels 
	against that in (\ref{gproof4}). Suppose now  
	for fixed $j \in \{1,\ldots, k\}$ there are $k_j$ isomorphic 
	subtrees $T_j$ (not separately named) attached to the $W_{l-m}^{(k)}$ 
	vertex, with $\sum_j k_j = k$. Then accounting for the $1/k!$ in 
	(\ref{GamvsW4}) and omitting the 
	$\prod_{i=1}^I 1/{\rm Sym}(L_i)$ from (\ref{gproof4}) 
	we obtain the full prefactor of the choice of 
	$W_{l-m}^{(k)}$ vertex
	\begin{equation}
	\label{prefac1}
	\frac{1}{|{\rm Perm}(B(v_0))|}
	\prod_{j=1}^{k}\bigg(\frac{(-)^{k_j (|\nu_1 (T_j)|+ 
			|\epsilon(T_j)|)}}{k_j!\,{\rm Sym}(T_j)^{k_j}}\bigg)\,.
	\end{equation}
	Let $T\in \mathcal{T}(B(v),n)$, $n = 1 + \sum_{j=1}^k n_j$, (in the 
	notation introduced after (\ref{gproof3})) be the graph reassembled 
	from the rooted subtrees $T_j$ at the vertex $o$ with weight 
	$\omega_{k+d(v_0)}$. The total weight is $\omega_{k+d(v_0)}$ times the 
	products of the weights of the subtrees, and is of the form 
	$\mu(T)$ in part (a) of the graph rule. The overall sign is 
	$(-)^{|\nu_1 (T)|+ |\epsilon(T)|}=(-)^{s(T)}$. A straightforward application 
	of the orbit stabilizer theorem shows that the modulus of 
	(\ref{prefac1}) equals
	\begin{equation}
	\frac{1}{|{\rm fix}(\nu_0^B)||{\rm Aut}(T^1)|},
	\end{equation}
	where $T^1$ is $T$ seen as rooted at the open circle vertex $o$. 
	Further, we used that ${\rm fix}(\nu_0^B)$ is a direct 
	product over cells and treated the ${\rm Perm}(B(v_0))$
	from (\ref{prefac1}) as the factor in ${\rm fix}(\nu_0^B)$ 
	associated with the $v_0$ cell. 
	There may be several identical choices for the $k$-valent open 
	circle vertex $o$ introduced after (\ref{gproof3}) and this is just 
	${\rm orb}(o)$, the orbit of $o$ under ${\rm Aut}(T)$.
	Taking into account (\ref{SymTdef}) and  the $(l-m)/l$ from 
	(\ref{GamvsW4}) the net coefficient is
	\begin{equation}
	\label{prefac2}
	\frac{(-)^{s(T)}}{{\rm Sym}(T)}|{\rm orb}(o)|\times \frac{l-m}{l}.
	\end{equation}
	Each of these $k$-valent open circle vertices $o$ is labeled
	by a cell containing the vertices $b_i(v)\in B_i$ of some 
	subset of blocks $L_i = (B_i,E_i),\,i\in I_o$, and 
	we usually omit the edge set $E_i$. Restoring some of the information
	we may attribute to $o$ the total number $l(o) = \sum_{i \in I_o} 
	|E_i|$ of solid lines in the blocks labeling it. Then 
	$\sum_{ o \in \nu_0^B} l(o) = |E|$ is the total number of edges 
	in the original graph with one articulation vertex.  
	
	We now apply this to the clover like 1LI graph $L^W$
	with $l-m$ edges. After the relabeling $T' \mapsto T$ 
	described at ($\ast$) the relevant vertex set is $\nu_0^B$, 
	$B = B(v)$, with the blocks attributed to $o$ such 
	that $l(o)=l-m$. The point of the re-interpretation is 
	each orbit in $\nu_0^B/{\rm Aut}(T)$ corresponds to a distinct 
	choice of the $W_{l-m}^{(k)}$ piece, and the 
	contribution (\ref{prefac2}) of $T$ to $\mu(v|L)$ involves 
	\begin{equation}
	\label{prefac3} 
	\sum_{[o]\in \nu^B_0/{\rm Aut}(T)}|{\rm orb}(o)|l(o)=
	\sum_{o\in \nu_0^B} l(o)=l\,.
	\end{equation}
	Summing (\ref{prefac2}) using (\ref{prefac3}) the 
	coefficient of $\mu(T)$ is $(-)^{s(T)}/{\rm Sym}(T)$. 
	Finally, restoring the 
	\begin{equation} 
	\label{prefac4} 
	\prod_{i=1}^I \frac{1}{{\rm Sym}(L_i)} = 
	\frac{|{\rm Perm}(B)|}{ {\rm Sym}(L)}\,,
	\end{equation}
	from (\ref{gproof4}) and the $\mu(T)$ itself the
	normalized contribution of the articulation vertex $v$ 
	is   
	\begin{equation}
	(-)^{s(T)}\frac{|{\rm Perm}(B(v))|}{{\rm Sym}(T)}\mu(T),
	\end{equation}
	as claimed by the graph rule.
	\end{proof}
	Next we show a `locality' result which allows one to reduce the 
	case with multiple articulation points to that with just one. 
	
	\begin{lemma} (Locality)  
		\label{localitylemma} 
		The recursion (\ref{GamvsW4}) implies that the weights 
		$\mu^{\Gamma}(v|L)$, $L \in \mathcal{L}_{l}$, depend only on the block decomposition 
		$B(v) = \{L_i(v) = (B_i,E_i):\, i \in I(v)\}$ of $L$ at $v$, 
		symbolically 
		\begin{equation}
		\mu^{\Gamma}(v|L) = \mu^{\Gamma}(v|B)\,, 
		\end{equation} 
	\end{lemma}
	where on the right hand side $\mu^{\Gamma}(v| \,\cdot\,)$ is 
	regarded as a map from $B(v)$ to the smooth functions in $\varphi_v$. 
	\begin{proof} 
	By (\ref{GamEuler}) we know the structure of $\mu^{\Gamma}(v|L)$ 
	but the coefficients could in principle 
	depend on all aspects of the graph $L$ to which $v$ belongs. 
	To exclude this, we retain the notation from the preceeding Lemma and 
	trace the changes that occur if the original $L$ has more 
	than one articulation point. We single out one articulation 
	point $v$ write $v_0$ for its copy in the 1LI graph $L^W$. 
	In the paragraph leading to 
	(\ref{gproof3}) then any number $r =1,\ldots, k$ of $h$ 
	derivatives can act on the $\omega_{d_0}(h)|_{h = H_v}$, 
	in which case $v_0$ should be viewed as a root with multiplicity $r$.  
	Attached to $v_0$ will be $r$ 1LI graphs selected from 
	the $L_j^{\Gamma} \in \mathcal{L}_{m_j}^{1 \bullet}$, $j=1,\ldots,k$. 
	Without loss of generality we may take $L_j^{\Gamma}$, $j=1,\ldots,r$,
	as the 1LI graphs attached to $v_0$. We write again $B(v_j)$ for 
	the vertex set blocks stemming from $L_j^{\Gamma}$ and  
	$B(v_0)$ for the vertex set of the $L^W$ blocks at $v_0$. 
	Then $B(v)$ is the union of $B(v_0)$ and $B(v_j), j=1,\ldots r$, 
	as $1 \leq r \leq k \leq [m/2]$ runs through all possible 
	values allowed by (\ref{GamvsW4}). For any fixed $r$ the weight 
	associated with $v$ is 
	\begin{equation} 
	\label{gproof6} 
	\omega_{d(v_0) +r } \prod_{j=1}^r \partial_{\varphi} \mu^{\Gamma}(v_j|B(v_j))\,. 
	\end{equation}
	With the replacement of $k$ by $r$ the reasoning after 
	(\ref{gproof3}) carries over and establishes in 
	particular $n_{\rm max} = |I(v)|$. The relabeling  
	($\ast$) from $T'$ to $T$ proceeds as before, with $k$ replaced by $r$. 
	Summing the contributions (\ref{gproof6}) over all 
	allowed $1 \leq r \leq k \leq [m/2]$ must result in 
	a weight of the form (\ref{gproof1}) with $n_{\rm max} = |I(v)|$. 
	The $\mu(T)$ obtained is solely determined by 
	block structure of $L$ at $v$ and adheres to the graph rule.
	The way the coefficient of $\mu(T)$ is computed by 
	the recursion (\ref{GamvsW4}), however, initially refers to 
	pieces of information not localized at $v$. 
	
	In pinning down the coefficient of $\mu(T)$ the key difference 
	to the previous Lemma is that the relevant $1$-rooted 1LI graphs 
	$L^W$ (multiplicity $r$) and $L_1^{\Gamma}, \ldots, L_r^{\Gamma}$ 
	(all multiplicity $1$) no longer have to have a clover like structure. 
	That is, in addition to articulation vertex $v_0$ and 
	$v_1,\ldots,v_r$ in focus these graphs can have other 
	articulation vertices. This complicates the reduction 
	of the symmetry factors ${\rm Sym}(L^W)$, 
	${\rm Sym}(L_1^{\Gamma}), \ldots,{\rm Sym}(L_r^{\Gamma})$, 
	to those of the constituent blocks. However
	(\ref{gproof4}) remains valid if the $L_i$ on the 
	right hand side are interpreted as the not necessarily 
	1VI subgraphs that arise by disassembling the 
	$L^W$, $L_1^{\Gamma}, \ldots, L_r^{\Gamma}$, 
	at $v_0$, $v_1, \ldots, v_r$, respectively. 
	With this reinterpretation the structure of 
	the $L^W$, $L_1^{\Gamma}, \ldots, L_r^{\Gamma}$, remains 
	clover-like at the vertex in focus.  
	The line of reasoning from (\ref{gproof5}) to (\ref{prefac2}) 
	carries over with $k$ replaced by $r$ and so does the remainder 
	of Lemma \ref{oneartlemma}. In summary, the pieces 
	of information (\ref{gproof4}), (\ref{prefac2}), (\ref{prefac4}) 
	referring to the global structure of the reassembled graph 
	$L$ cancel out in the final result for the coefficient 
	of $\mu(T)$, which has the form demanded in (\ref{gproof2}). 
	\end{proof}
	Combined, Lemma \ref{oneartlemma} and Lemma \ref{localitylemma} 
	imply Theorem \ref{graphrule}.  
	
	\newpage
	\section{Reduction to integer labeled trees} 
	
	Our formula (\ref{grule2}) for the $\mu^{\Gamma}(v|L)$ weight at $v$ 
	renders the `locality' of the data $B(v)$ determining it manifest. 
	The labeling of the tree graphs 
	$\mathcal{T}(B(v),n),  n =1,\ldots, |I(v)|$, by the set partitions of 
	$B(v)$ is a convenient way to account for the coefficients
	with which a certain monomial in $\gamma_2, \gamma_m, \omega_m,\,m \geq 3$, 
	occurs. Comparing with (\ref{gproof1}), (\ref{gproof2}) one may suspect 
	that the labeling by vertex set partitions $\{c_1,\ldots, c_n\} \in 
	\mathcal{S}(B(v),n)$ somewhat overspecifies the necessary data.  Indeed, 
	upon identifying the integers $d_i$ in (\ref{ddeg3}), (\ref{ddeg4}) 
	with $d_i = d(c_i)$, the sum of the vertex degrees in cell $c_i$, 
	one expects $\mu^{\Gamma}(v|B)$ to depend only on these integers,
	not on the details of the labeling cells $c_i$ themselves. 
	This turns out to be the case because subsums in 
	(\ref{grule2}) with fixed $\mu(T)$ can be performed and
	manifestly depend only the unlabeled tree and the $d(c_i)$. 
	To avoid complications due to accidental degeneracies we sum 
	over a subset of graphs whose defining criterion is sufficient 
	but not necessary for the constancy of $\mu(T)$. 
	
	We return to the projection (\ref{gproof1}) and note
	that the weight $\mu(T)$ depends only on ${\rm pr}(T) = \tau$. 
	Indeed, with $\mu(\tau)$ formed according to (\ref{ddeg3}), one 
	has $\mu(T) = \mu(\tau) = m(t) \prod_{i=1}^n \omega_{r_i}$,  
	where $m(t)$ collects the $\gamma_m, \,m \geq 2$, factors 
	that depend only on the unlabeled graph. The integers 
	$r_i = {\rm ddeg}(o_i, d(c_i))$ 
	lie in the range of the ddeg function (\ref{ddeg1}) and depend only 
	on the integer labeled vertex set ${\rm pr} \nu_0^{\pi} 
	= \nu_0(\tau)$, $\tau \in  \mathcal{T}_n^{d(\pi)}$. Here 
	$d(\pi)$ can be any element of the projected label set 
	$\mathcal{S}(D(v), n)$ as defined after (\ref{gproof1}). 
	It is convenient to introduce for given $t \in \mathcal{T}_n$, 
	$D_n \in \mathcal{S}(D(v),n)$, the range of ${\rm ddeg}$ 
	acting on the vertex set $\nu_0(\tau)$ of some $\tau \in \mathcal{T}_n^D$
	\begin{equation} 
	\label{Ddef1}
	\rho(t,D_n) = \{ {\rm ddeg} \nu_0(\tau) :\, \tau \in \mathcal{T}_n^D\}\,.
	\end{equation}  
	For $n \geq 2$ elements $\rho_n$ of $\rho(t,D_n)$ are 
	of the form $\rho_n = \{ 3 \leq r_i \in \mathbb{N}:\, i =1,\ldots, n\}$ 
	and the weight $\mu(\tau) = \mu(t,\rho_n)$ only depends on $t$ and $\rho_n 
	\in \rho(t,D_n)$. We seek to identify labeled graphs
	$T \in \mathcal{T}(B(v),n)$ with fixed $\mu(T)$; by the previous 
	considerations this requires to hold for given $t,D_n$ 
	the $\rho_n \in \rho(t,D_n)$ fixed. We thus write $\mathcal{T}(t,B(v),n)$ 
	for the set of topologically inequivalent dashed graphs that 
	arise by labeling $t \in \mathcal{T}_n$ with $\mathcal{S}(B(v),n)$. 
	Defining 
	\begin{equation} 
	\label{Ddef2} 
	\mathcal{T}^B(t,D_n,\rho_n) :=
	\{ T \in \mathcal{T}(t,B(v),n):\, {\rm pr} T = \tau \in \mathcal{T}_n^D\,,\;
	{\rm ddeg}({\rm pr} \nu_0^{\pi}) = \rho_n\}\,,
	\end{equation}
	all its elements have the same weight $\mu(t,\rho_n)$. Further, 
	the full set of labeled dashed graphs can be partitioned according to 
	\begin{equation} 
	\label{Ddef3} 
	\mathcal{T}(t,B(v),n) = \bigcup_{ D_n \in \mathcal{S}(D(v),n)} 
	\bigcup_{\rho_n \in \rho(t,D_n)} 
	\mathcal{T}^B(t,D_n,\rho_n) \,.
	\end{equation}  
	The image under ${\rm pr}$ of the union of (\ref{Ddef3}) over all $t\in \mathcal{T}_n$ is partitioned analogously, 
	\begin{equation} 
	\label{Ddef4} 
	\mathcal{T}(D(v),n) = \bigcup_{ D_n \in \mathcal{S}(D(v),n)} \bigcup_{t \in \mathcal{T}_n} 
	\bigcup_{\rho_n \in \rho(t,D_n)} 
	\mathcal{T}^D_n(\rho_n)\,, 
	\end{equation}  
	with $\mathcal{T}^D_n(\rho_n) := \{ \tau \in \mathcal{T}_n^D:\, {\rm ddeg} 
	\nu_0(\tau) = \rho_n\}$. In the graph rule formula (\ref{grule2}) 
	the decomposition (\ref{Ddef3}) allows one to  
	`pull in' the sub-sub over $\mathcal{T}^B(t,D_n,\rho_n)$. The evaluation 
	of this subsum is the the main result of this section:   
	
	\begin{theorem} 
		\label{Dtheorem} 
		In the graph rule (\ref{grule2}) the sum over graphs $T \in 
		\mathcal{T}(B(v),n)$ labeled by partitions of the vertex set $B(v)$ can 
		be replaced with a sum over {\it integer} labeled trees.  
		Specifically, $\mu^{\Gamma}(v|B) = \mu^{\Gamma}(v|D)$, with 
		$D(v) = d(B(v))$ and  
		\begin{eqnarray}
		\label{Dtheorem1} 
		\mu^{\Gamma}(v|D) & = & \sum_{n=1}^{|I(v)|} \sum_{D_n \in \mathcal{S}(D(v), n)} 
		\sum_{t \in \mathcal{T}_n} (-)^{s(t)} \sum_{\rho_n \in \rho(t,D_n)} 
		c(t,D_n,\rho_n) \mu(t,\rho_n)\,,
		\nonumber\\
		c(t,D_n,\rho_n) & = & \sum_{T \in \mathcal{T}^B(t,D_n,\rho_n)} 
		\frac{{\rm Perm}(B(v))}{{\rm Sym}(T)} = 
		\frac{|\nu_0(D_n,\rho_n)|}{|{\rm Aut}(t)|} 
		P(D(v), D_n)\,.
		\end{eqnarray}
		Here $P(D(v),D_n)$ is the number of partitions 
		of $|I(v)|$ distinct labels $\{b_1,\ldots, b_{|I(v)|}\}$ 
		into $n$ cells such the the sum of the $d(b_j)$ in 
		the $i$-th cell cell equals the given $d_i$, 
		and $|\nu_0(D_n,\rho_n)|$ is the cardinality of $\{\nu_0^{\pi} :\,
		{\rm ddeg}({\rm pr} \nu_0^{\pi}) = \rho_n, \;\pi \in 
		\mathcal{S}(B(v),D_n) \}$. For the latter one has explicitly 
		\begin{equation} 
		\label{Dtheorem2} 
		|\nu_0(D_n,\rho_n)| = \Big(\prod_{i=1}^{k} n_i!\Big)\prod_{j=1}^n 
		\frac{s_j!}{s_{j,1}!\ldots s_{j,k}!},
		\end{equation}
		where $n_1,\ldots,n_k$, $\sum_{i=1}^k n_i=n$, are the numbers of 
		equally valent open circle vertices in $t$,  and $s_{j,i}$ is the 
		number of equally valent open circle vertices of type $i$ labeled 
		by $d_j$, where $D_n=\{d_1^{s_1},\ldots, d_n^{s_n}\}$. 
	\end{theorem} 
	
	The notational complexity notwithstanding, the formula (\ref{Dtheorem1}) is in fact a simplification compared to (\ref{grule2}) as a much smaller set of labeled trees need to be considered. Before turning to the proof we illustrate the statement 
	in the examples from Sections 2.2 and 2.3. The possible $D(v)$'s 
	for the examples considered in Sections 2.2, 2.3 are 
	$D(v)=\{2,2\},\,\{2,3\},\,\{2,2,3\}$. 
	Theorem \ref{Dtheorem} produces the correct weights for each $D(v)$ 
	with cardinality $2$ by inspection. As a simple illustration of Theorem 
	\ref{Dtheorem}, we detail the constituents for $D(v) = \{2,2,3\}$.
	\begin{itemize}
		\itemindent +10mm	
		\item[$n=1$:] $D_1=\{7\},\quad P(D(v),D_1)=1$.
		\begin{eqnarray}
		\begin{tikzpicture}
		\draw[fill=white] (0,0)circle (3pt) ;
		\node at (-0.5,0) {$7$};
		\node at (3,0) {$\rho_1=\{7\}\quad \frac{|\nu_0(D_1,\rho_1)|}{|{\rm Aut}(t)|}=1$};
		\end{tikzpicture} 
		\end{eqnarray}
		
		\item[$n=2:$] $D_2=\{2,5\},\quad P(D(v),D_2)=2$.
		\begin{eqnarray}
		\begin{tikzpicture}
		\draw[black, dashed](0,0)--(1,0);
		\draw[fill=white] (0,0)circle (3pt) ;
		\draw[fill=white] (1,0)circle (3pt) ;
		\node at (-0.5,0) {$2$};
		\node at (1.5,0) {$5$};
		\node at (5,0) {$\rho_2=\{3,6\}\quad \frac{|\nu_0(D_2,\rho_2)|}{|{\rm Aut}(t)|}=\frac{2!}{2!}=1$};
		\end{tikzpicture} 
		\end{eqnarray}
		
		\item[] $D_2=\{3,4\},\quad P(D(v),D_2)=1$.
		\begin{eqnarray}
		\begin{tikzpicture}
		\draw[black, dashed](0,0)--(1,0);
		\draw[fill=white] (0,0)circle (3pt) ;
		\draw[fill=white] (1,0)circle (3pt) ;
		\node at (-0.5,0) {$3$};
		\node at (1.5,0) {$4$};
		\node at (5,0) {$\rho_2=\{4,5\}\quad \frac{|\nu_0(D_2,\rho_2)|}{|{\rm Aut}(t)|}=\frac{2!}{2!}=1$};
		\end{tikzpicture} 
		\end{eqnarray}
		
		\item[$n=3:$] $D_3=\{2,2,3\},\quad P(D(v),D_3)=1$.
		\begin{eqnarray}
		&&\begin{tikzpicture}
		\draw[black, dashed] (0,0) -- (0.5,0.5);
		\draw[black, dashed] (0.5,0.5) -- (1,0);
		\draw[fill=white]  (0,0) circle (3pt) ;
		\draw[fill=white] (0.5,0.5) circle (3pt) ;
		\draw[fill=white]  (1,0) circle (3pt) ;
		\node at (-0.5,0) { $2 $};
		\node at (0.5,1) { $3 $};
		\node at (1.5,0) { $2 $};
		\node at (5,0.25) {$\rho_3=\{3,3,5\}\quad \frac{|\nu_0(D_3,\rho_3)|}{|{\rm Aut}(t)|}=\frac{2!}{2!}=1$};
		\end{tikzpicture}
		\nonumber\\
		&&\begin{tikzpicture}
		\draw[black, dashed] (0,0) -- (0.5,0.5);
		\draw[black, dashed] (0.5,0.5) -- (1,0);
		\draw[fill=white]  (0,0) circle (3pt) ;
		\draw[fill=white] (0.5,0.5) circle (3pt) ;
		\draw[fill=white]  (1,0) circle (3pt) ;
		\node at (-0.5,0) { $2 $};
		\node at (0.5,1) { $2 $};
		\node at (1.5,0) { $3 $};
		\node at (5,0.25) {$\rho_3=\{3,4,4\}\quad \frac{|\nu_0(D_3,\rho_3)|}{|{\rm Aut}(t)|}=\frac{2!^2}{2!}=2$};
		\end{tikzpicture}
		\nonumber\\
		&&\begin{tikzpicture}
		\draw[black, dashed]  (0,0) -- (0.5,0.5);
		\draw[black, dashed]  (0.5,0.5) -- (1,0);
		\draw[black, dashed]  (0.5,0.5)-- (0.5,1);
		\draw[fill=white]  (0,0) circle (3pt) ;
		\draw[fill=white]  (1,0) circle (3pt) ;
		\draw[fill=white]  (0.5,1) circle (3pt) ;
		\node at (-0.5,0) { $2 $};
		\node at (0.5,1.5) { $2 $};
		\node at (1.5,0) { $3 $};
		\node at (5,0.25) {$\rho_3=\{3,3,4\}\quad \frac{|\nu_0(D_3,\rho_3)|}{|{\rm Aut}(t)|}=\frac{3!}{3!}=1$};
		\end{tikzpicture}
		\end{eqnarray}
	\end{itemize}	
	
	\noindent This yields $\omega_7 -2\gamma_2 \omega_3 \omega_6 -\gamma_2 \omega_4 \omega_5+\gamma_2^{2}\omega_3^2\omega_5 +2\gamma_2^{2}\omega_3\omega_4^2+\gamma_3\omega_3^2\omega_4$, in agreement with the result in (\ref{grule7.1a}) and (\ref{grule9.1}).
	
	In preparation of the proof of Theorem \ref{Dtheorem} we note 
	that ${\rm pr} T = \tau \in \mathcal{T}_n^D$ implies that the 
	set partitions $\pi$ labeling $T \in \mathcal{T}(B(v),n)$ 
	are constrained to lie in $\mathcal{S}(B(v), D_n) := 
	\{ \pi \in \mathcal{S}(B(v), n) \,| \, d(\pi) = D_n\}$. 
	These are viewed as (constrained) multiset partitions in the sense 
	explicated after (\ref{multiset1}). For the subsequent 
	proofs a realization of the multisets as sets of distinct 
	elements $B = \{b_1,\ldots, b_I\}$ modulo an equivalence 
	relation is convenient (to avoid further complicating the notation we 
	write $B(v)$ for the multiset and $B$ for 
	$\{b_1,\ldots, b_I\}$ equipped with an equivalence relation). For the moment we merely stipulate the 
	existence 
	of an equivalence relation ``$\,\sim$'' on $B$ compatible 
	with the degree assignments, i.e.~$b_i \sim b_j$ implies 
	$d(b_i) = d(b_j)$ but not necessarily vice versa. We denote 
	by ${\rm Perm}(B)$ the subgroup of $S_I$ that permutes equivalent 
	$b_i$'s. In this setting the counterpart of the constrained 
	multiset partitions $\mathcal{S}(B(v), D_n)$ is 
	$\bar{\mathcal{S}}(B, D_n)= \mathcal{S}(B, D_n)/{\rm Perm}(B)$, while $\mathcal{S}(\{b_1,\ldots, b_I\}, D_n)$ does not depend on the 
	equivalence relation and neither does its cardinality $P(D(v),D_n)$. 
	The counterpart of $\mathcal{T}^B(t,D_n,\rho_n)$ is $\overline{\mathcal{T}}^B(t,D_n,\rho_n)$,
	the set of labeled graphs obtained by labeling $t \in \mathcal{T}_n$ 
	with $C\in \bar{\mathcal{S}}(B,D_n)$ such that ${\rm ddeg}({\rm pr} \nu_0^C) 
	=\rho_n$. The latter condition defines the vertex set $\nu_0(D_n,\rho_n)$. 
	In this setting we later show: 
	
	\begin{proposition}
		\label{stabprop2}
		Let $B = \{b_1,\ldots,b_I\}$ be a set of distinct vertices 
		equipped with an equivalence relation ``$\,\sim$'' that is 
		compatible with the degrees, {\it i.e.\ } $b_i \sim b_j$ only if 
		$d(b_i) = d(b_j)$. Define $\mathcal{T}(D_n,\rho_n)$, $\nu_0(D_n,\rho_n)$,
		and $P(D(v),D_n)$, as above. Then: 
		\begin{equation}
		\label{prop2sum}
		\sum_{T \in \overline{\mathcal{T}}^B(t,D_n,\rho_n)}\frac{|{\rm Perm}(B)|}{{\rm Sym}(T)}
		= \frac{P(D(v),D_n)|\nu_0(D_n,\rho_n)|}{|{\rm Aut}(t)|},
		\end{equation}
		is independent of the equivalence relation on $B$.
	\end{proposition} 
	
	Theorem \ref{Dtheorem} is an easy consequence of Proposition 
	\ref{stabprop2}: Using (\ref{Ddef3}) in (\ref{grule2})  
	(and the fact that $s(T) = s(t)$ is manifestly labeling 
	independent) one finds (\ref{Dtheorem1}) with the $c(t,D_n,\rho_n)$ 
	given by the sum over $T \in \mathcal{T}^B(t,D_n,\rho_n)$. Since 
	$\sum_{T \in \mathcal{T}^B(t,D_n,\rho_n)} |{\rm Perm}(B(v))|/ {\rm Sym}(T) = 
	\sum_{T \in \overline{\mathcal{T}}^B(t,D_n,\rho_n)} |{\rm Perm}(B)|/ {\rm Sym}(T)$, 
	the second line of (\ref{Dtheorem1}) follows from (\ref{prop2sum}). 
	The formula (\ref{Dtheorem2}) is straightforward combinatorics:   
	
	Keeping the integer labels from $D_n$ fixed, the dummy labels of the 
	equally valent open circle vertices may be permuted while preserving 
	$\rho_n$. This contributes the factor $\prod_{i=1}^k n_i!$. The remaining 
	factor follows from the number of ways the $s_j$ labels with degree 
	$d_j$ can be distributed amongst the $n_1,\ldots ,n_k$ equally labeled 
	vertices. Application of the multinomial 
	theorem gives the contribution $\prod_{j=1}^n s_j!/
	(s_{j,1}!\ldots s_{j,k}!)$, and hence the result (\ref{Dtheorem2}).

	It remains to establish Proposition \ref{stabprop2}. 
	Its proof is broken up into several lemmas, some of them 
	more general than needed. 
	
	\begin{lemma}
		Let $B$, ``$\,\sim$'', and ${\rm Perm}(B)$ be as in 
		Prop.~\ref{stabprop2} and ${\rm Sym}(T)$ as in (\ref{SymTdef}). 
		Then the ratios $|{\rm Perm}(B)|/{\rm Sym}(T)$ are integers, and 
		so are the $d_{i_3,\ldots i_{m-1}}$ in (\ref{GamEuler}). 
	\end{lemma} 
	\begin{proof} 
	The second part follows trivially from the first. The $\mu(T)$ in 
	(\ref{grule2}) refer to a mixed basis of $\gamma_2,\omega_m,\gamma_m$,
	$m \geq 3$. By Appendix B the transition from $\omega_m(\varphi)$ to 
	$\gamma_n(\varphi)$'s involves integer coefficients only. It follows that 
	the coefficients in (\ref{GamEuler}) are also integers,
	$d_{i_3,\ldots i_{m-1}} \in \mathbb{Z}$. 
	
	For the first part, recall that labels are generated from 
	the set partitions $\mathcal{S}(B,n)$ of $B$ into $n$ cells. 
	Any resulting partition $\pi = \{c_1, \ldots, c_n\}$ carries an 
	induced equivalence relation defined by $c_i\sim c_{i'}$ 
	iff there is a (possibly non-unique) $\sigma \in {\rm Perm}(B)$ 
	such that $\sigma(c_i)=c_{i'}$. This implies that the subset 
	\begin{equation}	
	\label{Phidef}
	{\rm stab}(\pi) =\{\sigma \in{\rm Perm}(B):\,
	\forall i \,\exists i' \in \{1,\ldots, n\}\;\;{\rm s.t.} \;\;
	\sigma(c_i)=c_{i'}\},
	\end{equation}
	is a subgroup of ${\rm Perm}(B)$.  The case $i=i'$ in (\ref{Phidef}) 
	is allowed and gives rise to a subgroup ${\rm fix}(\pi) \subset 
	{\rm stab}(\pi)$, which in the multiset formulation corresponds to 
	${\rm fix}(\nu_0^B)$. In fact, 
	\begin{equation} 
	\label{Phi1def}
	{\rm fix}(\pi) =\{\sigma
	\in {\rm stab}(\pi): \forall i\,\sigma(C_i)=C_i\}
	\quad \mbox{is a normal subgroup of} \;\; {\rm stab}(\pi). 
	\end{equation}
	Recall, $H \subset G$ is a normal subgroup if $\forall g\in G$, 
	$g^{-1} H g=H$. Here, let $\sigma\in\Phi(\pi)$, $\sigma_1 \in\Phi_1(\pi)$. 
	For each $i$, $\sigma(c_i)=c_{i'}$, $\sigma^{-1}(c_{i'})=c_i$ 
	with $c_i\sim c_{i'}$. Note that $\sigma$ may not be unique but 
	any given $\sigma$ has a unique inverse. 
	Hence, $\sigma_1 \sigma(c_i)=c_{i'}$ and 
	$\sigma ^{-1} \sigma_1 \sigma(c_i)=c_i$, valid for all $\sigma$, 
	implies (\ref{Phi1def}). 
	
	Since ${\rm fix}(\pi)$ is a normal subgroup of ${\rm stab}(\pi)$, 
	the quotient group ${\rm stab}(\pi)/{\rm fix}(\pi)$ is well defined. 
	Moreover, as ``$\,\sim$'' induces an equivalence relation on the 
	partition $\pi=\{c_1,\ldots,c_n\}$, we may define ${\rm Perm}(\pi)$ 
	as the subgroup of $S_k$ comprising only those elements that 
	permute equivalent cells. Both groups are naturally isomorphic  
	\begin{equation} 
	\label{permpi} 
	{\rm stab}(\pi)/{\rm fix}(\pi) \cong {\rm Perm}(\pi)\,.
	\end{equation}
	We can set up an isomorphism as follows. Use the ${\rm fix}(\pi)$ 
	subgroup to permute in each cell $c_i$ the equivalent elements it 
	contains into some lexicographic order. Then cells $c_i,c_{i'}$ are 
	equivalent iff they contain lexicographically ordered strings 
	of equal cardinalities for each ``$\,\sim$'' equivalence class. 
	The quotient group permutes equivalent  cells while preserving 
	the lexicographic order of the strings. As such it gives one 
	realization of ${\rm Perm}(\pi)$ and hence (\ref{permpi}). Since 
	$|{\rm fix}(\pi)| = |{\rm fix}(\nu_0^B)|$ it follows
	that $|{\rm stab}(\pi)| = |{\rm Perm}(\pi)| |{\rm fix}(\nu_0^B)|$. 
	When treating $\pi$ as a label set for the graphs 
	$T \in \mathcal{T}(B(v),n)$ the automorphism group ${\rm Aut}(T)$ 
	is a subgroup of ${\rm Perm}(\pi)$. Lagrange's theorem 
	\begin{equation} 
	\label{BTratio2} 
	\frac{|{\rm Perm}(B)|}{|{\rm fix}(\nu_0^B)| |{\rm Aut}(T)|}
	=\frac{|{\rm Perm}(B)|}{|{\rm stab}(\pi)|}
	\frac{|{\rm Perm}(\pi)|}{|{\rm Aut}(T)|}
	\in \mathbb{N}\,,
	\end{equation}
	completes the argument. 
	\end{proof}
	
	We proceed with labeling the 
	dashed graphs $t \in \mathcal{T}_n$ by an abstract $n$ element label 
	set $C = \{c_1,\ldots, c_n\}$. Later on the $c_i$ will be identified 
	with the cells of a set partition in $\mathcal{S}(B(v),n)$, for now 
	the origin of the $c_i$'s is irrelevant. In order to model the 
	equivalence of cells we assume that $C$ carries an 
	equivalence relation ``$\sim$'' and that a subgroup 
	${\rm Perm}(C)$ of $S_n$ acts by permuting equivalent 
	$c_i$'s.  As before, only the open circle vertices $\nu_0$ of 
	$t \in \mathcal{T}_n$ are labeled, technically via the graph 
	of a bijection $\sigma: \nu_0 \rightarrow C$. Each graph is 
	referred to as a labeling set (or pairing) and corresponds to a permutation 
	$\sigma \in S_n$, so for $\nu_0 = \{o_1,\ldots, o_n\}$ and 
	$C = \{c_1,\ldots, c_n\}$ we write a labeling set as 
	$\nu_0^{\sigma}=\{(o_i, c_{\sigma(i)}):\, i =1,\ldots, n\}$, 
	by slight abuse of notation. For fixed $C$ we now consider the 
	set of all pairings 
	\begin{equation} 
	\label{nuCdef}
	\nu_0^C = \{ \nu_0^{\sigma}:\, \sigma \in S_n\}\,,
	\quad |\nu_0^C| = n!\,.
	\end{equation}
	Recall that an unlabeled graph $t\in\mathcal{T}_n$  may be written as 
	$t=(\nu_0\cup\nu_1,\epsilon)$, for one of its labeled counterparts 
	we write $T=(\nu_0^{\sigma}\cup\nu_1,\epsilon)$. As $\sigma$ runs through
	$S_n$ the set of labeled dashed graphs generated is denoted by $\mathcal{T}_n^{C}$. 
	
	The product group ${\rm Aut}(t)\times {\rm Perm}(C): 
	\nu_0^C\to \nu_0^C$  acts termwise on the elements of 
	$\nu_0^{\sigma}$: for $(g,h)\in {\rm Aut}(t)\times {\rm Perm}(C)$ 
	and $\nu_0^{\sigma} = \{ (o_i,c_{\sigma(i)}):\, i = 1,\ldots,n\}$ 
	define $(g,h)(o_i, c_{\sigma(i)}) :=\big(g(o_i),h(c_{\sigma(i)})\big) $. 
	We note that two distinct 
	labeling sets $\nu_0^{\sigma_1}, \nu_0^{\sigma_2}\in \nu_0^C$ 
	with $\sigma_1\neq \sigma_2$ can correspond to the same labeled 
	$T\in  \mathcal{T}_n^{C}$. This occurs if there is an element of 
	${\rm Aut}(t)\times {\rm Perm}(C)$ that maps between the labeling sets. 
	As an illustration consider $t\in \mathcal{T}_3$ with open circle vertex 
	set $\nu_0=\{o_1,o_2,o_3\}$
	\begin{eqnarray}
	&& \begin{tikzpicture}
	\draw[black, dashed] (0,0) -- (0.5,0.5);
	\draw[black, dashed] (0.5,0.5) -- (1,0);
	\draw[fill=white]  (0,0) circle (3pt) ;
	\draw[fill=white] (0.5,0.5) circle (3pt) ;
	\draw[fill=white]  (1,0) circle (3pt) ;
	\node at (-0.3,-0.2) { $o_1 $};
	\node at (0.5,0.9) { $o_3 $};
	\node at (1.3,-0.2) { $o_2$}; 
	\node at (-2,0.5) { $t$};
	\end{tikzpicture}
	\end{eqnarray}
	With labels $C=\{c_1,c_2,c_3\}$ two distinct labeling sets are 
	$\nu_0^{\sigma_1}=\big\{(o_1,c_1),(o_2,c_2),(o_3,c_3)\big\}$ and $\nu_0^{\sigma_2}=
	\big\{(o_1,c_2),(o_2,c_1),(o_3,c_3)\big\}$, and the resulting labeled graphs 
	$T_1$ and $T_2$ are shown in (\ref{stabex1}). On inspection it is clear 
	that $T_1$ and $T_2$ are the same labeled graph as one can be mapped 
	into the other by interchanging $o_1$ and $o_2$.
	
	\begin{eqnarray}
	\label{stabex1}
	\begin{tikzpicture}
	\draw[black, dashed] (0,0) -- (0.5,0.5);
	\draw[black, dashed] (0.5,0.5) -- (1,0);
	\draw[fill=white]  (0,0) circle (3pt) ;
	\draw[fill=white] (0.5,0.5) circle (3pt) ;
	\draw[fill=white]  (1,0) circle (3pt) ;
	\node at (-0.5,-0.4) { $(o_1,c_1)$};
	\node at (0.5,0.9) { $(o_3,c_3) $};
	\node at (1.4,-0.4) { $(o_2,c_2)$}; 
	\node at (-2,0.4) { $T_1$};
	\end{tikzpicture}
	\makebox[1cm]{} 
	\begin{tikzpicture}
	\draw[black, dashed] (0,0) -- (0.5,0.5);
	\draw[black, dashed] (0.5,0.5) -- (1,0);
	\draw[fill=white]  (0,0) circle (3pt) ;
	\draw[fill=white] (0.5,0.5) circle (3pt) ;
	\draw[fill=white]  (1,0) circle (3pt) ;
	\node at (-0.5,-0.4) { $(o_1,c_2)$};
	\node at (0.5,0.9) { $(o_3,c_3) $};
	\node at (1.4,-0.4) { $(o_2,c_1)$}; 
	\node at (-2,0.4) { $T_2$};
	\end{tikzpicture}
	\end{eqnarray}
	Generally, labeling sets related by the above action of 
	${\rm Aut}(t) \times {\rm Perm}(C)$ give rise to the same labeled graph. 
	This underlies the following

	\begin{lemma}
		\label{stablemma1}
		Let $t\in\mathcal{T}_n$ be an unlabeled graph and $C=\{c_1,\ldots,c_n\}$ 
		be a set of distinct labels equipped with an equivalence relation ``$\,\sim$''. 
		Let Perm$(C)$ be the subgroup of $S_n$ that permutes equivalent $c_i$'s,
		and let $\mathcal{T}^{C}_n(t)$ be the set of all topologically distinct 
		labeled dashed graphs obtained by labeling $t$ with $C$. Then: 
		\begin{equation}
		\sum_{T \in \mathcal{T}_n^{C}(t)}\frac{|{\rm Perm}(C)|}{|{\rm Aut}(T)|}
		=\frac{n!}{|{\rm Aut}(t)|},
		\end{equation}
		i.e.~this sum is independent of the equivalence relation 
		``$\,\sim$'' on $C$.
	\end{lemma}
	
	\begin{proof}
		We consider the orbit ${\rm orb}(\nu_0^{\sigma})$ of some 
		$\nu_0^{\sigma}\in \nu_0^C$ under the action of 
		${\rm Aut}(t)\times {\rm Perm}(C)$. By the comment after (\ref{stabex1}) 
		the orbit is the subset of $\nu_0^C$ whose elements correspond to the same 
		labeled graph $T$. Hence there exists a bijection between labeled graphs 
		in $\mathcal{T}_n^C(t)$ and equivalence classes in 
		$\nu_0^C/[{\rm Aut}(t)\times {\rm Perm}(C)]$, {\it i.e.}~orbits. 
		The orbits are disjoint and their union is $\nu_0^C$. 
		A sum over $T\in \mathcal{T}_n^C(t)$ may be reexpressed as a sum over orbits 
		$[\nu_0^{\sigma}]\in \nu_0^C/[{\rm Aut}(t)\times {\rm Perm}(C)]$. 
		
		Next we claim that ${\rm Aut}(T)$ for a labeled $T\in \mathcal{T}_n^C(t)$ is isomorphic 
		to some subgroup ${\rm Aut}(t)\times {\rm Perm}(C)$. Suppose an element of 
		${\rm Aut}(T)$ permutes two labeled vertices $(o,c)$ and $(o',c')$ while 
		preserving adjacency. This is possible iff there is a $g\in{\rm Aut}(t)$ 
		that exchanges $v,v'$,  {\it and} there is a $h\in {\rm Perm}(C)$ that 
		exchanges $c, c'$. 
		For a labeling set $\nu_0^{\sigma}$ corresponding to $T$, then 
		$(g\times h)(\nu_0^{\sigma})= \nu_0^{\sigma}$. 
		Conversely, suppose there is an element of ${\rm Aut}(t)\times {\rm Perm}(C)$ 
		that leaves $\nu_0^{\sigma}$ invariant. This is a permutation of the pairs 
		in $\nu_0^{\sigma}$ labeling $T$ that preserves adjacency in $t$, and so 
		there is a corresponding element in ${\rm Aut}(T)$. Thus ${\rm Aut}(T)$ is 
		isomorphic to the subgroup ${\rm stab }(\nu_0^{\sigma})$
		of ${\rm Aut}(t)\times {\rm Perm}(C)$ that leaves any labeling set 
		$\nu_0^{\sigma}$ corresponding to $T$ invariant. 
		
		The stabilizer subgroups of two elements $\nu_0^{\sigma}, \nu_0^{\sigma'}$ 
		of the same orbit are related by conjugation with the group element 
		linking them. In particular, 
		$|{\rm stab}(\nu_0^{\sigma})|=|{\rm stab}(\nu_0^{\sigma'})|$,
		for $\nu_0^{\sigma}, \nu_0^{\sigma'} \in {\rm orb}(\nu_0^{\sigma})$.  
		We may write
		\begin{equation}
		\sum_{T \in \mathcal{T}_n^{C}(t)}\frac{|{\rm Perm}(C)|}{|{\rm Aut}(T)|}=
		\sum_{[\nu_0^{\sigma}]\in \nu_0^C/[{\rm Aut}(t)\times{\rm Perm}(C)]}
		\frac{|{\rm Perm}(C)|}{|{\rm stab}(\nu_0^{\sigma}|}.
		\end{equation}
		The orbit-stabilizer theorem implies $ |{\rm Aut}(t)\times {\rm Perm}(C)|
		=|{\rm stab}(\nu_0^{\sigma})||{\rm orb}(\nu_0^{\sigma})|$, i.e.
		\newline
		$|{\rm Perm}(C)|/|{\rm stab}(\nu_0^{\sigma})|
		= |{\rm orb}(\nu_0^{\sigma})|/|{\rm Aut}(t)|$.
		Thus 
		\begin{equation}
		\sum_{T \in \mathcal{T}_n^{C}(t)}\frac{|{\rm Perm}(C)|}{|{\rm Aut}(T)|} = 
		\frac{1}{|{\rm Aut}(t)|}\sum_{[\nu_0^{\sigma}]\in 
			\nu_0^C/[{\rm Aut}(t)\times{\rm Perm}(C)]}
		\!\!\!
		|{\rm orb}(\nu_0^{\sigma})|
		= \frac{|\nu_0^C|}{|{\rm Aut}(t)|}\,,
		\end{equation}
		as claimed.
	\end{proof}
	
	We proceed to a variant of Lemma \ref{stablemma1} where 
	the equivalence relation on $C$ is compatible with an 
	integer grading $d:C \rightarrow \mathbb{N}^n$. Each element of the label set 
	$C=\{c_1,\ldots,c_n\}$ is assigned an integer $d(c_i)\in \mathbb{N}$. 
	The range $d(C)=\{d(c_1),\ldots,d(c_n)\}$ will in general be a 
	multiset $D_n=\{d_1^{s_1},\ldots,d_n^{s_n}\}$,  with 
	$\sum_{i=1}^{n}s_i=n$, $s_i\in\mathbb{N}_0$. If $C$ is used to label some 
	$t\in \mathcal{T}_n$, the weight assignment to its open circle vertices 
	will by (\ref{ddeg1}), (\ref{ddeg3}) depend only 
	on the valency of the $o \in \nu_0$ and some integers which  
	we will now draw drom the range $d(C)$. To this end
	we extend the ddeg function in (\ref{ddeg1}) to the labeled 
	vertices $(o_i, c_{\sigma(i)})$ by ${\rm ddeg}( o_i , c_{\sigma(i)}) 
	= |o_i| + d(c_{\sigma(i)})$. In other words,  the sum $|o_i| + 
	d(c_{\sigma (i)})$ is viewed as an instance of (\ref{ddeg1}) where 
	the integers arise from the degrees of the labeling set. 
	This carries over to ${\rm ddeg}\nu_0^{\sigma} := 
	\{ {\rm ddeg}(o_i, c_{\sigma(i)}) :\, i =1,\ldots, n\}$ and 
	we define
	\begin{eqnarray} 
	\label{rhosubsets} 
	\mathcal{T}^C(\rho_n) &:=& \{ T \in \mathcal{T}_n^C(t) :\, 
	{\rm ddeg} \nu_0^{\sigma} = \rho_n\}\,,
	\nonumber\\
	\nu_0(\rho_n) &:=& \{ \nu_0^{\sigma} \in \nu_0^C :\,
	{\rm ddeg} \nu_0^{\sigma} = \rho_n\}\,,
	\end{eqnarray} 
	for some fixed $\rho_n \in \rho(t,d(C))$ in the range 
	of the ${\rm ddeg}$ function
	\begin{equation}
	\label{rhoweight}
	\rho(t,d(C)) := \{ {\rm ddeg} \nu_0^{\sigma}:\, 
	\sigma \in S_n\}\,.  
	\end{equation}
	By the weight assigments (\ref{ddeg3}) 
	all $T \in \mathcal{T}^C(\rho_n)$ have the same $\mu(T)$. 
	Equivalently, elements $\nu_0^{\sigma},\,\nu_0^{\sigma'}$ 
	of the same orbit in $\nu_0^C/[{\rm Aut}(t)\times {\rm Perm}(C)]$, 
	have the same $\rho_n$ and hence lie in the same 
	$\nu_0(\rho_n)$. Clearly, $\mathcal{T}^C_n(t)$ is partitioned by $\mathcal{T}^C(\rho_n)$ 
	as $\rho_n$ runs through $\rho(t,d(C))$.

	\begin{lemma}
		\label{stablemma2}
		Let $t\in\mathcal{T}_n$ be an unlabeled graph and 
		$C=\{c_1,\ldots,c_n\}$ 
		be a set of distinct labels equipped with a grading 
		$d: C \rightarrow \mathbb{N}^n$ and a compatible equivalence relation 
		``$\,\sim$'', {\it i.e.\ } $c_i\sim c_j$ only if $d(c_i)=d(c_j)$. Then:
		\begin{equation}
		\sum_{T \in \mathcal{T}^C(\rho_n)}\frac{|{\rm Perm}(C)|}{|{\rm Aut}(T)|}
		= \frac{|\nu_0(\rho_n)|}{|{\rm Aut}(t)|}\,,
		\end{equation}
		{\it i.e.\ } the sum is independent of the equivalence relation 
		``$\,\sim$'' on $C$.
	\end{lemma}
	
	\begin{proof}
		As noted in the proof of Lemma \ref{stablemma1}, there is a bijection 
		between the labeled graphs in $\mathcal{T}_n^C(t)$ and the orbits in 
		$\nu_0^C/[{\rm Aut}(t)\times {\rm Perm}(C)]$. Therefore we may 
		write
		\begin{eqnarray}
		\label{stab2.1}
		&& \sum_{T \in  \mathcal{T}(t,\rho_n)}\frac{|{\rm Perm}(C)|}{|{\rm Aut}(T)|} = 
		\sum_{[\nu_0^{\sigma}]\in \nu_0(\rho_n)/[{\rm Aut}(t)\times{\rm Perm}(C)]}
		\frac{|{\rm Perm}(C)|}{|{\rm stab}(\nu_0^{\sigma})|}
		\nonumber\\
		&& \quad = \frac{1}{|{\rm Aut}(t)|}\sum_{[\nu_0^{\sigma}] \in 
			\nu_0(\rho_n)/[{\rm Aut}(t)\times{\rm Perm}(C)]}|{\rm orb}(\nu_0^{\sigma})|
		= \frac{|\nu_0(\rho_n)|}{|{\rm Aut}(t)|}.
		\end{eqnarray}
		In the first identity the constancy of $\rho_n$ 
		within orbits entered, in the second the orbit-stabilizer theorem 
		was used as in the proof of Lemma \ref{stablemma1}. The 
		elements of $\nu_0(\rho_n)$ depend on the grading but 
		not on the specific equivalence relation ``$\sim$'' compatible with it. 
	\end{proof}
	
	We now return to the graph rule, where the label set $C$ originates 
	from partitioning the vertex set $B$ into $n$ cells. We adopt the 
	equivalence class setting from Proposition \ref{stabprop2}: 
	given a vertex set $B=\{b_1,\ldots,b_I\}$ of $I$ distinct elements 
	its $n$-cell set partitions $\{c_1,\ldots, c_n\} \in \mathcal{S}(B,n)$ are 
	formed. The cardinality $|\mathcal{S}(B,n)|=S(I,n)$ is the second Stirling number. 
	We stipulate the existence of an equivalence relation ``$\,\sim$'' 
	on $B$, and take ${\rm Perm}(B)$ to permute equivalent elements of $B$. 
	An action ${\rm Perm}(B):\mathcal{S}(B,n)\to \mathcal{S}(B,n)$ is induced, and we 
	write ${\rm orb}(\pi)$ for the orbit of $\pi \in \mathcal{S}(B,n)$ under 
	${\rm Perm}(B)$. 
	Observe that for given $\pi \in \mathcal{S}(B,n)$, all elements of ${\rm orb}(\pi)$ 
	correspond to the same label set $C$. We omit a formal proof and  
	instead present an illustrative example: let $B=\{b_1, b_2,b_3,b_4,b_5\}$, 
	with $b_1\sim b_2$. The partitions  $\pi_1=\big\{\{b_1,b_4\},
	\{b_2,b_3,b_5\}\big\}$ and $\pi_2=\big\{\{b_2,b_4\},\{b_1,b_3,b_5\}\big\}$ 
	are distinct, but they correspond to the same label set $C$ by virtue of 
	$b_1\sim b_2$. We define $\bar{\mathcal{S}}(B,n):=\mathcal{S}(B,n)/{\rm Perm}(B)$, 
	the set of distinct label sets $C$.
	
	\begin{lemma}
		\label{stablemma3}
		Let $B=\{b_1,\ldots,b_I\}$ be a set of distinct vertices 
		equipped with an equivalence relation ``$\,\sim$'', and let 
		${\rm Perm}(B)$ be the subgroup of $S_I$ that permutes 
		equivalent vertices. For given $t\in \mathcal{T}_n$ let 
		$\overline{\mathcal{T}}_n^B(t)$ be the set of topologically distinct 
		labeled dashed graphs obtained by labeling $t$ with $C\in \bar{\mathcal{S}}(B,n)$. 
		Then: 
		\begin{equation}
		\label{stab3.0}
		\sum_{T \in \overline{\mathcal{T}}_n^B(t)}\frac{|{\rm Perm}(B)|}{{\rm Sym}(T)}
		=S(I,n)\frac{n!}{|{\rm Aut}(t)|}\,.
		\end{equation}
	\end{lemma}
	
	\begin{proof}
		We may trivially rewrite the left hand side of (\ref{stab3.0})
		\begin{equation}
		\label{stab3.1}
		\sum_{T \in \overline{\mathcal{T}}_n^B(t)}
		\frac{|{\rm Perm}(B)|}{{\rm Sym}(T)}=
		\sum_{C = [\pi]\in \mathcal{S}(B,n)/{\rm Perm}(B)}
		\sum_{T \in \mathcal{T}_n^C}\frac{|{\rm Perm}(B)|}{{\rm Sym}(T)}.
		\end{equation}
		From (\ref{BTratio2}) we know 
		\begin{equation}
		\label{stab3.2}
		\frac{|{\rm Perm}(B)|}{{\rm Sym}(T)}
		=\frac{|{\rm Perm}(B)|}{|{\rm stab}(\pi)|}
		\frac{|{\rm Perm}(C)|}{|{\rm Aut}(T)|},
		\end{equation}
		where ${\rm stab}(\pi)$ is the subgroup of ${\rm Perm}(B)$ leaving 
		$\pi\in \mathcal{S}(B,n)$ invariant. It follows from the definition of 
		$\mathcal{S}(B,n)/{\rm Perm}(B)$  that if $\pi_1,\pi_2\in {\rm orb}(\pi)$ 
		then $|{\rm stab}(\pi_1)|=|{\rm stab}(\pi_2)|$.
		Combining successively (\ref{stab3.2}), Lemma \ref{stablemma1}, 
		and the orbit-stabilizer theorem gives the assertion: 
		\begin{eqnarray}
		\label{stab3.3}
		&\!\!\!& \sum_{C=[\pi]\in \mathcal{S}(B,n)/{\rm Perm}(B)}
		\sum_{T \in \mathcal{T}_n^C}\frac{|{\rm Perm}(B)|}{{\rm Sym}(T)}
		=\sum_{C=[\pi]\in \mathcal{S}(B,n)/{\rm Perm}(B)}\!\!
		\bigg(\,\frac{|{\rm Perm}(B)|}{|{\rm stab}(\pi)|}
		\sum_{T \in \mathcal{T}_n^C}\frac{|{\rm Perm}(C)|}{|{\rm Aut}(T)|}\bigg)
		\nonumber\\
		&\!\!\!& = \frac{n!}{|{\rm Aut}(t)|}
		\sum_{[\pi]\in \mathcal{S}(B,n)/{\rm Perm}(B)}|{\rm orb}(\pi)|
		=\frac{n!}{|{\rm Aut}(t)|}S(I,n).
		\end{eqnarray}
	\end{proof}
	
	Note that Lemma \ref{stablemma3} is the counterpart of 
	Lemma \ref{stablemma1} for $C$ induced by set partitions. Similarly, 
	Proposition \ref{stabprop2} is the counterpart of Lemma 
	\ref{stablemma2}. Instead of holding $C$ fixed we take it to 
	range over all $C \in \overline{\mathcal{S}}(B,D_n)$, as defined before 
	Proposition \ref{stabprop2}. Indicating all dependencies 
	in the notation we set
	\begin{eqnarray} 
	 \overline{\mathcal{T}}^B(t, D_n ,\rho_n) &:=&  \bigcup_{C \in
		\overline{\mathcal{S}}(B,D_n)} \mathcal{T}^C(\rho_n) 
	\nonumber\\
	& = &
	\{ T \in \mathcal{T}^B_n(t) :\, {\rm ddeg} \nu_0^{\sigma} = \rho_n \,,
	\;\nu_0^{\sigma} \in \nu_0^C, \;C \in \overline{\mathcal{S}}(B,D_n) \}\,.
	\end{eqnarray}
	
	\begin{proof}[Proof of Proposition \ref{stabprop2}]
		We begin as in the proof of Lemma \ref{stablemma3} and  
		rewrite the left hand side of (\ref{prop2sum}) as 
		\begin{eqnarray}
		&& \sum_{T \in \overline{\mathcal{T}}^B(t,D_n,\rho_n)}
		\frac{|{\rm Perm}(B)|}{{\rm Sym}(T)}=
		\sum_{C= [\pi]\in \mathcal{S}(B,D_n)/{\rm Perm}(B)}
		\sum_{T\in \mathcal{T}^C(\rho_n)}\frac{|{\rm Perm}(B)|}{{\rm Sym}(T)}
		\nonumber\\
		&& = \sum_{C= [\pi]\in \mathcal{S}(B,D_n)/{\rm Perm}(B)}
		\bigg(\,\frac{|{\rm Perm}(B)|}{|{\rm stab}(\pi)|}
		\sum_{T \in \mathcal{T}^C(\rho_n)}\frac{|{\rm Perm}(C)|}{|{\rm Aut}(T)|}\bigg).
		\end{eqnarray}
		To the subsum in round brackets we apply Lemma \ref{stablemma2} to obtain 
		\begin{equation}
		\sum_{T \in \overline{\mathcal{T}}^B(t,D_n,\rho_n)}\frac{|{\rm Perm}(B)|}{{\rm Sym}(T)}=
		\frac{|\nu_0(\rho_n)|}{|{\rm Aut}(t)|}
		\sum_{[\pi]\in \mathcal{S}(B,D_n)/{\rm Perm}(B)}\frac{|{\rm Perm}(B)|}{|{\rm stab}(\pi)|},
		\end{equation}
		using that $|\nu_0(\rho_n)|$ is independent of equivalence 
		relation on $C = [\pi]$. On account of the orbit-stabilizer theorem
		$|{\rm Perm}(B)|/|{\rm stab}(\pi)| = |{\rm orb}(\pi)|$ the 
		sum over $[\pi]$ produces the cardinality of the set 
		$\mathcal{S}(\{b_1,\ldots,b_I\}, D_n)$, i.e.~$P(D(v),D_n)$ and establishes 
		(\ref{prop2sum}). Its right hand side is manifestly independent of
		the (degree compatible) equivalence relation ``$\,\sim$'' on $B$. 
	\end{proof}
	
	\section{Conclusions} 
	
	Motivated by the widespread use of the FRG equation (\ref{i1}) 
	we formulated a program for its graph theoretical solution. 
	Subject to ultralocal initial conditions (\ref{i1}) 
	can be replaced by the iteratively soluble (\ref{i3}) 
	producing a long range hopping expansion (LRH) for 
	$\Gamma_{\kappa} = \Gamma_0 + \sum_{l \geq 1} \kappa^{l} 
	\Gamma_{l}$, from which a solution $\Gamma_k$
	of (\ref{i1}) can be obtained by substitution,
	$\Gamma_k = \Gamma_{\kappa}|_{ \ell \rightarrow \ell(k)}$. As the 
	iteration of (\ref{rec2}), or its equivalent mixed form (\ref{GamvsW4}), is only feasible to moderate 
	orders we formulated graph rules for the direct 
	evaluation of an arbitrary order $\Gamma_{l}$.   
	The derivation, computational test, and proof of these graph 
	rules constitute the main result of the paper. 
	
	By the results of Section IV the subsums over vertex labeled
	trees $T \in \mathcal{T}(B(v),n)$ with fixed weight $\mu(T)$ have a 
	combinatorial meaning in terms of the number of integer labeled
	tree graphs of the same topology as $T$. The graph 
	rule could therefore optimized once explicit results 
	for the number of set partitions $P(D(v),D_n)$ are available; 
	see \cite{multisets1} for some related results. 
	
	The construction so far only holds in the formal series 
	sense. Guided by a variety of convergence results for hopping 
	expansions in the literature (see \cite{BK,ReiszLRH,Viral} 
	and the references therein) we expect that the LRH expansion 
	for $\Gamma_{\kappa}$ has finite radius of convergence under 
	natural conditions. 
	From a computational perspective it would also be desirable 
	to identify subclasses of one-line irreducible graphs that 
	can be analytically summed and lead to controlled 
	approximate solutions of (\ref{i1}), replacing the traditional 
	ad-hoc Ans\"{a}tze. 
	
	\newpage 

\begin{appendix}	
	
	\section{Recursive results to fifth order} 
	
	Here we present explicit results for $\Gamma_{l},\,l = 2,\ldots,5$, 
	and various checks on them. A closed recursion arises from the expansion 
	of the $\kappa$-flow equation in (\ref{i3}). In preparation we define 
	polynomials $m_{l} = m_{l}(u_1,\ldots, u_{l})$ in non-commuting 
	variables $u_n,\,n \in \mathbb{N}$, by 
	\begin{eqnarray}
	\label{rec1} 
	&& \bigg( 1\! + \!\sum_{n \geq 1} \kappa^n u_n \bigg)
	\bigg( 1\! -\! \sum_{l \geq 1} m_{l}(u) \kappa^{l} \bigg) = 1 =
	\bigg( 1 \!-\! \sum_{l \geq 1} m_{l}(u) \kappa^{l} \bigg)
	\bigg( 1 \!+\! \sum_{n\geq 1} \kappa^n u_n \bigg)\,,
	\nonumber\\
	&& m_{l}(u)  = \sum_{n =1}^{l} \sum_{i_1 + \ldots + i_n = l, i_j \in \mathbb{N}} (-)^{n+1} 
	\,u_{i_1} \ldots u_{i_n}\,.
	\end{eqnarray}
	At low orders: $m_1 = u_1$, $m_2 = u_2 - u_1^2$, $m_3 = 
	u_3 - u_1 u_2 - u_2 u_1 + u_1^3$. Inserted into (\ref{i3}) one has 
	$\Gamma_1 \equiv 0$ and 
	\begin{equation}
	\label{rec2}
	\Gamma_{l}=\frac{1}{2l} \sum_{n=1}^{l-1} 
	\sum_{i_1 + \ldots + i_n = l-1}(-)^n 
	\,{\rm Tr}[u_1 u_{i_1} \ldots u_{i_n}]\,,\quad l \geq 2\,,
	\end{equation}
	with $\Gamma_0^{(2)} \cdot u_1 = \ell$, $\Gamma_0^{(2)} \cdot u_i = 
	\Gamma_i^{(2)}$, $i \geq 2$. Here $\Gamma_0^{(2)}[\phi]$ is invertible  
	\begin{equation}
	\label{rec3} 
	\Gamma_0^{(2)}[\phi]_{x,y} = \gamma_2(\phi_x)\delta_{x,y}\,,\quad 
	\gamma_2(\varphi)^{-1} = \omega_2|_{h = h(\varphi)}\,. 
	\end{equation}
	In slight abuse of notation we set $\omega_i(\varphi) := \omega_i(h(\varphi))$, 
	$\omega_i(h) = \partial^i \omega/\partial h^i$, $i \geq 2$, and find: 
	\begin{eqnarray} 
	\label{gammresults234}
	\Gamma_2[\phi] &=& - \sum_{x_1,x_2}\,\frac{1}{4} 
	\omega_2(\phi_{x_1})\omega_2(\phi_{x_2}) (\ell_{x_1 x_2})^2 
	\nonumber\\
	 \Gamma_3[\phi] &=& \sum_{x_1,x_2}\,
	\frac{1}{12} \omega_3(\phi_{x_1})\,\omega_3(\phi_{x_2})\,
	(\ell_{x_1 x_2})^3 
	\\[2mm]
	&+&\sum_{x_1,x_2,x_3}\,\frac{1}{6}\omega_2(\phi_{x_1})\omega_2(\phi_{x_2})\omega_2(\phi_{x_3})
	\ell_{x_1 x_2}\, \ell_{x_2 x_3}\,\ell_{x_1 x_3}\,
	\nonumber\\
	\Gamma_4[\phi] &=& - \sum_{x_1,x_2}\,\frac{1}{48} \omega_4(\phi_{x_1}) \omega_4(\phi_{x_2}) (\ell_{x_1 x_2})^4 
	\nonumber\\
	&-&\sum_{x_1,x_2,x_3}\,\frac{1}{4} \omega_3(\phi_{x_1})\,
	\omega_3(\phi_{x_2})\,\omega_2(\phi_{x_3}) (\ell_{x_1 x_2})^2\,\ell_{x_1 x_3}\, \ell_{x_2 x_3}
	\nonumber\\
	&-&\sum_{x_1,x_2,x_3}\,\frac{1}{8} \omega_2(\phi_{x_1})\,[ \omega_4 - \omega_3 \gamma_2 \omega_3](\phi_{x_2})
	\,\omega_2(\phi_{x_3}) (\ell_{x_1 x_2})^2\, (\ell_{x_2 x_3})^2
	\nonumber\\
	&-&\sum_{x_1,x_2,x_3,x_4}\,\frac{1}{8}\omega_2(\phi_{x_1})\omega_2(\phi_{x_2})\omega_2(\phi_{x_3})\omega_2(\phi_{x_4})
	\ell_{x_1 x_2}\, \ell_{x_2 x_3}\,\ell_{x_3 x_4}\,\ell_{x_4 x_1}\,.
	\nonumber
	\end{eqnarray} 

	\begin{eqnarray}
	\label{gammresults5}
	&&\Gamma_5[\phi]= \sum_{x_1,x_2}\,\frac{1}{120}\omega_5(\phi_{x_1})\omega_5(\phi_{x_2})(\ell_{x_1 x_2})^5
	\nonumber\\
	&&+\!\!\sum_{x_1,x_2,x_3}\frac{1}{12}\omega_4(\phi_{x_1})\omega_4(\phi_{x_2})\omega_2(\phi_{x_3})
	(\ell_{x_1 x_2})^3\ell_{x_1 x_3}\ell_{x_2 x_3}
	\nonumber\\
	&&+\!\!\sum_{x_1,x_2,x_3}\frac{1}{12}\omega_2(\phi_{x_1})
	\big[\omega_5\!-\!\omega_3\gamma_2\omega_4\big](\phi_{x_2})
	\omega_3(\phi_{x_3})(\ell_{x_1 x_2})^2(\ell_{x_2 x_3})^3
	\nonumber\\
	&&+\!\!\sum_{x_1,x_2,x_3}\frac{1}{8}\omega_3(\phi_{x_1})\omega_3(\phi_{x_2})\omega_4(\phi_{x_3})
	\ell_{x_1 x_2}(\ell_{x_1 x_3})^2(\ell_{x_2 x_3})^2
	\\[2mm]
	&&+\!\!\sum_{x_1,x_2,x_3,x_4}\frac{1}{4}\omega_2(\phi_{x_1})\omega_2(\phi_{x_2})\omega_3(\phi_{x_3})\omega_3(\phi_{x_4})
	\ell_{x_1 x_2}\ell_{x_2 x_3}(\ell_{x_3 x_4})^2\ell_{x_4 x_1}
	\nonumber\\
	&&+\!\!\sum_{x_1,x_2,x_3,x_4}\frac{1}{4}\omega_3(\phi_{x_1})\omega_2(\phi_{x_2})\omega_3(\phi_{x_3})\omega_2(\phi_{x_4})
	\ell_{x_1 x_2}\ell_{x_2 x_3}\ell_{x_3 x_4}\ell_{x_4 x_1}\ell_{x_1 x_3}
	\nonumber\\
	&&+\!\!\sum_{x_1,x_2,x_3,x_4}\frac{1}{4}\omega_2(\phi_{x_1})\omega_2(\phi_{x_2})\big[\omega_4
	\!-\!\omega_3 \gamma_2 \omega_3 \big](\phi_{x_3})\omega_2(\phi_{x_4})
	\ell_{x_1 x_2}\ell_{x_2 x_3}\ell_{x_3 x_1}(\ell_{x_3x_4})^2
	\nonumber\\
	&&+\!\!\sum_{x_1,x_2,x_3,x_4,x_5}\frac{1}{10}\omega_2(\phi_{x_1})\omega_2(\phi_{x_2})\omega_2(\phi_{x_3})
	\omega_2(\phi_{x_4})\omega_2(\phi_{x_5})\ell_{x_1 x_2}\ell_{x_2 x_3}\ell_{x_3 x_4}\ell_{x_4 x_5}\ell_{x_5 x_1}\,.
	\nonumber
	\end{eqnarray}
	A computational point worth mentioning is that the $(\Gamma^{(2)}_m)_{xy}$, 
	$m \geq 2$,  have in general diagonal elements. In evaluating the traces 
	one has to split off at intermediate steps subsums containing 
	$(\Gamma^{(2)}_m)_{xx}$ contributions. Such contributions combine with 
	others and lead to unresticted sums in the final result, but with 
	modified coefficients. 
	
	The corresponding $W_l$'s are readily obtained from the Wortis graph 
	rule and are not displayed explicitly. In line with table 1 the 
	expressions for the $\Gamma_{l}$ are (at matching orders) more concise 
	than the $W_{l}$'s and yet code the same information. 
	These results have been tested and compared with partial results 
	in the literature in various ways. (i) $W_1,\ldots, W_4$ and 
	$\Gamma_2,\ldots, \Gamma_5$ are related by the mixed recursion 
	(\ref{GamvsW3}). (ii) Specialized to the Ising model
	$\Gamma_2,\ldots,\Gamma_4$ agree with the resuls of \cite{GammIsing1}. 
	
	The $W_1,\ldots, W_4$ themselves can be specialized to a nearest neighbor 
	hopping matrix and matched to results in the literature. (iii) for 
	$H=0$ the textbook result for the free energy is reproduced. 
	(iv) The 2-point susceptibility $\chi_2 = \sum_x W^{(2)}_{x,0}|_{H=0}$ 
	and the 4-point susceptibility $\chi_4 = \sum_{x_1,x_2,x_3} 
	W^{(4)}_{x_1,x_2,x_3,0}|_{H=0}$ match (in $d=2$) the results in 
	\cite{BakerKin}. 
	
	\newpage
	\section{Single site data and zero-dimensional Legendre transform} 
	
	The function $\gamma(\varphi)$ entering the ultralocal initial functional 
	$\Gamma_0[\phi] = \sum_x \gamma(\phi_x)$ can be characterized by the 
	functional relation 
	\begin{equation} 
	\label{0lege0} 
	\exp - \gamma(\varphi) = \int_{-\infty}^{\infty} \! d\chi \,
	\exp \Big\{\! - s(\chi) + (\chi\! -\! \varphi) \frac{\partial \gamma}{\partial \varphi}\Big\}\,,
	\end{equation}
	where $s$ is the single site action. By shifting the argument $\varphi \mapsto \varphi + \alpha$
	and expanding in powers if $\alpha$ one can express the derivatives $\gamma_n(\varphi) = 
	\partial^n \gamma/\partial \varphi^n$ in terms of the cumulants $\omega_n(h) = \partial^n \omega/\partial h^n$ 
	of the measure $d\chi e^{-s(\chi)}$. The latter have generating function 
	$e^{\omega(h+k)} = \int\! d\chi \exp \{ - s(\chi) + (h+k) \chi\}$ upon expansion 
	in powers of $k$. To low orders one finds 
	\begin{equation}
	\label{0lege0a}
	\gamma_2(\varphi) = \omega_2^{-1}|_{h = h(\varphi)}\,, 
	\quad 
	\gamma_3(\varphi) = - \omega_2^{-3} \omega_3|_{h = h(\varphi)}\,,  
	\quad
	\gamma_4(\varphi) = 
	[- \omega_4 \omega_2^{-4} + 3 \omega_3^2 \omega_2^{-5}]_{h = h(\varphi)}\,. 
	\end{equation} 
	Augmented by $\omega_1|_{h = h(\varphi)} = \varphi$ and $\gamma_1(\varphi) = h(\varphi)$ 
	the inverse relations are obtained by flipping the roles of the 
	$\omega_m$'s and $\gamma_m$'s. 
	This reflects the fact that the generating functions $\omega(h\!+\!k)$ and 
	$\gamma(\varphi\!+\!\alpha)$ are Legendre transforms of each other. In 
	quantum field theoretical terminology the $\gamma_{l}(\varphi)$ are 
	zero-dimensional vertex functions with non-zero mean field and the 
	$\omega_{l}(h)$ are the zero-dimensional cumulants with non-zero source. 
	
	The combinatorial patterns arising through the Legendre transform 
	can be analyzed in closed form in the zero dimensional case. 
	The integral realization does not enter, so $\omega:\mathbb{R} \rightarrow \mathbb{R}$, can 
	be any smooth function with nonzero second derivative, $\omega^{(2)}(h) >0$, 
	say. We assume that $\omega^{(1)}(h) = \varphi$ can be solved for $h(\varphi)$, 
	where $h$ is likewise smooth and $h^{(1)}(\varphi) >0$. 
	We define the Legendre transform by $\gamma(\varphi) := 
	\varphi h(\varphi) - \omega(h(\varphi))$. Then $\gamma^{(1)}(\varphi) = h(\varphi)$ 
	and the primary assertion is that $\omega^{(1)}$ 
	and $\gamma^{(1)}$ are compositional inverses of each other
	\begin{equation} 
	\label{0lege1} 
	\omega^{(1)}(\gamma^{(1)}(\varphi)) = \varphi\,, \quad 
	\gamma^{(1)}(\omega^{(1)}(h)) = h\,. 
	\end{equation}
	By repeated differentiation of these basic formulas one can 
	generate relations between the derivatives $\gamma_{l}(\varphi) := 
	\gamma^{(l)}(\varphi)$ and the $\omega_{l}(h) := \omega^{(l)}(h)$. 
	For now we focus on 
	$\omega^{(1)}(\gamma^{(1)}(\varphi)) = \varphi$, where the first few relations 
	generated are (\ref{0lege0a}). Since the resulting $\omega(h)$ derivatives 
	are always evaluated at $h = h(\varphi) = \gamma^{(1)}(\varphi)$ 
	it is convenient to set $\omega_m(\varphi) := \omega_m(h)|_{h = h(\varphi)}$ by 
	slight abuse of notation. In this notation $\partial_{\varphi}$ is a linear 
	derivation acting via $\partial_{\varphi} \omega_n = \omega_{n+1} \gamma_2$ and 
	$\partial_{\varphi} \gamma_n = \gamma_{n+1}$ on the constituents. 
	The differentiation rule implies that the $l$-th order relation has 
	the form  
	\begin{equation} 
	\label{0lege2} 
	\omega_2^{l} \gamma_{l} =  - \omega_l \;+ \!\!\!
	\sum_{-2 i_2 + 3 i_3 + \ldots + (l-1) i_{l -1} =l}
	(-)^{i_3 + \ldots + i_{l-1}} c_{i_3\ldots i_{l-1}} \,(\omega_2^{-1})^{i_2} 
	\omega_3^{i_3} \ldots \omega_{l-1}^{i_{l-1}}\,,\quad l \geq 4\,,
	\end{equation}
	for $i_k \in \mathbb{N}$ and integer coefficients $c_{i_3\ldots i_{l-1}} \in \mathbb{N}$.
	Unless noted otherwise the $\omega_m$ in the following are the
	$\omega_m(\varphi) = \omega_m(h)|_{h = h(\varphi)}$ regarded as functions of $\varphi$. 
	
	The coefficients $c_{i_3\ldots i_{l-1}}$ are indirectly characterized 
	by the duality property (\ref{0lege1}): solving (\ref{0lege2}) recursively 
	for the $\omega_{l}$ in terms of $\gamma_2, \ldots, \gamma_{l}$, 
	the same formula arises. An explicit formula for them arises 
	from the known combinatorial expressions for the compositional inverse: 
	\begin{equation} 
	\label{0lege3} 
	c_{i_3\ldots i_{l-1}} = 
	\Big( l\! -\!2 + \mbox{$\sum_{j=3}^{l -1} i_j $} \Big)! 
	\prod_{j=3}^{l-1} \frac{1}{ i_j! (j\!-\!1)!^{\,i_j}} \,.
	\end{equation}
	This has been obtained in \cite{Jackson} in a setting that 
	mimics perturbation theory; the functions are power series in 
	the fields and the numerical coefficients are related as in 
	(\ref{0lege3}). A little thought shows that the coefficients arising 
	through repeated differentiation of arbitrary smooth functions 
	without setting the argument to zero are the same. For later 
	reference we sketch the argument. 
	
	The basic input is an explicit expression for the compositional 
	inverse of a formal power series. There are several variants of such 
	formulas and their tree-graph interpretation; the version most 
	directly leading to (\ref{0lege3}) is Eq.~(4.6) in \cite{Sokal}. 
	Given $a(z) = \sum_{n \geq 1} a_n \,z^n$, $a_1 \neq 0$, the 
	series $b(w) = \sum_{n \geq 1} \frac{b_n}{n!} \,w^n$ is the 
	compositional inverse of $a$, i.e.~$a(b(w)) =w$, iff 
	\begin{equation}
	\label{0lege4} 
	b_n = \sum_{k_2,k_3, \ldots \geq 0,\sum_j (j\!-\!1) k_j = n\!-\!1} 
	\!\!\!(-)^{\sum_j k_j} \, a_1^{-(1 + \sum_j j k_j)} 
	\Big( \sum_j j k_j \Big)! \,\prod_j \frac{a_j^{k_j}}{k_j!}\,,
	\end{equation}    
	where all sums and products range over $j \geq 2$ and are rendered finite 
	by the Euler relation $\sum_j (j\!-\!1) k_j = n\!-\!1$. For the application 
	here we shift the arguments of $\omega$ and $\gamma$ and re-expand. For the 
	first derivatives this gives 
	\begin{eqnarray} 
	\label{0lege5} 
	\omega^{(1)}(h + k) & = & \sum_{n \geq 0} \frac{k^n}{n!} \omega_{n+1}(h) =: \omega^{(1)}(h;k)\,, 
	\nonumber\\
	\gamma^{(1)}(\varphi + \alpha) & = & \sum_{n \geq 0} 
	\frac{\alpha^n}{n!} \gamma_{n+1}(\varphi) =: 
	\gamma^{(1)}(\varphi;\alpha)\,.
	\end{eqnarray}
	By (\ref{0lege1}) we require both series to be compositional inverses 
	of each other as series in $k,\alpha$. The result (\ref{0lege3}) 
	then follows from (\ref{0lege4}). 
	
	The formula (\ref{0lege4}) has several known combinatorial and tree graph interpretations, see \cite{multisets1,Sokal,Jackson} and the references therein. In the remainder of this appendix we present a graph theoretical 
	interpretation of (\ref{0lege2}), (\ref{0lege3}) which, together 
	with its proof, mirrors some aspects of its quantum field theoretical 
	counterpart in Section III. 
	
	{\bf Graph rules for $\gamma_{l}$:}

	\begin{itemize}
		\item  [(i)] At order $l \geq 4$ draw  all topologically distinct 
		connected tree graphs $t \in \mathcal{T}_{l}$ with $l = |\nu_0|$ 
		external vertices of order $1$ and any number $|\nu_1|$ of $k$-valent 
		vertices, $k=3,\ldots, l$, joined by dashed lines. Multiply by 
		$l!/|{\rm Aut}(t)| \in \mathbb{N}$, where ${\rm Aut}(t)$ is the 
		automorphism group of the graph. 
		\item[(ii)] Attribute to each $t \in \mathcal{T}_{l}$ a weight 
		$(-)^{|\nu_1|}\mu(t)$ 
		as follows: a factor $\omega_2^{-1}$ to each dashed line, $1$ to an 
		$1$-valent vertex, and $-\omega_k$ to an $k$-valent vertex, $k \geq 3$.
		\item[ (iii)] Sum over all contributions to obtain
		\begin{equation} 
		\label{0graphrule} 
		\gamma_{l} = \sum_{t \in \mathcal{T}_{l}} 
		(-)^{|\nu_1|} \frac{l!}{ |{\rm Aut}(t)|} \mu(t) \,, \quad
		\mbox{i.e.}\quad  
		c_{i_3\ldots i_{l-1}} = 
		l!\sum_{t \in \mathcal{T}_{i_3\ldots i_{l-1}} } \frac{1}{|{\rm Aut}(t)|}\,,  
		\end{equation}
		where the $c$'s are those in (\ref{0lege2}) and the graphs 
		contributing to a fixed $\mu(t)$ with labels $(i_3\ldots i_{l-1})$ 
		are denoted by $\mathcal{T}_{i_3\ldots i_{l-1}} \subset \mathcal{T}_{l}$. 
	\end{itemize}
	The automorphism group in (i) is defined as in Section II.A, with 
	$\nu_0$ the set of  external vertices, $\nu_1$ the set of 
	multi-valent vertices, and edge list $\epsilon \subset 
	(\nu_0 \cup \nu_1)_2$. 
	The Euler relations holds in the form $|\nu_0| + |\nu_1| = 
	|\epsilon| +1$ and $\sum_{k=1}^{l} k |v|_k = 2 |\epsilon|$, where 
	$|v|_k$ is the number of $k$-valent vertices, $k=1,3,4,\ldots$. 
	The second form ensures that the degrees and signs in (\ref{0lege2}) are 
	correctly reproduced by (ii) and only the coefficients 
	$c_{i_3\ldots i_{l-1}}$ in (iii) need to be understood in graph 
	theoretical terms. Importantly, for $l \geq 6$ 
	several topologically distinct tree graphs contributing to a 
	fixed a $(i_3\ldots i_{l-1})$ configuration can occur. 
	As an example, the tree graphs contributing to $\gamma_5$ are 
	displayed below 
	\begin{eqnarray}
	\label{0lege7}  
	&& \makebox[5mm]{} 
	\begin{tikzpicture}[scale=0.2]
	\draw[black, dashed] (2.5,0) -- (5.5,0.0);
	\draw[black, dashed] (2.5,0) -- (3.5,3.0);
	\draw[black, dashed] (2.5,0) -- (3.5,-3.0);
	\draw[black, dashed] (2.5,0) -- (0,1.7);
	\draw[black, dashed] (2.5,0) -- (0,-1.7);
	\end{tikzpicture}
	\makebox[2cm]{} \;\;\omega_2^{-5} \omega_5 \makebox[1cm]{} |{\rm Aut}(t)| = 5!
	\nonumber
	\\[5mm] 
	&& \makebox[6mm]{}   
	\begin{tikzpicture}[scale=0.4]
	\draw[black, dashed] (1.5,0) -- (3.5,0.0);
	\draw[black, dashed] (2.5,-1) -- (2.5,1);
	\draw[black, dashed] (2.5,1) -- (3.5,1.5);
	\draw[black, dashed] (2.5,1) -- (1.5,1.5);
	\end{tikzpicture}
	\makebox[2cm]{} \quad \omega_2^{-6} \omega_3 \omega_4 \quad \;\;|{\rm Aut}(t)| = 3!2!
	\nonumber
	\\[5mm]
	&& 
	\begin{tikzpicture}[scale=0.4]
	\draw[black, dashed] (1,0) -- (4,0.0);
	\draw[black, dashed] (1,0) -- (0,1);
	\draw[black, dashed] (1,0) -- (0,-1);
	\draw[black, dashed] (2.5,0) -- (2.5,-1);
	\draw[black, dashed] (4,0) -- (5,1);
	\draw[black, dashed] (4,0) -- (5,-1);
	\end{tikzpicture}
	\makebox[2cm]{} \omega_2^{-7} \omega_3^3 \quad \quad \;|{\rm Aut}(t)| = 2^3
	\end{eqnarray} 
	
	A proof of (\ref{0graphrule}) can be based on the known tree graph 
	interpretation of (\ref{0lege4}), see \cite{Jackson,Sokal} and 
	the references therein. Below we provide an alternative ab-initio 
	proof without reference to the compositional inverse formula. 
	The key ingredient is the following mixed recursion relation 
	\begin{equation}
	\label{0Lprf4}
	\gamma_l (\varphi)=
	-\sum_{j=2}^{l} \omega_j(\varphi) \frac{l!}{j!}
	\sum_{k_1+\ldots + k_j =l,\,1\leq k_i\leq l-2}
	\frac{\gamma_{k_1}^{(1)}(\varphi)}{k_1!}\cdots 
	\frac{\gamma_{k_j}^{(1)}(\varphi)}{k_j!}\,, \quad l \geq 3\,.
	\end{equation}
	This is the zero-dimensional counterpart of the recursions 
	(\ref{GamvsW3}), (\ref{GamvsW4}) instrumental for our 
	analysis of the $\Gamma_{\kappa}$ graph rules. It can 
	be derived along similar lines starting from 
	$\gamma(\varphi + \alpha) = (\varphi + \alpha) 
	\gamma^{(1)}(\varphi;\alpha) - \omega(\gamma^{(1)}(\varphi;\alpha))$ 
	and (\ref{0lege5}). 
	
	\begin{proof}[Ab-initio proof of $\gamma_l$ graph rule based on (\ref{0Lprf4}).] 
	We proceed by induction in $l$, assuming that (\ref{0graphrule}) 
	is known to produce the correct coefficients (\ref{0graphrule}) 
	for $k=1,\ldots,l\!-\!1$. 
	To obtain the result at order $l$ we first note a simple 
	generation recipe ($\ast$): the set of tree graphs in $\mathcal{T}_{k-1}$ can be 
	obtained from those in $\mathcal{T}_{k-2}$ by insertion of a line 
	in all possible ways either at a multi-valent vertex or in the middle 
	of an existing line. In fact, differentiating 
	a weight of order $k\!-\!2$ from part (ii) of the graph 
	rule, $\partial_{\varphi} \mu(t)$, produces a sum of terms whose 
	interpretation as order $k\!-\!1$ tree graphs follows the pattern 
	($\ast$). The terms occur with integer multiplicities which by 
	the origin of (\ref{0lege2}) from (\ref{0lege1}) must be 
	compatible with (\ref{0graphrule}). 
	
	The recursion (\ref{0Lprf4}) also mirrors the pattern ($\ast$). 
	Fix some $t \in \mathcal{T}_{l}$ generated from order $l\!-\!1$ graphs 
	as indicated. The contribution of $t$ to $\gamma_{l}/l!$ can 
	be matched to terms on the right hand side of (\ref{0Lprf4}) in 
	the following way.
	Case 1: any $3 \leq j$-valent internal 
	vertex can be seen as the $\omega_j$ piece, and the $j$ subtrees it 
	connects to as the $1$-rooted $\gamma_{k_i}^{(1)}/k_i!$ pieces. 
	Case 2: the middle of an internal line with adjacent vertices of 
	weights $\omega_{n_1},\omega_{n_2}$ can be seen 
	as an $\omega_2$ pseudo-vertex via $\omega_{n_1} \omega_2^{-1} \omega_{n_2} = 
	\partial_{\varphi} \omega_{n_1 -1} \omega_2 \partial_{\varphi} \omega_{n_2 -1}$, 
	and the two subtrees it connects to as $1$-rooted $\gamma_{k_i}^{(1)}/k_i!$ 
	pieces. The full contribution of $t$ obtained from (\ref{0Lprf4}) 
	is then the sum of reassembled rooted graph weights produced by 
	each distinct choice of the $\omega_j$, $j \geq 2$, 
	and $\gamma_{k_i}^{(1)}$ pieces. Our task is to keep track of
	the coefficients.

	Case 1: $j$-valent vertex as $\omega_j$, $j\geq3$. By induction 
	hypothesis each of the $\gamma_{k_i}$ has a graph realization
	in $\mathcal{T}_{k_1}$ via (i),(ii),(iii). Its derivative $\gamma^{(1)}_{k_i}$ has the 
	same structure where initially the differentiated weights 
	$\partial_{\varphi} \mu(t)$ occur. By the remark following ($\ast$) 
	each $\partial_{\varphi} \mu(t)$ expands into tree graphs of one order 
	higher which we regard as $1$-rooted, $t' \in \mathcal{T}_{k_i+1}^{1\bullet}$
	(with the rooted vertex always an internal one).  
	The regrouping leads to coefficients of the $\mu(t')$'s that must 
	by the differentiation compatibility be given by the 
	graph rule (at lower orders) applied to rooted trees.  
	In summary, each term in the graph expansion 
	of $\gamma_{k_i}^{(1)}/k_i!$ carries the coefficient 
	\begin{equation}
	\label{case1.1}
	\frac{(-1)^{|\nu_1(t')|}}{|{\rm Aut}(t')|}\,,
	\quad t' \in \mathcal{T}_{k_i+1}^{1\bullet}\,.
	\end{equation}
	Suppose that there are $j_i$ isomorphic subtrees $t'_i$, 
	$i=1,\ldots n$ attached to the $\omega_j$ vertex. Then, accounting 
	for the $1/j!$ in (\ref{0Lprf4}) we obtain the full prefactor 
	for the choice of $\omega_j,\,j \geq 3$, as vertex 
	\begin{equation}
	\label{case1.2}
	\prod_{i=1}^{n} \bigg( \frac{(-1)^{j_i|\nu_1(t'_i)|}}{j_i! 
		|{\rm Aut}(t'_i)|^{j_i}}\bigg)\,.
	\end{equation}
	Let $t \in \mathcal{T}_{l}$ be the graph reassembled from the 
	rooted subtrees $t'_i$ at the vertex with weight $\omega_j$. 
	The total weight is $\omega_j$ times 
	the product of the weights of the subtrees and is of the 
	form $\mu(t)$ as in part (ii) of the graph rule. 
	The overall sign $(-)^{|\nu_1(t)|}$, with $|\nu_1(t)|$ the 
	number of internal vertices of $t$. A straightforward 
	application of the orbit stabilizer theorem shows that 
	the modulus of (\ref{case1.2}) equals the symmetry factor of 
	$t$ rooted at our choice of $\omega_j$ vertex. As an unrooted 
	graph the overall coefficient is $(-)^{|\nu_1(t)|}/|{\rm Aut}(t)|$. 
	There may be several choices of $\omega_j$ vertices contributing 
	equally, so the net coefficient for Case 1 is 
	\begin{equation}
	\label{case1.3}
	\frac{(-)^{|\nu_1(t)|}}{|{\rm Aut}(t)|} \times 
	\textrm{\# of $\omega_j$ choices with fixed $t$}.
	\end{equation}
	Case 2: middle of an internal line as $\omega_2$ pseudo-vertex. As before, 
	each of the two subtrees attached to $\omega_2$ 
	contributes with coefficient (\ref{case1.1}). 
	While two subtrees may be distinct or identical, their contribution 
	to the overall symmetry factor will be accounted for by the 
	$1/2!$ prefactor in (\ref{0Lprf4}). Again we write $t \in \mathcal{T}_{l}$ 
	for the graph  obtained by reassembling the two subtrees at 
	the $\omega_2$ pseudo-vertex.
	The overall symmetry factor obtained is that for $t$ 
	rooted at the two ends of the internal line. 
	When reassembled to $t$ via $\partial_{\varphi} \omega_{n_1 -1} \omega_2 
	\partial_{\varphi} \omega_{n_2 -1} =\omega_{n_1} \omega_2^{-1} \omega_{n_2}$ (with 
	$\omega_{n_1},\omega_{n_2}$ the weights of the rooted vertices)
	the overall coefficient is $-(-)^{|\nu_1(t)|}/|{\rm Aut}(t)|$. 
	The extra sign accounts for the fact that in the graph rule
	$\omega_2^{-1}$ carries no sign while in (\ref{0Lprf4}) the 
	$\omega_2$ term does. There may be several equivalent internal 
	lines in $t$ that are reassembled in this way. The net 
	coefficient for Case 2 then is  
	\begin{equation}
	\label{case2.1}
	-\frac{(-)^{|\nu_1(t)|}}{|{\rm Aut}(t)|} \times 
	\textrm{\# of equivalent internal lines in $t$}.
	\end{equation}
	The full contribution to $\gamma_{l}/l!$ associated with $t$ 
	is obtained from (\ref{0Lprf4}) by summing over the contributions 
	from Case 1 and Case2 with weight $\mu(t)$ and coefficients 
	(\ref{case1.3}),(\ref{case2.1}). This gives 
	\begin{equation}
	\label{0Lprf8}
	\frac{(-)^{|\nu_1(t)|}}{|{\rm Aut}(t)|} 
	\times \big(|\nu_1| - |\epsilon_1|\big)\,,
	\end{equation}
	where $|\nu_1|$, $|\epsilon_1|$ are the total number of internal vertices
	and internal lines of $t$, respectively. For the tree graphs considered
	the number of external lines and vertices coincide, $|\nu_0| = l = 
	|\epsilon_0|$, so that the Euler relation reduces to 
	$|\nu_1| - |\epsilon_1| =1$. \end{proof} 
	
	A multi-dimensional version of the above graph rule would 
	similarly relate the vertex functions of a lattice 
	quantum field theory to its connected correlation functions
	(even at non-zero mean field or source).  This is implicit in 
	many text books; a proof can be read off from \cite{Jackson, BP} and 
	also the above derivation carries over. We briefly comment 
	here on this multi-dimensional version in order to highlight
	that the trees invoked are unrelated to those in Section III and IV. 
	We denote the standard (unmodified) Legendre transform by 
	$\tilde{\Gamma}[\phi] := \phi \cdot H[\phi] - W[H[\phi]]$, 
	with $W^{(1)}[H[\phi]] = \phi$. Then, $\tilde{\Gamma}^{(1)}[\phi] = H[\phi]$ 
	and the counterpart of (\ref{0lege1}) reads 
	\begin{equation} 
	\label{lege1}   
	W^{(1)}\big[\tilde{\Gamma}^{(1)}[\phi]\big] = \phi \,, \quad 
	\tilde{\Gamma}^{(1)}\big[W^{(1)}[H]\big] = H \,. 
	\end{equation}
	Throughout a superscript $(n)$ denotes $n$-fold differentiation of 
	a functional of one field with respect to its argument. 
	By repeated differentiation with respect to $\phi$ or $H$ 
	one obtains in principle mutually equivalent relations between 
	the $\tilde{\Gamma}^{(l)}[\phi]$ (vertex functions in non-zero mean field) 
	and $W^{(l)}[H]$ (cumulants with non-zero source). These coincide 
	essentially with those in the zero-dimensional case (\ref{0lege2}), 
	just that different lattice sum contractions will remove most
	of the degeneracies that give rise to non-unit coefficients. 
	That is, in the QFT counterpart of (\ref{0lege2}) there will be 
	$c_{i_3\ldots i_{l-1}}$ structurally similar terms 
	(with $W^{(m)}[H[\phi]], \tilde{\Gamma}^{(m)}[\phi]$ replacing $\omega_m,\gamma_m$,
	respectively) where the indices in the lattice sums are 
	contracted differently. The graph rule producing these
	correctly contracted terms in the $\Gamma^{(l)}[\phi]$ expansion 
	invokes the previous tree graphs $\mathcal{T}_{l}$, but now  labeled by lattice 
	points. The external points $x_1,\ldots, x_{l}$ will be taken 
	distinct but the lattice points summed over 
	in the products of $W^{(k)}_{y_1,\ldots,y_k}$, $k \geq 3$, vertices
	may coincide. This may occasionally produce coinciding labels
	for the internal vertices but the tree structure precludes 
	nontrivial automorphisms. A counterpart of the above 
	graph rule can then easily be formulated, see \cite{MO, BP} for related Hopf algebraic constructions. Despite the occurrence 
	of labeled tree graphs in this context it is $\tilde{\Gamma}^{(l}[\phi]$, 
	the $l$-th functional derivative of $\tilde{\Gamma}[\phi]$, 
	that is related to its $W^{(k)}$ counterparts, not the 
	order in a $\ell_{xy}$ expansion. Performing a $\kappa$ 
	expansion of both sides of (\ref{0lege2})'s multi-dimensional
	counterpart is of no immediate help in understanding the 
	graph rule underlying $\Gamma$'s hopping expansion.

\end{appendix}	
	
	\bigskip
	
	{\it Acknowledgements:} This work was supported by the PITT-PACC. R.B. also acknowledges support by the A\&S PITT-PACC fellowship.




\begin{thebibliography}{0}%
\makeatletter
\providecommand \@ifxundefined [1]{%
 \@ifx{#1\undefined}
}%
\providecommand \@ifnum [1]{%
 \ifnum #1\expandafter \@firstoftwo
 \else \expandafter \@secondoftwo
 \fi
}%
\providecommand \@ifx [1]{%
 \ifx #1\expandafter \@firstoftwo
 \else \expandafter \@secondoftwo
 \fi
}%
\providecommand \natexlab [1]{#1}%
\providecommand \enquote  [1]{``#1''}%
\providecommand \bibnamefont  [1]{#1}%
\providecommand \bibfnamefont [1]{#1}%
\providecommand \citenamefont [1]{#1}%
\providecommand \href@noop [0]{\@secondoftwo}%
\providecommand \href [0]{\begingroup \@sanitize@url \@href}%
\providecommand \@href[1]{\@@startlink{#1}\@@href}%
\providecommand \@@href[1]{\endgroup#1\@@endlink}%
\providecommand \@sanitize@url [0]{\catcode `\\12\catcode `\$12\catcode
  `\&12\catcode `\#12\catcode `\^12\catcode `\_12\catcode `\%12\relax}%
\providecommand \@@startlink[1]{}%
\providecommand \@@endlink[0]{}%
\providecommand \url  [0]{\begingroup\@sanitize@url \@url }%
\providecommand \@url [1]{\endgroup\@href {#1}{\urlprefix }}%
\providecommand \urlprefix  [0]{URL }%
\providecommand \Eprint [0]{\href }%
\providecommand \doibase [0]{http://dx.doi.org/}%
\providecommand \selectlanguage [0]{\@gobble}%
\providecommand \bibinfo  [0]{\@secondoftwo}%
\providecommand \bibfield  [0]{\@secondoftwo}%
\providecommand \translation [1]{[#1]}%
\providecommand \BibitemOpen [0]{}%
\providecommand \bibitemStop [0]{}%
\providecommand \bibitemNoStop [0]{.\EOS\space}%
\providecommand \EOS [0]{\spacefactor3000\relax}%
\providecommand \BibitemShut  [1]{\csname bibitem#1\endcsname}%
\let\auto@bib@innerbib\@empty
\end{thebibliography}%


\newpage

\section{References}

\begin{thebibliography}{9}
		
		\bibitem{FRGbook1} P.~Kopietz, L.~Bartosch, and F.~Sch\"{u}tz, {\it
			Introduction to the Functional Renormalization Group}, Springer, 2010.
            
		\bibitem{FRGbook2} A.~Wipf, {\it Statistical Approach to Quantum Field Theory}, Springer 2013.
        
	\bibitem{FRGbook3}R.~Percacci, {\it An introduction to covariant Quantum Gravity and 
			Asymptotic Safety}, World Scientific, 2017.   
		
		
		\bibitem{DupuisMachado}
		T. Machado and N. Dupuis, From local to critical fluctuations in 
		lattice models: a nonperturbative renormalization group approach, 
		Phys. Rev. {\bf E82}, 041128 (2010).
		
		\bibitem{Wortis} M.~Wortis, Linked Cluster Expansions, in: {\it Phase Transitions and 
			Critical Phenomena}, Vol 3, eds.~C.~Domb and M.~Green, Academic Press, 1974. 
		
		\bibitem{ReiszLRH} A.~Pordt and T.~Reisz, Linked cluster expansions beyond nearest 
		neighbor interactions: convergence and graph classes, 
		Int.\ J.\ Mod.\ Phys.\ {\bf A12} (1997) 3739. 
		
		\bibitem{Clusterbook1} J.~Oitmaa, C.~Hamer, and W.~Zheng,
		{\it Series Expansion Methods for strongly interacting 
			lattice models}, Cambridge UP, 2006.
            
		\bibitem{Clusterbook2}C.~Itsykson and J.~Drouffe, {\it Statistical Field Theory},
		Vol.~2, Cambridge UP, 1989. 
		
		\bibitem{BK} D.~Brydges and T.~Kennedy, Mayer expansions 
		and the Hamilton-Jacobi equation, J.\ Stat.\ Phys.\ {\bf 481} (1987) 19.
		
		\bibitem{multisets1} J.~Engbers, D.~Galvin, and C.~Smyth, Restricted Stirling and Lah number matrices and their inverses, arXiv: 1610.05803.
		
		\bibitem{Jackson} D.~Jackson, A.~Kempf, and A.~Morales, 
		A robust generalization of the Legendre transform for QFT,
		J.\ Phys.\ (Math.\ Theor.) {\bf A50} (2017) 225201. 
		
		\bibitem{MO} A.~Mestre and R.~Oeckl, Generating loop graphs via Hopf algebra in quantum field theory, J.\ Math.\ Phys.\ {\bf 47} (2006) 122302.
		
		\bibitem{BP} C.~Brouder and F.~Patras, One particle irreducibility with initial conditions, in: Contemporary Math.\ {\bf 539} (Combinatorics and Physics) (2011) 1-25.
		
		\bibitem{Sokal} A.~Sokal, A ridiculously simple and explicit implicit 
		function theorem, S\'{e}minaire Lotharingien de Combinatoire {\bf 61 A}
		(2009) B61Ad.  	
		
		\bibitem{GammIsing1} A.~Vasilev and A.~Radzhabov, Legendre transforms in 
		the Ising model, Theor.\ Math.\ Phys. {\bf 21} (1974) 963.
        
		\bibitem{GammIsing2}A.~Vasilev and A.~Radzhabov, Analysis of nonstar graphs of the Legendre transform 
		in the Ising model, Theor.\ Math.\ Phys. {\bf 23} (1975) 575. 
		
		\bibitem{BakerKin} G.~Baker and J.~Kincaid, The continuous-spin Ising model, 
		$g_0 :\phi_0^4:_d$ field theory, and the renormalization group, 
		J.\ Stat.\ Phys.\ 
		{\bf 24} (1981) 469.  
		
		\bibitem{GamFeyn} M.~de la Plata and L.~Salcedo, Feynman 
		diagrams with the effective action, J.\ Phys.\ {\bf A31} (1998) 4021.   
		
		\bibitem{BrydgesLeroux} D.~Brydges and P.~Leroux, Note on Legendre 
		transform and line-irreducible graphs, preprint, 2003. 
		
		\bibitem{Faris} W.~Faris, Combinatorial Species and Feynman 
		diagrams, S\'{e}minaire Lotharingien de Combinatoire 
		{\bf 61A} (2011) B61An. 
		
		\bibitem{Viral} S.~Jansen, S.~Tate, D.~Tsagkarogiannis, 
		and D.~Ueltschi, Multispecies Viral Expansions, Commun.\ Math.\ Phys.\
		{\bf 330} (2014) 801.   
		
		
	

\end{thebibliography}

\end{document}